\documentclass{aa}  

\usepackage{color,xcolor}
\usepackage[normalem]{ulem}
\newcommand{\gaia}{{\it Gaia}~}
\usepackage{graphicx}
\usepackage{pdflscape}
\usepackage{float}
\usepackage{placeins}    
\usepackage{txfonts}
\usepackage{hyperref}
\begin{document} 

   \title{The Kinematics of 30 Milky Way Globular Clusters and the Multiple Stellar Populations within}

   \author{E. I. Leitinger \inst{1,2,3}
          \and
          H. Baumgardt \inst{3}
          \and
          I. Cabrera-Ziri \inst{4}
          \and
          M. Hilker \inst{1}
          \and
          J. Carbajo-Hijarrubia \inst{5,6,7}
          \and
          M. Gieles \inst{5,7,8}
          \and
          T.O. Husser \inst{9}
          \and 
          S. Kamann \inst{10}
          }
          
    \institute{European Southern Observatory, Karl-Schwarzschild-Str. 2, 85748 Garching, Germany
       \and
       Dipartimento di Fisica e Astronomia, Universit{\'a} degli Studi di Bologna, Via Gobetti 93/2, I-40129 Bologna, Italy
        \and
        School of Mathematics and Physics, The University of Queensland,        St. Lucia, QLD, 4072, Australia
        \and 
        Astronomisches Rechen-Institut, Zentrum f\"ur Astronomie der Universit\"at Heidelberg, M\"onchhofstra{\ss}e 12-14, D-69120 Heidelberg, Germany
        \and
        Institut de Ciències del Cosmos (ICCUB), Universitat de Barcelona (UB), Martí i Franquès, 1, 08028 Barcelona, Spain
        \and
        Departament de Física Quàntica i Astrofísica (FQA), Universitat de Barcelona (UB), Martí i Franquès, 1, 08028 Barcelona, Spain
        \and
        Institut d’Estudis Espacials de Catalunya (IEEC), Gran Capità, 2-4, 08034 Barcelona, Spain
        \and
        ICREA, Pg. Lluís Companys 23, E08010 Barcelona, Spain
        \and
        Institut für Astrophysik und Geophysik, Georg-August-Universität Göttingen, Friedrich-Hund-Platz 1, 37077 Göttingen, Germany
        \and
        Astrophysics Research Institute, Liverpool John Moores University, 146 Brownlow Hill, Liverpool L3 5RF, UK
        }

   \date{Received XXXX; accepted XXXX}

% \abstract{}{}{}{}{} 
% 5 {} token are mandatory
 
\abstract
{}
{The spectroscopic and photometric classification of multiple stellar populations (MPs) in Galactic globular clusters (GCs) has enabled comparisons between contemporary observations and formation theories regarding the initial spatial configurations of the MPs. However, the kinematics of these MPs is an aspect that requires more attention. We investigated the 3D kinematics of 30 Galactic GCs, extending to 3-5 half-light radii, as well as their MPs, in order to uncover clues of the initial conditions of GCs and the MPs within.}
% Methods
{We have combined Hubble Space Telescope and \gaia DR3 proper motions together with a comprehensive set of line-of-sight velocities to determine the 3D rotation amplitudes, rotation axes, and anisotropy profiles of the clusters. We include additional radial velocities from new IFU observations of NGC 5024 and an analysis of archival MUSE data of NGC 6101. We compare our kinematic results with structural and orbital parameters of each cluster, reporting the most significant correlations and common features.}
% Results
{We find significant ($>3\sigma$) rotation in 21 GCs, with no significant differences between the total rotational amplitudes of the MPs, except for NGC 104. We find no significant differences in the position angles of the rotation axis or inclination angles. We find that the 3D rotational amplitude of the clusters in our sample is strongly correlated with their mass, relaxation time, enriched star fraction and concentration. We determine the anisotropy profiles of each cluster and the MPs where possible. We investigate correlations with the structural parameters, orbital parameters and accretion history of the clusters from their progenitor systems, finding that the dynamically young clusters with the highest central concentrations of primordial stars exhibit radial anisotropy in their outer regions ($>2$ half-light radii). The dynamically young clusters with a central concentration of enriched stars show significant tangential anisotropy or isotropy in their outer regions.}
{}

   \maketitle
%
%-------------------------------------------------------------------

\section{Introduction}
A well established feature of globular clusters (GCs) is that the vast majority host multiple stellar populations (MPs), which are identifiable due to differences in the light-element abundances between stars, shown both in spectroscopy and photometry (see reviews by \citealt{2018Bastian} and \citealt{2022Milone}). In general, the MPs consist of a `primordial' population of stars (`P1' stars) with light-element abundance patterns similar to surrounding field stars of the same metallicity. There is also at least one `enriched' population of stars (`P2' stars), which are commonly enriched in elements such as N, Na and Al, but depleted in C, O and sometimes Mg, in comparison to P1 stars. It is currently unknown which initial conditions are responsible for the birth of these MPs, which occurs during the early stages of GC formation. Many formation theories suggest a generational gap between the populations, in which P1 stars were born first, while P2 stars were created from enriched material at some stage afterwards (i.e. the AGB scenario \citep{1981cottrell,1983Dantona,2001Ventura}, fast rotating massive stars (FRMS) \citep{2007decressin,2007charbonnel} super-massive stars (SMS) \citep{2014denissenkov,2018gieles} or massive interacting binaries \citep{2009demink,2022renzini}). Different formation mechanisms imply different kinematic imprints on the MPs, but as most of the GCs that exist today within our Milky Way are very old ($>11$ Gyr), long-term dynamical evolution will most likely have erased traces of the initial conditions of their birth. However, GCs that have not undergone significant dynamical evolution during their lifetime can be described as 'dynamically young' and are expected to retain their initial conditions, particularly in the outer regions \citep{2013Vesperini}. In the absence of direct observations of MPs as they form in high redshift galaxies, young massive clusters and dynamically young GCs serve as the next best proxies for studying the initial conditions of the formation of MPs.\\

The dynamics of GCs can be separated into two aspects: a wider view of the overall motion of GCs within the host galaxy and a closer view of the internal dynamics as determined by the interactions of individual stars within the cluster. Within the host galaxy, the orbital information of GCs can provide insight into whether specific GCs formed in-situ or were accreted into the host galaxy \citep{2019Massari,2024Belokurov}, with dynamical modelling providing insight into the effect of tidal interactions on the clusters \citep{2017Sollima}. Within the cluster itself, gravitational interactions influence the motions of the stars and two-body relaxation causes the cluster to become isotropic over time. Each cluster can also exhibit varying levels of rotation, in which the stars exhibit ordered motions. This present-day rotation can also be indicative of the rotation of the cluster at birth, depending on the degree of dynamical evolution the cluster has undergone within its lifetime \citep{2020Sollima}. Long-term dynamical evolution within GCs causes the initial orbital properties of the multiple stellar populations to mix in phase space \citep{2008Decressin,2013Vesperini,2015HenaultGieles}. The two-body relaxation timescale is inversely proportional to the local density, so since the density of a cluster increases towards the centre \citep{1987Spitzer}, the relaxation timescale will be longer in the outer regions. This means that the initial configurations in phase-space will become mixed on a faster timescale in the central regions, whereas the outer regions are more likely to be representative of the initial configurations. It is therefore crucial to extend any kinematic analysis into the outer regions of GCs in order to obtain the best chance of observing any remaining initial conditions.\\

Some formation theories imply that MPs within GCs have different ages, initial radial configurations or kinematics. Studying the rotation of stars within a GC, and in particular any differences in the rotational signatures of the multiple populations, can help to distinguish between the possible formation scenarios. For formation theories which assume 1) that P1 stars are located primarily in the outer regions and 2) that $\sim$ 90\% of P1 stars are removed from a cluster within its lifetime (i.e. as expected in the AGB scenario \citep{2007decressin,2008Dercole,2011Schaerer,2015CabreraZiri}), the result is that a large amount of angular momentum would be removed with them. In this case, one would expect P1 stars (if they are predominantly in the outer regions) to have lower angular momentum \citep{2015HenaultGieles} than the enriched stars. If P2 stars form with high angular momentum in the core of a cluster with low angular momentum, these P2 stars are predicted to have higher initial rotational amplitudes than the P1 stars located further out, due to the conservation of angular momentum. Observationally, \cite{2024Dalessandro} found this to be the case after analysing the 3D rotation profiles of 16 Galactic GCs, concluding that P2 stars rotate faster than P1 stars for the majority of clusters in their sample. \cite{2017Cordero} also discovered the same result for NGC 6205 after analysing the spectra of 111 giant stars. They found that the P2 stars in their sample have higher rotational amplitudes than the P1 stars, and found differences of 0-45 degrees between the spin axes of the populations. In addition, P2 stars were found to rotate faster than P1 stars in NGC 5272 \citep{2021Szigeti}, NGC 6093 \citep{2020kamann} and NGC 6362 \citep{2021Dalessandro}. Similarly, \cite{2023Martens} analysed the rotational differences between populations in 25 Milky Way GCs using a combination of HST photometry and MUSE spectroscopy and found that only two GCs in their sample had P2 stars rotating faster than P1 stars; however, they also discovered one GC in their sample that showed the opposite. This was also observed by \cite{2018Bellini} using HST photometry of $\omega$ Centauri, where they discovered P1 stars with higher rotational amplitudes than P2 stars. A possible explanation for this is that if P2 stars - regardless of where they formed in the cluster - initially passed through the cluster core, they would be on radial orbits with low angular momentum and exhibit lower rotational amplitudes \citep{2015Hentault}. In many GCs, no differences in the rotational amplitudes of their MPs have been observed, such as the clusters NGC 104 \citep{2018Milone,2020Cordoni}, NGC 6121 and NGC 6838 \citep{2020Cordoni}, NGC 6352 \citep{2019Libralato}, as well as NGC 7078 \citep{2021Szigeti}. Many formation theories predict differences between the rotational amplitudes of the MPs, but the absence of a difference has not been explicitly predicted by any formation theories besides the FRMS scenario \citep{2013Krause}, which predicts that P1 and P2 stars should share the same kinematics, since P2 stars are formed within the decretion disks of massive P1 stars.\\

Theoretically, anisotropy profiles should also provide insight into the possible initial conditions and evolution of GCs \citep{2021Pavlik,2022Pavlik,2024Pavlik}. It is predicted that if a GC is born with tangential anisotropy, the relaxation processes are accelerated in comparison to GCs that were born isotropically or with radial anisotropy. It is also expected that the outer regions of a GC have a higher chance of retaining any initial anisotropy of a GC, with radial anisotropy expected for GCs that are tidally underfilled and tangential anisotropy expected for tidally filled GCs \citep{2016Zocchi,2024Pavlik}. Observationally, the outer regions of dynamically young GCs should therefore be ideal for determining clues about the initial configuration.\\
The anisotropy of MPs are also expected to provide insight into the kinematic evolution and potential initial conditions of their GCs according to \textit{N}-body simulations \citep{2019Tiongco,2021Vesperini}. Theoretically, if P2 stars are more centrally concentrated and diffuse in the outer regions, then they should exhibit radial anisotropy in the outskirts, which has been observed in a handful of MW GCs \citep{2013Richer,2015Bellini,2018Milone,2018Libralato}. This is especially expected for clusters which begin compact in comparison to their tidal radii, as the expansion of the cluster forces stars into strongly radial orbits in the outer regions, until many of these highly radial stars escape the cluster \citep{2016Tiongco,2019Tiongco}. On the other hand, tangential anisotropy or isotropy is expected for clusters which have already filled their Roche lobe during formation and therefore have not expanded significantly in their lifetime.\\
The present-day signature of radial anisotropy in the outer regions is not specific to any particular formation theory, so it cannot help to narrow down the possible options. All generational scenarios predict that P2 stars are centrally concentrated during their formation, leading them to be scattered on wider radial orbits as the cluster evolves \citep{2015HenaultGieles}. In fact, in any scenario where the gas cools and collects in the centre of the cluster and P2 stars are formed, \textit{or} the P2 stars are the first to traverse the cluster core on radial orbits, the resulting anisotropy profiles of both populations will end up being the same in the present day according to simulations by \cite{2015HenaultGieles}. \\

The rotational orientation of GCs and their MPs can also be useful for comparisons between observations and formation theories. For the SMS scenario described in \cite{2018gieles}, it is theorised that P2 stars are formed from the winds of an SMS and therefore are not expected to show coherent motion relative to the P1 stars, nor are their rotational orientations expected to match those of the P1 stars. This is because the SMS is formed by stellar collisions, so its angular momentum increases through a random walk, causing the SMS to exhibit a random spin direction relative to the pre-existing P1 stars. Although the rotational velocity of the SMS can reach the order of $\sim$ 100 km/s, this velocity is not inherited by the P2 stars, which instead formed from the stellar winds of the SMS, with the rotational velocity of the SMS becoming negligible once the wind reaches $\sim$ 1 pc. Using 3D hydrodynamical simulations that follow the formation process of the AGB scenario, \cite{2021Mckenzie} found that after formation, P2 stars will rotate faster than P1 stars and exhibit highly coherent rotation in a disc-like structure parallel to the plane of the parent galaxy. Similarly, the 3D hydrodynamical simulations of \cite{2022Lacchin}, also based on the AGB formation scenario, result in P2 stars rotating faster than P1 stars, with both sets of simulations showing that the P1 and P2 stars rotate in phase with each other (i.e. their rotation peaks are aligned). However, these simulations differ in terms of the spatial concentration of the MPs after formation. At the end of the 370 Myr timeframe of the \cite{2021Mckenzie} simulation, P1 stars are found to be centrally concentrated within the cluster, while \cite{2022Lacchin} found the opposite. Current simulations have yet to accurately match observations of MPs in terms of both the present day spatial distributions and kinematic signatures.\\

In this paper, we have homogeneously analysed 30 Galactic GCs and explored the differences in the dynamics of each cluster, as well as their MPs. Unveiling clues about the formation history of these GCs is a task best approached by investigating a large, diverse sample of resolved GCs, as the observable trends will have higher significance than if the focus is to thoroughly investigate one or two clusters. The GCs in our sample are diverse in terms of the range of masses, dynamical ages, metallicities and locations within the halo of the Milky Way, with every GC in our sample found to contain multiple stellar populations.

%--------------------------------------------------------------------
\section{Observational Data}
\label{sec:data}
\begin{figure}
    \centering
    \includegraphics[width=\hsize]{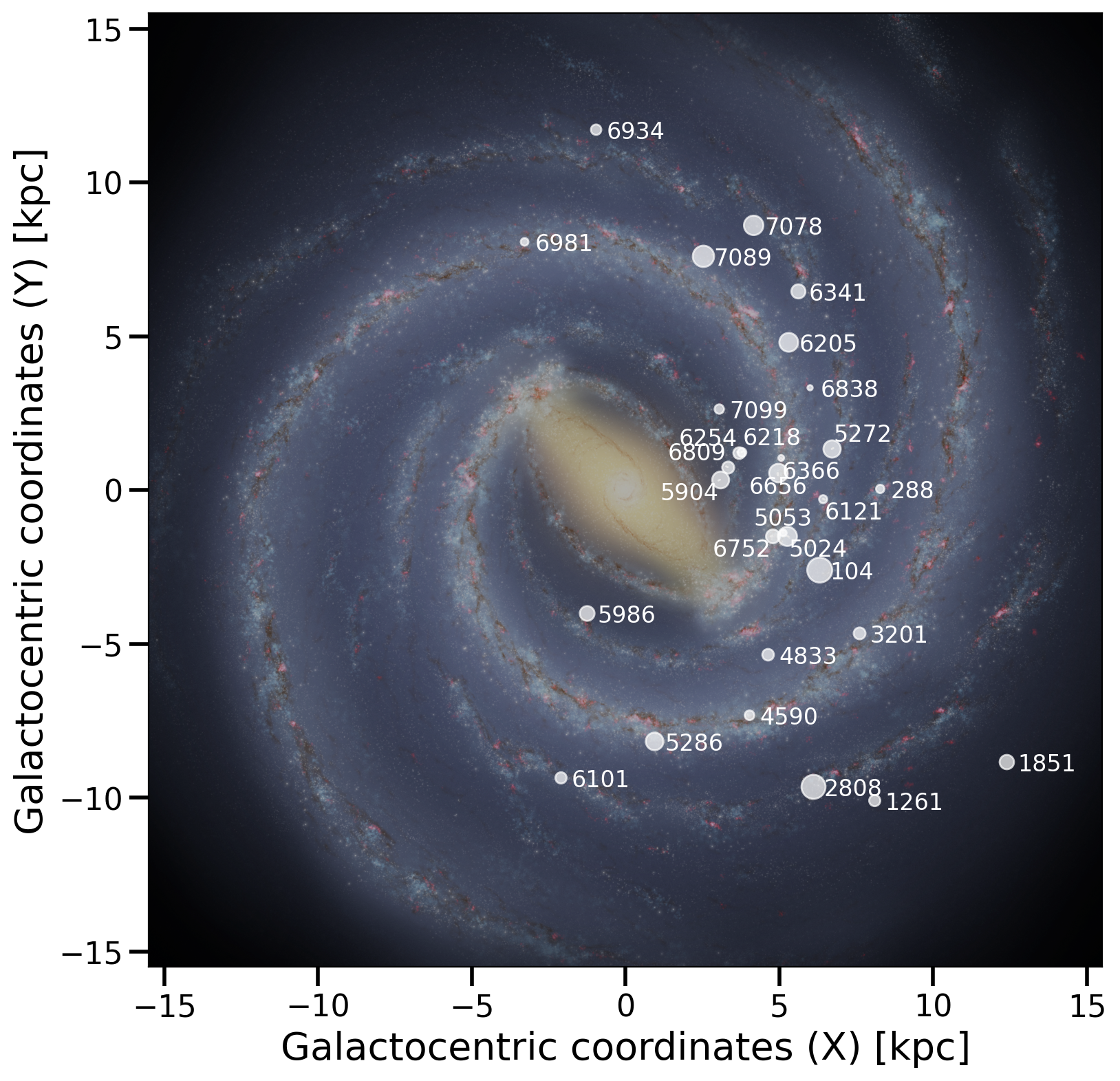}
    \vspace{-0.4cm}
    \caption{The sample of 30 Galactic globular cluster in galactocentric coordinates plotted over an artist's conception image of the Milky Way (R. Hurt: NASA/JPL-Caltech/SSC) using {\tt mw-plot}. The size of the points are proportional to the mass of the clusters.}
    \label{fig:positions_of_GCs}
\end{figure}

We investigated 30 Galactic globular clusters with distances from the Sun between $1.85 \pm 0.02 \leq R_\odot \rm{[kpc]} \leq 18.5 \pm  0.18$ \citep{2021BaumgardtVasiliev}, masses between $(3.78 \pm 0.14) \cdot 10^4 \leq M_\odot \leq (8.53 \pm 0.05) \cdot 10^5$ \citep{2018BaumgardtHilker}\footnote{The exact values used for the global parameters first derived in \cite{2018BaumgardtHilker} were taken from the Galactic Globular Cluster Database: \url{https://people.smp.uq.edu.au/HolgerBaumgardt/globular/}, which contains updates of their initially published values.} and metallicities between $ -2.37 \leq \rm{[Fe/H]} \leq -0.59$ \citep{2010Harris}. Our GC sample is listed in Table \ref{tab:kinematics} and contains the 28 GCs for which we identified multiple stellar populations in \cite{2023Leitinger}, plus the addition of NGC 104 and NGC 6656 as discussed in \ref{sec:photometry_data}. The positions of our clusters imposed on an artist's conception image of the Milky Way are shown in Figure \ref{fig:positions_of_GCs}. The cluster parameters such as position, distance from the Galactic centre, proper motion of the cluster, etc. used throughout this paper were taken from the Galactic Globular Cluster Database \cite{GGCD}, unless specified otherwise in the text.\\

\subsection{Photometric catalogues}
\label{sec:photometry_data}

In our first paper \citep{2023Leitinger}, we used space-based HST \citep{2015Piotto,2018Nardiello} and ground-based \citep{2019Stetson} photometric catalogues to obtain a wide-field view of 28 clusters and classified their multiple stellar populations using RGB stars. The normalised, cumulative radial distributions of the P1 and P2 stars were then calculated in order to determine which population was more centrally concentrated in the cluster. Quantitatively, this was expressed by the $A^+$ parameter \citep{2016Alessandrini,2019Dalessandro}, with negative(positive) values indicating the P2(P1) stars were more centrally concentrated, or a value close to zero indicating the populations were spatially mixed. Using the same method, we added NGC 104 and NGC 6656 to the sample and determined the $A^+$ values and enriched star fractions $\rm{P_2 / P_{tot}}$, which are shown in Table \ref{tab:kinematics}. Figures 15 and 16 of \cite{2023Leitinger} showed the combined $A^+$ values as a function of both the dynamical age (age/relaxation time) and mass loss ratio ($\rm{M_{current}/M_{initial}}$), so an updated version of these plots is described and included in Appendix \ref{Appendix_updated_Aplus}, in order to reflect the addition of NGC 104 and NGC 6656 and any changes to the global structural parameters of our sample. \\

\subsection{Proper motions}
\label{sec:pm_data}
In order to determine the kinematics of the globular clusters in the plane of the sky, we used proper motions from \gaia DR3 \citep{2016gaia1,2022gaia2} which cover mainly the outer regions of each cluster up to the tidal radius, combined with proper motions from the Hubble Space Telescope Atlases of Cluster Kinematics (HACKS) \citep{2022Libralato} which covers the central regions of the clusters.\\

For each globular cluster, we searched for stars in \gaia DR3 out to the tidal radius ($r_{\rm{t}}$). The list of candidate members found in \gaia DR3 was then cleaned of non-members following the method described in Section 2 of \cite{2021Vasiliev}, as well as proper motion cleaning using the $\chi^2$ test described by Equation 3 from \cite{2023Leitinger}. By using the \gaia BP-RP and G magnitudes, we isolated stars in the resulting CMD which were one magnitude fainter than the turn-off ($\rm{G_{RGBturn-off}}$) and performed $N-\sigma$ clipping to keep only stars which belonged to the RGB, AGB, HB and MS-TO of each cluster. This created a cleaned catalogue of \gaia DR3 proper motions, which we used for the remainder of the analysis.

 \cite{2021Vasiliev} found that the proper motion uncertainties given in the \gaia DR3 catalogue \citep{2016gaia1,2022gaia2} are underestimated by 10-20$\%$, particularly in the dense central regions of globular clusters. To correct for this, we first calculated the number density ($\Sigma$) of all stars and followed the method described in \cite{2021Vasiliev}, applying their Equation 3 to the original \gaia proper motion errors in order to correct this underestimation. We selected the value of $\epsilon_{\bar{\mu} \rm{,sys}}\simeq 0-0.02$ mas for each cluster based on Figure 2 of \cite{2021Vasiliev}. Throughout the paper we used these externally calibrated proper motion errors $\epsilon_{\bar{\mu} \rm{,ext}}$ rather than the errors given in the \gaia catalogue.\\

In the HACKS astro-photometric catalogue \citep{2022Libralato}, we removed stars with unreliable photometry and proper motions before using them in our kinematic analysis. To do this, we selected well-measured objects in the HACKS photometric catalogues for each cluster in our sample, following the method described in points (i) to (vi) in Section 4 of \cite{2022Libralato}. We then selected reliable stars in the corresponding HACKS proper motion catalogues by removing all objects with $N_u^{\rm{PM}} / N_f^{\rm{PM}} < 0.8$ to $0.9$ (with the value dependent on each cluster), $\chi^2_{\mu_\alpha \rm{cos} \delta}, \chi^2_{\mu_\delta} > 1.25$ to $1.5$ and proper motion errors $> 0.5$ mas $\rm{yr^{-1}}$, as also detailed by \cite{2022Libralato}.

For each cluster, stars with reliable photometry were matched with the corresponding cleaned proper motion catalogue in order to create CMDs using the combinations of available $F606W$ and $F814W$ filters. The resulting $F606W$ - $F814W$ vs. $F606W$ CMDs served as a means to isolate the RGB, AGB, HB and MS-TO stars for the kinematic analysis. We implemented a cut to remove stars fainter than one magnitude below $\rm{F606W}_{\rm{RGB turn-off}}$, as well as removing blue straggler stars from the final sample, so as not to affect the velocity dispersions of the remaining stars. We also removed stars located too far from the RGB, AGB, HB and MS-TO using iterative $N-\sigma$ clipping along the main distribution of the stars.

The proper motions included in the HACKS catalogue are reported in a relative reference system in which the motion of a star is measured relative to the surrounding stars, unlike the absolute reference system used by the \gaia DR3 catalogue. The rotation of the clusters can therefore not be observed in HACKS, and it was not possible to perform a rotational analysis using the HACKS data. However, we did use the HACKS proper motions to calculate the velocity dispersion and anisotropy profiles of each cluster in Section \ref{sec:vdisp} and Section \ref{sec:anisotropy}.

\subsection{Line-of-sight velocities}
\label{sec:los_data}
We obtained line-of-sight (LOS) velocities from the homogeneous catalogue created by \cite{2017Baumgardt} and \cite{2018BaumgardtHilker}, which included velocities derived from ESO FLAMES/UVES, Keck HIRES/DEIMOS spectra and published literature values. We also used the additional line-of-sight velocities of globular cluster stars which \cite{2023Baumgardt} compiled from the APOGEE DR17 \citep{2022APOGEE}, \gaia DR3 \citep{2016gaia1,2022gaia2}, Galah DR3 \citep{2021Buder}, LAMOST DR8 \citep{2012Cui}, MIKIS \citep{2018Mikis} and RAVE DR5 \citep{2017Rave} surveys. In total, this data set contains about 37,000 LOS velocity measurements for about 22,000 unique cluster stars in the 30 studied clusters, with an average of 1.7 LOS measurements per star. Additionally, MUSE LOS velocities from \cite{2023Martens} were used for stars in the inner regions of the clusters. We removed stars with line-of-sight velocity errors $> 5$ km/s, and averaged the remaining velocities and associated errors for each star. The distribution of LOS velocities as a function of projected distance from the centre of each cluster was then fitted with a linear function and cleaned of $>3\sigma$ outliers using an iterative process.

We began with the same sample of GCs as included in \cite{2023Leitinger}, with the addition of NGC 104 and NGC 6656. However, from this combined sample of GCs, 3 GCs did not have a sufficient ($>$100) number of stars in the LOS velocity catalogues to perform the rotation analysis described in Section \ref{sec:rotation_axes} and were therefore not included in the multiple stellar population analysis of Section \ref{sec:rotation_axes_MP}. These clusters were NGC 5053 ($\rm{N_{LOS}} = 70$), NGC 6934 ($\rm{N_{LOS}} = 77$) and NGC 6981 ($\rm{N_{LOS}} = 65$), where the quoted $\rm{N_{LOS}}$ of each cluster refers to the full sample of stars with available LOS velocities \textit{before} any cleaning was applied. Despite having insufficient LOS velocities to determine rotation, the number of proper motion measurements allowed us to include these clusters in the velocity dispersion and anisotropy analysis of Sections \ref{sec:vdisp} through to  \ref{sec:vdisp_anisotropy_MPs}.\\

We added archival MUSE observations of NGC 6101 from proposal ID 099.D-0824 (PI: Peuten), covering the inner $\sim$ $3\times3$ arcmin of the cluster with eight pointings of the MUSE Wide Field Mode. The data were reduced with the standard MUSE pipeline \citep{2020Weilbacher}, resulting in a single reduced data cube for each of the  pointings. Each cube contained the signal from four exposures of 600~s each, and small spatial dithers and derotator offsets of 90~degrees between them. Individual stellar spectra were extracted from the cubes using \textsc{PampelMuse} \citep{2013Kamann}. The code uses an astrometric reference catalogue in order to determine the positions of resolvable stars and the point-spread function in the integral field data as a function of wavelength. This information is then used to optimally extract the spectra of the resolvable stars while accounting for blends between adjacent stars. The photometric reference catalogue required by \textsc{PampelMuse} was taken from the ACS survey of Galactic globular clusters \citep{2007Sarajedini,2008Anderson}. The extracted spectra were processed with \textsc{Spexxy} \citep{2016Husser}, which performed a full-spectrum fit of each input spectrum against the \textsc{Glib} \citep{2013Husser} synthetic spectral library. The required initial guesses for the surface gravity and effective temperature of each star were obtained by comparing the aforementioned photometry to an isochrone from the database of \citet{2012Bressan}, selected to match the metallicity of NGC~6101. Initial guesses for the line-of-sight velocity of each spectrum were obtained by cross-correlating it with a \textsc{GLib} template of matching stellar parameters. In total, the \textsc{Spexxy} fits yielded 427 velocity measurements that passed our quality cuts.\\

We included new observations of the central part of the cluster NGC 5024. We used a mosaic of 8 pointings with the MEGARA@GTC IFU. The data were reduced using the MEGARA Data Reduction Pipeline \citep[MDRP,][]{Pascual2022}. The process began by subtracting the \textit{bias} level present in the images. To achieve this, we utilized the \textit{bias} calibration and the {\tt MegaraBiasImage} task. The \textit{bias} level varies slightly between the top and bottom of the image (by approximately 100 counts) due to the MEGARA CCD (an E2V CCD231-84) being read out through two diagonally opposite amplifiers. At this stage, defective pixels were automatically masked using the {\tt master\_bpm.fits} file provided by the MDRP. Next, the spectra were processed using the {\tt Trace} and {\tt ModelMap} tasks, which trace the fibre spectra from the halogen lamp images. The MDRP was then used to perform wavelength calibration using ThNe lamp images. Flat-field correction was carried out using the halogen lamp images. For each calibrated exposure, we computed the median of the images within each observing block. Additionally, the MDRP subtracts the mean sky spectrum, derived from the dedicated sky fibres, to produce the fully reduced spectra.
The reduced MEGARA data of NGC~5024 consisted of raw-stacked spectra (RSS) files, with each one-dimensional spectrum containing the flux recorded by a single hexagonal fibre. Again, we extracted individual stellar spectra from the data using \textsc{PampelMuse} \citep{2013Kamann}. To avoid having to resample the MEGARA data to a rectangular grid, \textsc{PampelMuse} was modified such it accepts the native (hexagonal) MEGARA sampling as an input format. As basis for the reference catalogue, we used the F814W photometry of NGC~5024 from the HACKS survey \citep{2022Libralato}. As many bright stars are missing in the catalogue, likely because they were saturated in the underlying HST images, we complemented the catalogue with Gaia DR3.

After extraction of the spectra, we determined stellar radial velocities with the help of the IRAF\footnote{IRAF is distributed by the National Optical Astronomy Observatories, which are operated by the Association of Universities for Research
in Astronomy, Inc., under cooperative agreement with the National Science Foundation.} {\tt fxcor} task, using as templates the spectra of cool giant stars of a metallicity that is comparable to the cluster metallicities. We created the template spectra with the stellar synthesis program {\tt SPECTRUM} \citep{1994Gray} using {\tt ATLAS9} stellar model atmospheres \citep{2003Castelli}. For the later analysis, we only use stars with radial velocity errors $<3$ km/sec, magnitude errors $<0.5$ and signal-to-noise ratio S/N$>3$.

\section{Method}
\label{sec:rotation_outline}

Our approach was to first calculate the rotation and anisotropy profiles of the 30 Galactic globular clusters as a whole, using the full sample of RGB, AGB, HB and MS-TO stars available in the observational data described in Section \ref{sec:data}. In order to then investigate the differences in 3D kinematics between the multiple stellar populations of each cluster, we cross-match the cleaned, observational data for each cluster with the photometrically classified multiple stellar populations of RGB stars presented in \cite{2023Leitinger} and completed the same analysis again on this smaller subset.

We analysed the 3D rotation of stars in each cluster by combining the cleaned proper motions from \gaia DR3 with the LOS velocity catalogues, determining the rotation axes and 3D total rotational amplitudes for all RGB, AGB, HB and MS-TO stars available. Section \ref{sec:rotation_axes} describes the analysis of this large sample of stars, for each cluster. Section \ref{sec:rotation_axes_MP} then describes the analysis of the RGB subset of this sample, which were classified into multiple stellar populations.

Similarly, in order to analyse the anisotropy of each cluster, we combined the HACKS proper motions with \gaia DR3 proper motions for all RGB, AGB, HB and MS-TO stars available. Section \ref{sec:vdisp} describes the velocity dispersion analysis and Section \ref{sec:anisotropy} describes the anisotropy analysis for this large sample. Then, Section \ref{sec:vdisp_anisotropy_MPs} describes the analysis of the subset which was cross-matched with the photometric RGB sample of stars classified into multiple stellar populations.

\subsection{Rotation axes and total rotational amplitudes}
\label{sec:rotation_axes}

We began with all cleaned RGB, AGB, HB and MS-TO stars available in the \gaia DR3 and LOS catalogues for the clusters in our sample, except NGC 5053, NGC 6934 and NGC 6981, due to the low number of LOS velocities for those clusters.

In order to account for perspective rotation, which can add an artificial rotation signal to the data, we corrected the proper motions and LOS velocities. A quick estimate shows that for our sample, perspective rotation is only important for NGC 3201 due to its large LOS velocity of $R_V = 495.39 \pm 0.06$ km/sec and close proximity to the observer at $R_{\odot} = 4.74 \pm 0.04$ kpc. We therefore only correct the motion of stars in NGC 3201. The method for correcting the velocities is described in \cite{2006Vandeven} and \cite{2021Wan} using Equations \ref{eq:vandeven1} and \ref{eq:vandeven2} for the components of the proper motion and Equation \ref{eq:vandeven3} for the LOS velocities. In these equations, $\mu_{x'}^{sys}, \mu_{y'}^{sys}$ and $v_{z'}^{sys}$ are the systematic velocities, $x'$ and $y'$ are the projected coordinates as calculated by Equation 2 of \cite{2006Vandeven} and $D$ is the distance to the cluster. \\
\begin{eqnarray}
    \label{eq:vandeven1}
    \mu_{x'}^{\rm{pr}} &=& -6.1363 \times 10^{-5} x' v_{z'}^{sys} / D \quad \rm{mas/yr}\\
    \label{eq:vandeven2}
    \mu_{y'}^{\rm{pr}} &=& -6.1363 \times 10^{-5} y' v_{z'}^{sys} / D \quad \rm{mas/yr}\\
    \label{eq:vandeven3}
    v_{z'}^{\rm{pr}} &=& 1.3790 \times 10^{-3} (x' \mu_{x'}^{sys} + y' \mu_{y'}^{sys}) D \quad \rm{km/s}
\end{eqnarray}

\begin{figure}
    \centering
    \includegraphics[width=\hsize]{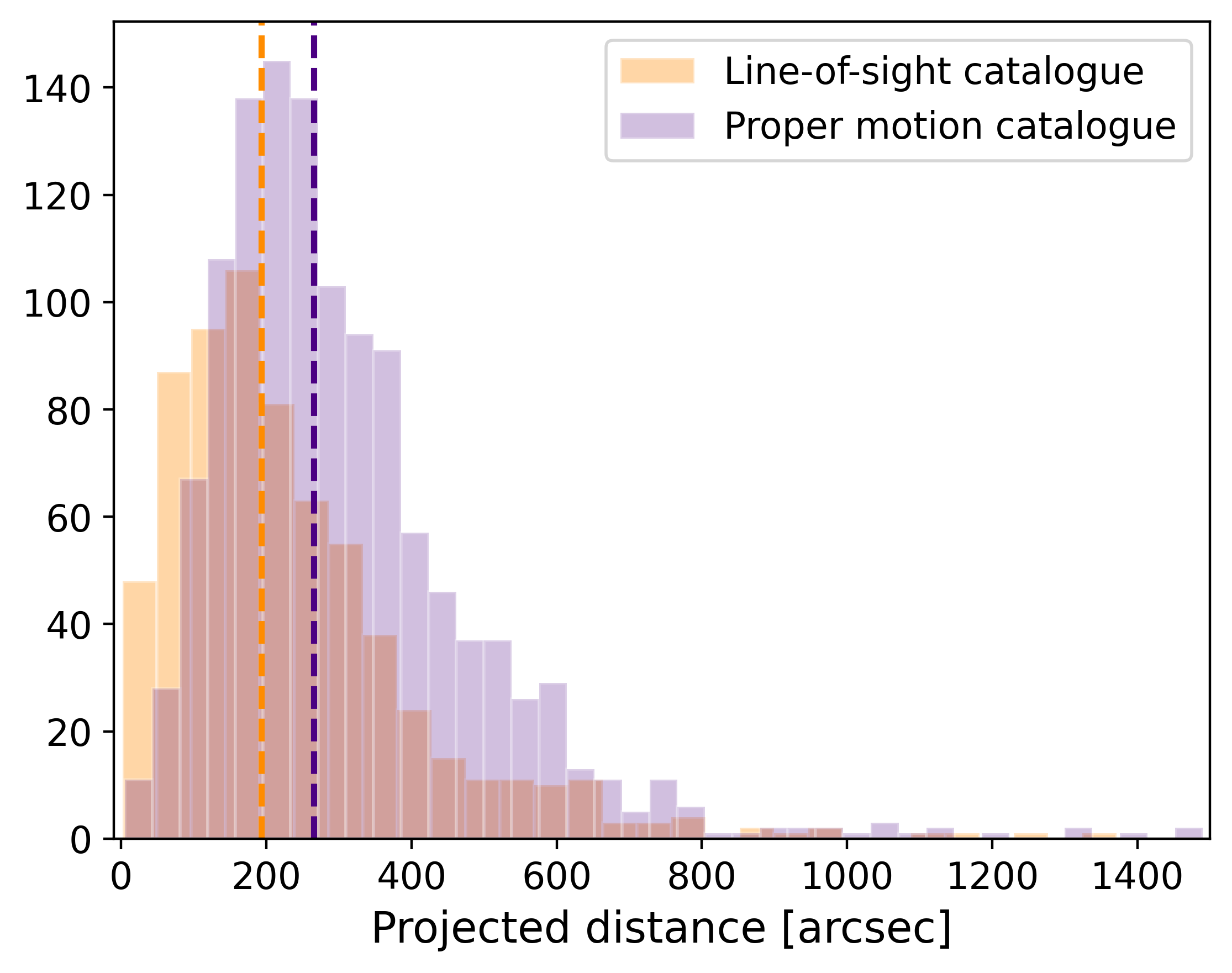}
    \vspace{-0.4cm}
    \caption{The number of stars with measured proper motions (indigo) and LOS velocities (orange) as a function of projected distance from the cluster centre for NGC 6809. The median values are shown by dashed lines.}
    \label{fig:PM_LOS_histogram}
\end{figure}

We compared the projected distances of stars with measured proper motions and radial velocities for all clusters, as shown in Figure \ref{fig:PM_LOS_histogram} for NGC 6809, to ensure the median distances are within $100''$ of each other for all clusters. This guarantees that both samples are taken at similar radii on average so that the proper motions and LOS velocities can directly be compared with each other. In this way, we avoided the requirement that each star be matched in both the proper motion and LOS catalogue, which would decrease the number of stars available for analysis. If the median distances of the stars with LOS measurements disagreed by more than $100''$ from that of the stars with PM measurements, we cross-matched stars from the MUSE catalogue with \gaia DR3, keeping only the stars which could be identified in \gaia DR3. This cross-match allowed the medians to align for all clusters except NGC 104, which instead required a cross-match between the proper motions and LOS catalogues in order to ensure we were observing rotation at comparable radii. Due to the large number of observations for stars in NGC 104, we were still left with $1409$ stars after cross-matching in this way.\\

When using \gaia data for globular clusters, it is good practice to adjust the celestial coordinates and proper motions to account for projection effects due to the curvature of the sky, which is especially necessary for large and nearby clusters. We used the standard orthographic projection shown in Equation 2 of \cite{2018Helmi} to transform the celestial coordinates and corresponding proper motions into Cartesian coordinates on a tangent plane. We then subtracted the average cluster proper motions ($\rm{\mu_{\alpha*,cluster}}$, $\rm{\mu_{\delta,cluster}}$) and LOS velocity ($\rm{v_{LOS,cluster}}$) using the values determined in \cite{2018BaumgardtHilker}. Using the transformed coordinates and proper motions, we calculated the radial ($\rm{v_{rad}}$) and tangential ($\rm{v_{tan}}$) components of the proper motions, such that:
\begin{align}
    \rm{v_{rad}} &= \frac{x.\mu_x + y.\mu_y}{r}\\
    \rm{v_{tan}} &= \frac{-y.\mu_x + x.\mu_y}{r},
\end{align}
for stars with positions $(x,y)$ relative to the cluster centre, proper motions $(\mu_x,\mu_y)$, and the projected distance of each star to the centre of the cluster $r$, as described by \cite{2000Vanleeuwen}.\\ 

We calculated the position angle of each star ($\theta_i$) using the projected distance of the star from the centre of the cluster ($x,y$), with values varying between $0 < \theta_i < 360$. In general, the radial component ($\rm{v_{rad}}$) as a function of position angle is expected to be close to zero, as a non-zero value would indicate the cluster is experiencing radial fluctuations, implying the cluster is `breathing'. Negative $\rm{v_{rad}}$ values indicate stars approaching the cluster centre, while positive values indicate stars moving away from the centre. The tangential component ($\rm{v_{tan}}$) can be negative, zero or positive, with negative values indicating clockwise rotation of the stars and positive values indicating counter-clockwise rotation. In a rotating cluster, the LOS component ($\rm{v_{LOS}}$) can display a sinusoidal rotation signal as a function of position angle, with the amplitude of this sinusoidal signal dependent on both the strength of the cluster rotation and the orientation of the cluster with regard to the observer.\\

\begin{figure}
    \centering
    \includegraphics[width=\hsize]{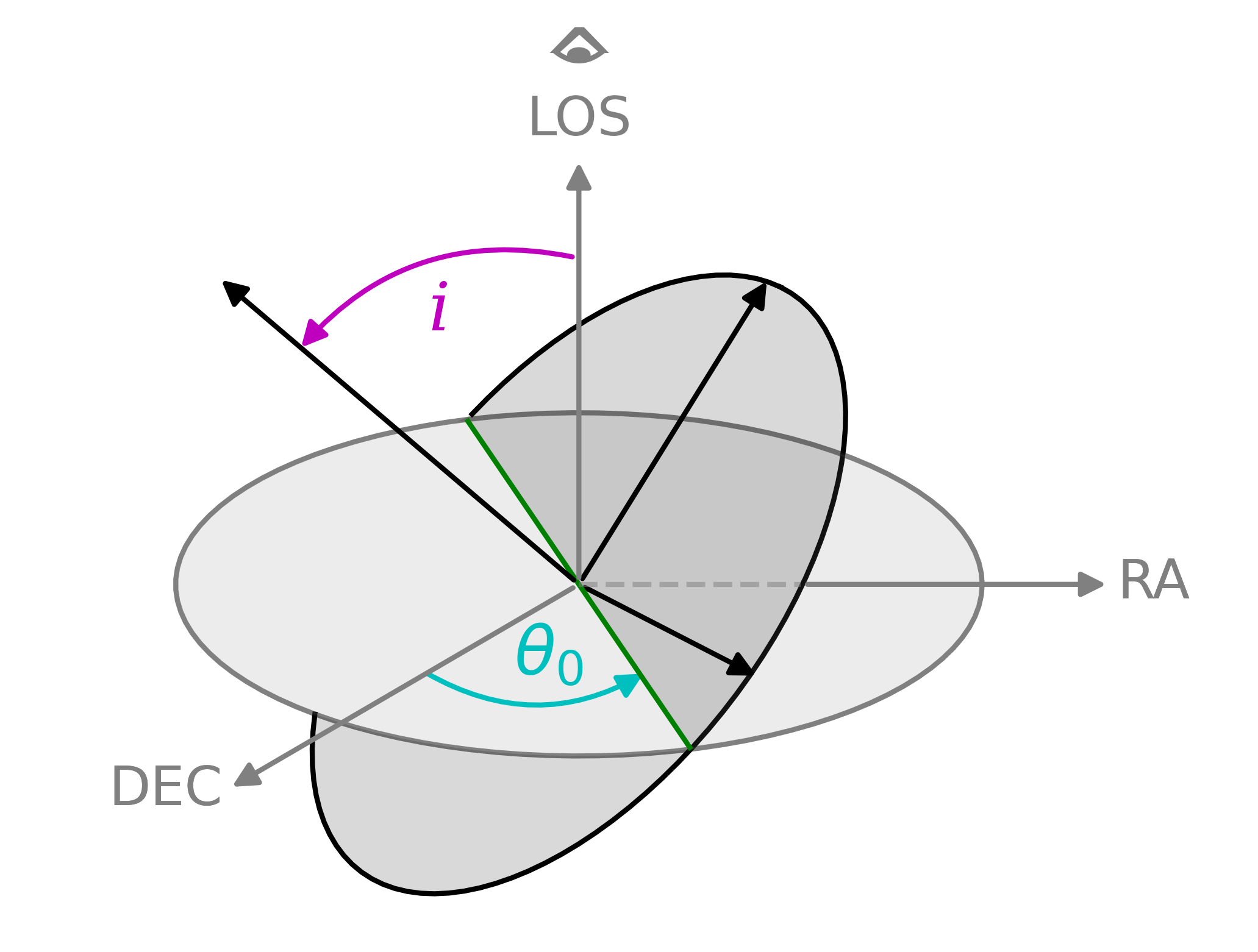}
    \vspace{-0.4cm}
    \caption{A schematic of the rotation axis angles for a globular cluster with an inclined plane. The light grey plane shows a face-on cluster with position vectors in the plane of the sky: RA, DEC and the line-of-sight (LOS) vector towards/away from the observer. In dark grey is an inclined plane, with projected position vectors and a rotation axis described by two angles with respect to the face-on cluster: position angle of the rotation axis $\theta_0$ and inclination angle $i$.}
    \label{fig:rotation_axes_diagram}
\end{figure}
The rotation axis of stars within a rotating cluster can be expressed using the two angles shown in Figure \ref{fig:rotation_axes_diagram} - the position angle of the rotation axis ($0^\circ \leq \theta_0 < 360^\circ$) and inclination angle ($0^\circ \leq i \leq 180^\circ$). The position angle describes the rotation axis of stars in the plane of the sky and was calculated for each cluster by plotting $\Delta \rm{v_{LOS}}$ as a function of position angle $\theta_i$, fitting a sin curve to the rotation signal and determining the angle in which the sin function amplitude reached zero before increasing. The inclination angle $i = \rm{arctan}(\rm{v_{tan}} / \rm{A_{LOS}})$, describes the inclination of the rotation axis in relation to the observer. An inclination of $i=0^\circ$ corresponds to a face-on view of the cluster and therefore a strong $\rm{v_{tan}}$ signal, whereas $i=90^\circ$ provides an edge-on view and a strong $\rm{v_{LOS}}$ signal. In order to clarify that these angles were recoverable from the combination of $\rm{v_{rad}}$, $\rm{v_{tan}}$ and $\rm{v_{LOS}}$ as functions of position angle $\theta_i$, we created mock clusters using 100,000 stars with position and velocity components. Using different combinations of $\theta_0$ and $i$ input values and using only the projected positions and 3D velocities, we were able to reliably recover the input parameters within reasonable ($<5^\circ$) uncertainties.\\
\begin{figure}
    \centering
    \includegraphics[width=\hsize]{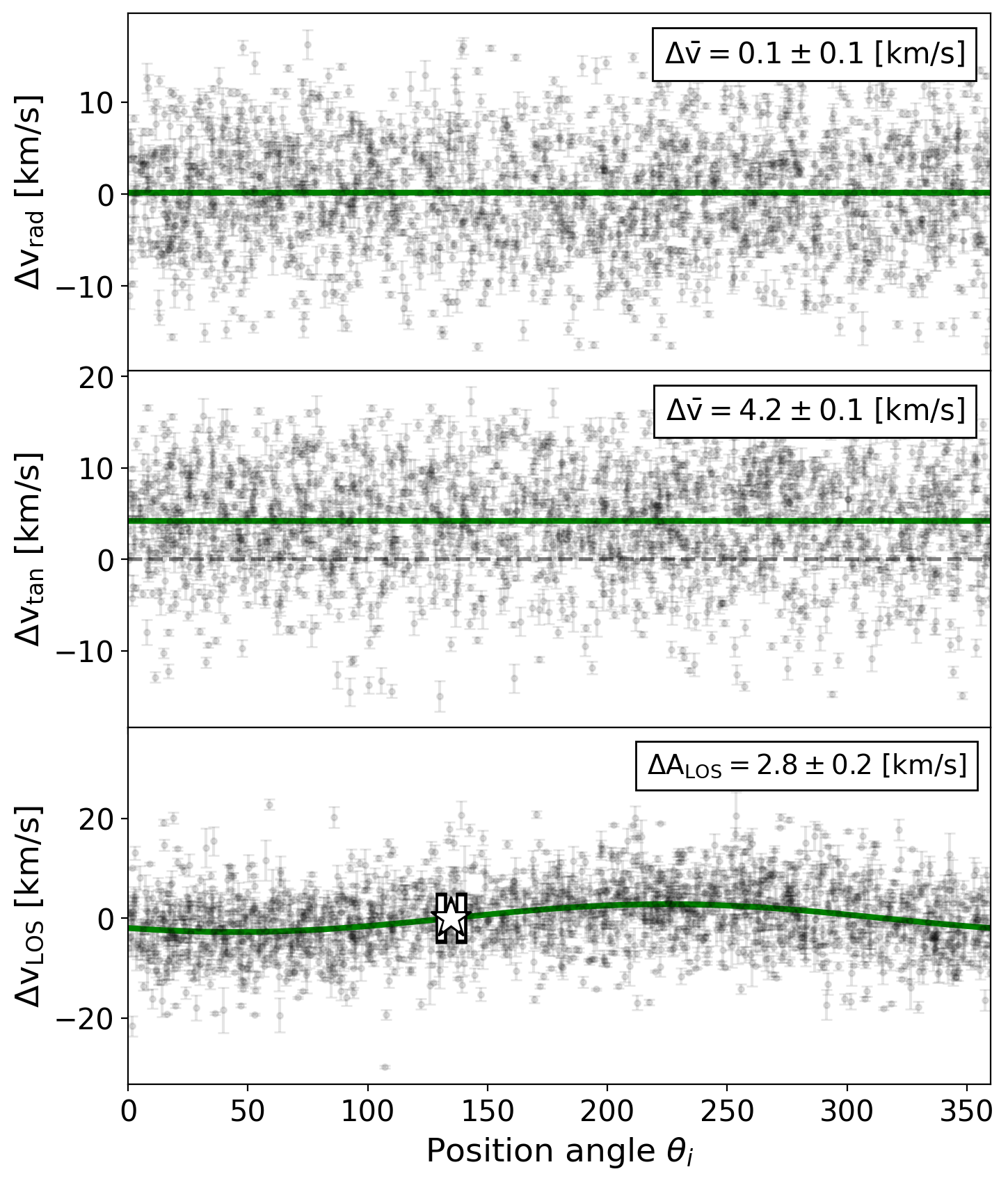}
    \vspace{-0.4cm}
    \caption{The 3-dimensional velocity components of NGC 104 with respect to the mean cluster velocity, as a function of position angle $\theta_i$. The combined sample of RGB, AGB, HB and MSTO stars in each velocity catalogue are shown as black points with errorbars. \textit{Top panel:} the radial component of the proper motion, with the mean value shown in green. \textit{Middle panel:} the same as the top panel, but for the tangential component of the proper motion. \textit{Bottom panel:} the LOS velocity as a function of $\theta_i$, with the sinusoidal fit to the distribution shown in green and the position angle of the rotation axis $\theta_0$ shown as a white star with errorbars.}
    \label{fig:rotation_amp_all_stars}
\end{figure}

To recover the rotation axis angles of all stars in the clusters within the sample, we determined $\Delta \rm{v_{rad}}$, $\Delta \rm{v_{tan}}$ and $\Delta \rm{v_{LOS}}$ as functions of position angle $\theta_i$, as shown in Figure \ref{fig:rotation_amp_all_stars}. We used the Markov Chain Monte Carlo method (\texttt{emcee} in Python \citep{emcee}) to simultaneously fit the radial, tangential and LOS velocity components of the stars in each cluster, using $32$ walkers, $10,000$ steps and a burn-in of 2000 samples. For the radial and tangential components, we fit for the means ($\Delta \rm{\bar{v}_{rad}}$,$\Delta \rm{\bar{v}_{tan}}$) and dispersions ($\sigma_{{\rm{v_{rad}}}}$,$\sigma_{{\rm{v_{tan}}}}$), while for the LOS component, we fit for the amplitude ($\Delta \rm{A_{LOS}}$), phase shift ($\theta_0$) and dispersion ($\sigma_{{\rm{v_{LOS}}}}$). Figure \ref{fig:rotation_amp_all_stars} shows the result of this analysis for NGC 104: the top and middle panels show the distribution of stars (black) in each position angle bin for the radial and tangential velocities, respectively, with the averaged mean velocity ($\Delta \rm{\bar{v}}$) shown in the legends. The bottom panel shows the sinusoidal rotation signal recovered from the LOS velocities, with the position angle of the rotation axis ($\theta_0$) shown as a white star with error bars.

\begin{figure*}
    \centering
    \includegraphics[width=\hsize]{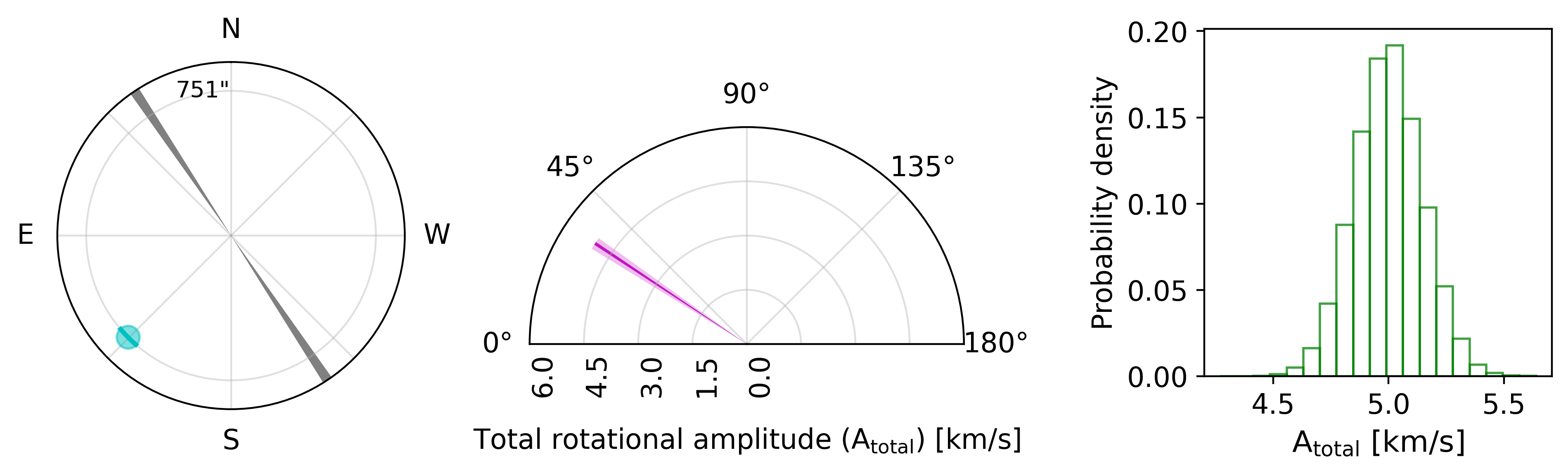}
    \vspace{-0.4cm}
    \caption{\textit{Left panel:} The position angle of the rotation axis ($\theta_0$) of NGC 104 is shown as a cyan circle with associated uncertainty, at the average projected distance of the velocity catalogues ($r = 751$ arcsec). In grey is the photometric major axis of the cluster derived in this work, with the uncertainty shown by the spread of the line. \textit{Middle panel:} The inclination angle is shown in magenta, with associated uncertainty shown by the width of the line. An inclination angle of $i=0^\circ$ indicates the cluster is face-on with respect to the observer, while $i=90^\circ$ indicates an edge-on orientation. The length of the inclination angle line corresponds to the total rotational amplitude of the cluster ($\rm{A_{total}}$), shown on the x-axis. \textit{Right panel:} The probability distribution of the total rotational amplitudes calculated for each MCMC sample.}
    \label{fig:rotation_axes}
\end{figure*}

The inclination angle $i$ of each cluster was calculated using Equation \ref{eq:inclination} from the amplitude of the rotation signal in the LOS velocity and the mean tangential velocity. The uncertainty was calculated analytically using the associated errors of the LOS amplitude ($\Delta \rm{A_{LOS,err}}$) and mean tangential velocity ($\Delta \rm{\bar{v}_{tan,err}}$), shown in Equation \ref{eq:inclination_err}.

\begin{align}
i ={}& \rm{arctan} \left(\frac{\Delta \rm{A_{LOS}}}{\Delta \rm{\bar{v}_{tan}}} \right) \label{eq:inclination}\\ 
\begin{split}
    \Delta i ={}& \sqrt{ \left( \frac{\Delta \rm{\bar{v}_{tan}}}{ \Delta \rm{A^2 _{LOS}}+ \Delta \rm{\bar{v}_{tan}}^2 } \cdot \Delta \rm{A_{LOS,err}} \right)^2 +}\\
    &  \quad \quad \quad \quad \quad \quad \quad \quad \overline{\left( \frac{-\Delta \rm{A_{LOS}}}{ \Delta \rm{A^2_{LOS}} + \Delta \rm{\bar{v}_{tan}}^2} \cdot \Delta \rm{\bar{v}_{tan,err}} \right)^2  } \label{eq:inclination_err}
\end{split}
\end{align}
We calculated the total rotational amplitude using the mean tangential velocity and amplitude of the LOS velocity signal $\rm{A_{total}} = \sqrt{ \Delta \rm{A_{LOS}}^2 + \rm{\bar{v}_{tan}}^2 }$, as well as the total velocity dispersion $\rm{\sigma} = \sqrt{ \rm{\sigma_{LOS}}^2 + \rm{\sigma_{tan}}^2 }$, in order to determine the total rotation over dispersion ratio ($\rm{A_{total}/\sigma}$).\\

In order to calculate the photometric semi-major axis orientation of each cluster, we fit the spatial distribution of its members with a bivariate Gaussian distribution, using only stars brighter than the turn-off. To obtain a robust estimate of the position angle (PA) of the semi-major axis, we took the median and the standard deviation of 100 bootstrap experiments. In Table \ref{tab:pa}, we report the estimated PA of the photometric semi-major axis and eccentricity of each cluster. These values were compared with those calculated in \cite{1987white}, as well as \cite{2018Kamann} for clusters overlapping with our sample. We found consistency between the orientations from all 3 sources for clusters NGC 104, NGC 1851 and NGC 2808, but although we found good agreement in general with the values calculated by \cite{2018Kamann}, we did not find our values aligned with those of \cite{1987white} and therefore we opted to only use our values for this analysis.\\

In the left panel of Figure \ref{fig:rotation_axes} we show the position angle of the rotation axis of NGC 104 ($\theta_0 = 134^\circ \pm 4^\circ$) in cyan, with the corresponding photometric major axis orientation of the cluster shown in grey and the uncertainty represented by the width of the line. We found that for NGC 104, the position angle in the plane of the sky is perpendicular to the photometric major axis orientation, which is to be expected for a cluster flattened by initial rotation. The inclination angle ($i = 33^\circ \pm 2^\circ$) for NGC 104 shown in magenta in the middle panel of Figure \ref{fig:rotation_axes} indicates the orientation of the cluster is approximately mid-way between face-on and edge-on, which is also reflected in the strong signal of both the observed tangential and LOS velocities of Figure \ref{fig:rotation_amp_all_stars}. The length of the inclination angle vector corresponds to the total rotational amplitude ($\rm{A_{total}}$) of the cluster as displayed on the x-axis, while the width of the inclination angle vector indicates the uncertainty in the calculated value. Finally, the right panel of Figure \ref{fig:rotation_axes} demonstrates the probability distribution of the total rotational amplitudes $\rm{A_{total}}$ calculated from the samples of the MCMC analysis for all stars.

\begin{table*}
\centering
\caption{The kinematic properties of the 30 Galactic globular clusters in our sample. The columns are as follows: (1) cluster name, (2) number of stars with proper motions, (3) number of stars with line-of-sight velocities, (4) the $A^+$ value calculated in \protect\cite{2023Leitinger}, (5) the enriched star fraction calculated in \protect\cite{2023Leitinger}, (6) the dynamical age calculated with values from \protect\cite{GGCD}, (7) the rotational amplitude of the line-of-sight velocity, (8) the median velocity of the tangential component of the proper motions with respect to the cluster velocity, (9) the median velocity of the radial tangential of the proper motions with respect to the cluster velocity, (10) the total rotation over dispersion ratio as described in Section \ref{sec:rotation_axes}, (11) the position angle of the rotation axis (12) the inclination angle.}
\resizebox{\hsize}{!}{
\begin{tabular}{lccccccccccc}
\hline
\hline\\[-0.7em]
Cluster  & $\rm{N_{PM}}$ & $\rm{N_{LOS}}$ & $A^+$            & $\rm{P_2/P_{tot}}$ & $\rm{age/T_{rh}}$ & $\rm{\Delta A}$ & $\rm{\Delta v_{tan}}$ & $\rm{\Delta v_{rad}}$ & $\rm{A_{total}/\sigma}$ & $\theta_0$ & $i$        \\
& & & & & & [km/s] & [km/s] & [km/s] & & [deg] & [deg]\\[0.4em]
\hline
NGC 104  & 12894         & 1409           & -0.22 $\pm$ 0.03 & 0.73 $\pm$ 0.02    & 2.5 $\pm$ 0.1     & 2.8 $\pm$ 0.2   & 4.2 $\pm$ 0.1         & 0.1 $\pm$ 0.1         & 0.60 $\pm$ 0.02         & 134 $\pm$ 4  & 33 $\pm$ 2   \\
NGC 288  & 2446          & 593            & 0.21 $\pm$ 0.08  & 0.37 $\pm$ 0.04    & 3.7 $\pm$ 0.3     & 0.3 $\pm$ 0.2   & -0.4 $\pm$ 0.1        & 0.4 $\pm$ 0.1         & 0.10 $\pm$ 0.04         & 6 $\pm$ 49   & 144 $\pm$ 20 \\
NGC 1261 & 4034          & 373            & 0.02 $\pm$ 0.06  & 0.58 $\pm$ 0.03    & 6.7 $\pm$ 0.4     & 0.8 $\pm$ 0.3   & 0.4 $\pm$ 0.4         & 0.7 $\pm$ 0.4         & 0.14 $\pm$ 0.05         & 108 $\pm$ 20 & 66 $\pm$ 25  \\
NGC 1851 & 6797          & 610            & -0.08 $\pm$ 0.07 & 0.70 $\pm$ 0.03    & 10.5 $\pm$ 0.6    & 1.5 $\pm$ 0.2   & 0.7 $\pm$ 0.3         & -0.2 $\pm$ 0.2        & 0.19 $\pm$ 0.03         & 0 $\pm$ 8    & 65 $\pm$ 8   \\
NGC 2808 & 12131         & 1070           & -0.08 $\pm$ 0.03 & 0.75 $\pm$ 0.02    & 5.3 $\pm$ 0.2     & 2.8 $\pm$ 0.3   & -0.1 $\pm$ 0.2        & 0.0 $\pm$ 0.2         & 0.24 $\pm$ 0.03         & 321 $\pm$ 6  & 92 $\pm$ 3   \\
NGC 3201 & 3456          & 1125           & 0.46 $\pm$ 0.12  & 0.49 $\pm$ 0.04    & 2.5 $\pm$ 0.2     & 0.8 $\pm$ 0.1   & -0.4 $\pm$ 0.1        & -1.2 $\pm$ 0.1        & 0.19 $\pm$ 0.03         & 296 $\pm$ 9  & 116 $\pm$ 5  \\
NGC 4590 & 1904          & 284            & -0.13 $\pm$ 0.10 & 0.53 $\pm$ 0.05    & 4.0 $\pm$ 0.2     & 0.5 $\pm$ 0.3   & -0.1 $\pm$ 0.2        & 0.1 $\pm$ 0.2         & 0.12 $\pm$ 0.06         & 266 $\pm$ 31 & 98 $\pm$ 21  \\
NGC 4833 & 2662          & 189            & -0.07 $\pm$ 0.07 & 0.51 $\pm$ 0.03    & 9.0 $\pm$ 0.3     & 0.4 $\pm$ 0.3   & -0.2 $\pm$ 0.2        & -0.1 $\pm$ 0.1        & 0.08 $\pm$ 0.05         & 34 $\pm$ 83  & 119 $\pm$ 26 \\
NGC 5024 & 5306          & 438            & -0.42 $\pm$ 0.05 & 0.56 $\pm$ 0.03    & 1.5 $\pm$ 0.1     & 0.7 $\pm$ 0.3   & -1.1 $\pm$ 0.3        & 0.0 $\pm$ 0.3         & 0.19 $\pm$ 0.05         & 29 $\pm$ 32  & 147 $\pm$ 14 \\
NGC 5053 & 464           & 61             & 0.07 $\pm$ 0.16  & 0.46 $\pm$ 0.07    & 1.6 $\pm$ 0.2     & -               & -                     & -                     & -                       & -            & -            \\
NGC 5272 & 6979          & 546            & -0.36 $\pm$ 0.05 & 0.64 $\pm$ 0.02    & 3.8 $\pm$ 0.2     & 1.0 $\pm$ 0.3   & 1.6 $\pm$ 0.1         & 0.2 $\pm$ 0.1         & 0.24 $\pm$ 0.03         & 9 $\pm$ 9    & 32 $\pm$ 8   \\
NGC 5286 & 4687          & 816            & 0.09 $\pm$ 0.05  & 0.61 $\pm$ 0.02    & 9.0 $\pm$ 0.3     & 2.7 $\pm$ 0.4   & 0.4 $\pm$ 0.3         & -0.6 $\pm$ 0.3        & 0.26 $\pm$ 0.04         & 276 $\pm$ 9  & 80 $\pm$ 7   \\
NGC 5904 & 6915          & 787            & -0.10 $\pm$ 0.06 & 0.70 $\pm$ 0.03    & 3.6 $\pm$ 0.2     & 2.7 $\pm$ 0.2   & -2.7 $\pm$ 0.2        & -0.4 $\pm$ 0.1        & 0.50 $\pm$ 0.03         & 341 $\pm$ 16 & 134 $\pm$ 2  \\
NGC 5986 & 4959          & 260            & -0.08 $\pm$ 0.04 & 0.59 $\pm$ 0.03    & 7.2 $\pm$ 0.4     & 0.7 $\pm$ 0.5   & -1.3 $\pm$ 0.4        & 0.5 $\pm$ 0.4         & 0.15 $\pm$ 0.04         & 167 $\pm$ 53 & 150 $\pm$ 18 \\
NGC 6101 & 2109          & 453            & 0.66 $\pm$ 0.13  & 0.51 $\pm$ 0.05    & 1.4 $\pm$ 0.1     & 0.8 $\pm$ 0.4   & 0.6 $\pm$ 0.2         & -0.6 $\pm$ 0.2        & 0.17 $\pm$ 0.05         & 334 $\pm$ 28 & 54 $\pm$ 15  \\
NGC 6121 & 1024          & 3005           & -0.09 $\pm$ 0.10 & 0.63 $\pm$ 0.05    & 14.0 $\pm$ 0.7    & 0.4 $\pm$ 0.1   & 0.2 $\pm$ 0.1         & -0.1 $\pm$ 0.1        & 0.10 $\pm$ 0.02         & 140 $\pm$ 15 & 63 $\pm$ 12  \\
NGC 6205 & 8202          & 431            & -0.04 $\pm$ 0.04 & 0.68 $\pm$ 0.03    & 4.1 $\pm$ 0.2     & 2.1 $\pm$ 0.4   & 0.1 $\pm$ 0.1         & 0.6 $\pm$ 0.1         & 0.27 $\pm$ 0.05         & 197 $\pm$ 10 & 87 $\pm$ 3   \\
NGC 6218 & 3373          & 534            & 0.20 $\pm$ 0.07  & 0.57 $\pm$ 0.04    & 15.2 $\pm$ 0.5    & 0.2 $\pm$ 0.1   & -0.6 $\pm$ 0.1        & -0.0 $\pm$ 0.1        & 0.11 $\pm$ 0.02         & 83 $\pm$ 75  & 164 $\pm$ 12 \\
NGC 6254 & 3509          & 420            & 0.02 $\pm$ 0.05  & 0.62 $\pm$ 0.03    & 7.3 $\pm$ 0.5     & 0.2 $\pm$ 0.2   & -0.5 $\pm$ 0.1        & -0.1 $\pm$ 0.1        & 0.09 $\pm$ 0.02         & 104 $\pm$ 80 & 158 $\pm$ 16 \\
NGC 6341 & 4780          & 401            & -0.16 $\pm$ 0.07 & 0.57 $\pm$ 0.03    & 10.4 $\pm$ 0.3    & 0.9 $\pm$ 0.4   & 1.7 $\pm$ 0.2         & 0.1 $\pm$ 0.2         & 0.26 $\pm$ 0.03         & 314 $\pm$ 29 & 26 $\pm$ 10  \\
NGC 6366 & 1977          & 271            & 0.02 $\pm$ 0.14  & 0.45 $\pm$ 0.05    & 8.8 $\pm$ 0.8     & 0.2 $\pm$ 0.2   & 0.0 $\pm$ 0.1         & 0.1 $\pm$ 0.1         & 0.06 $\pm$ 0.06         & 134 $\pm$ 84 & 82 $\pm$ 20  \\
NGC 6656 & 5632          & 757            & 0.08 $\pm$ 0.04  & 0.50 $\pm$ 0.03    & 4.1 $\pm$ 0.2     & 2.5 $\pm$ 0.3   & -2.1 $\pm$ 0.1        & 0.3 $\pm$ 0.1         & 0.36 $\pm$ 0.03         & 279 $\pm$ 7  & 129 $\pm$ 3  \\
NGC 6752 & 4257          & 1683           & 0.08 $\pm$ 0.08  & 0.71 $\pm$ 0.04    & 6.1 $\pm$ 0.2     & 0.4 $\pm$ 0.2   & 0.7 $\pm$ 0.1         & 0.2 $\pm$ 0.1         & 0.11 $\pm$ 0.02         & 180 $\pm$ 30 & 27 $\pm$ 10  \\
NGC 6809 & 1816          & 664            & -0.51 $\pm$ 0.07 & 0.56 $\pm$ 0.04    & 4.1 $\pm$ 0.2     & 0.2 $\pm$ 0.2   & -0.8 $\pm$ 0.1        & -0.2 $\pm$ 0.1        & 0.13 $\pm$ 0.02         & 131 $\pm$ 72 & 162 $\pm$ 14 \\
NGC 6838 & 1502          & 384            & 0.04 $\pm$ 0.12  & 0.34 $\pm$ 0.05    & 15.6 $\pm$ 1.5    & 0.2 $\pm$ 0.1   & 0.0 $\pm$ 0.1         & 0.0 $\pm$ 0.1         & 0.04 $\pm$ 0.04         & 110 $\pm$ 97 & 74 $\pm$ 30  \\
NGC 6934 & 2453          & 72             & 0.21 $\pm$ 0.07  & 0.61 $\pm$ 0.04    & 7.9 $\pm$ 0.3     & -               & -                     & -                     & -                       & -            & -            \\
NGC 6981 & 1071          & 62             & -0.00 $\pm$ 0.09 & 0.40 $\pm$ 0.05    & 8.7 $\pm$ 0.9     & -               & -                     & -                     & -                       & -            & -            \\
NGC 7078 & 9792          & 519            & -0.03 $\pm$ 0.06 & 0.77 $\pm$ 0.03    & 8.0 $\pm$ 0.2     & 1.8 $\pm$ 0.3   & -2.6 $\pm$ 0.2        & 0.2 $\pm$ 0.2         & 0.35 $\pm$ 0.03         & 127 $\pm$ 10 & 146 $\pm$ 5  \\
NGC 7089 & 2344          & 475            & -0.06 $\pm$ 0.05 & 0.64 $\pm$ 0.02    & 4.3 $\pm$ 0.2     & 2.9 $\pm$ 0.3   & 1.5 $\pm$ 0.3         & -0.0 $\pm$ 0.3        & 0.33 $\pm$ 0.03         & 29 $\pm$ 7   & 62 $\pm$ 5   \\
NGC 7099 & 2020          & 1407           & 0.14 $\pm$ 0.09  & 0.55 $\pm$ 0.05    & 12.8 $\pm$ 0.4    & 0.2 $\pm$ 0.2   & -0.7 $\pm$ 0.2        & -0.0 $\pm$ 0.2        & 0.12 $\pm$ 0.03         & 176 $\pm$ 58 & 162 $\pm$ 13\\
\hline
\label{tab:kinematics}
\end{tabular}
}
\end{table*}

\subsection{Rotation of the multiple stellar populations}
\label{sec:rotation_axes_MP}

\begin{figure}
    \centering
    \includegraphics[width=\hsize]{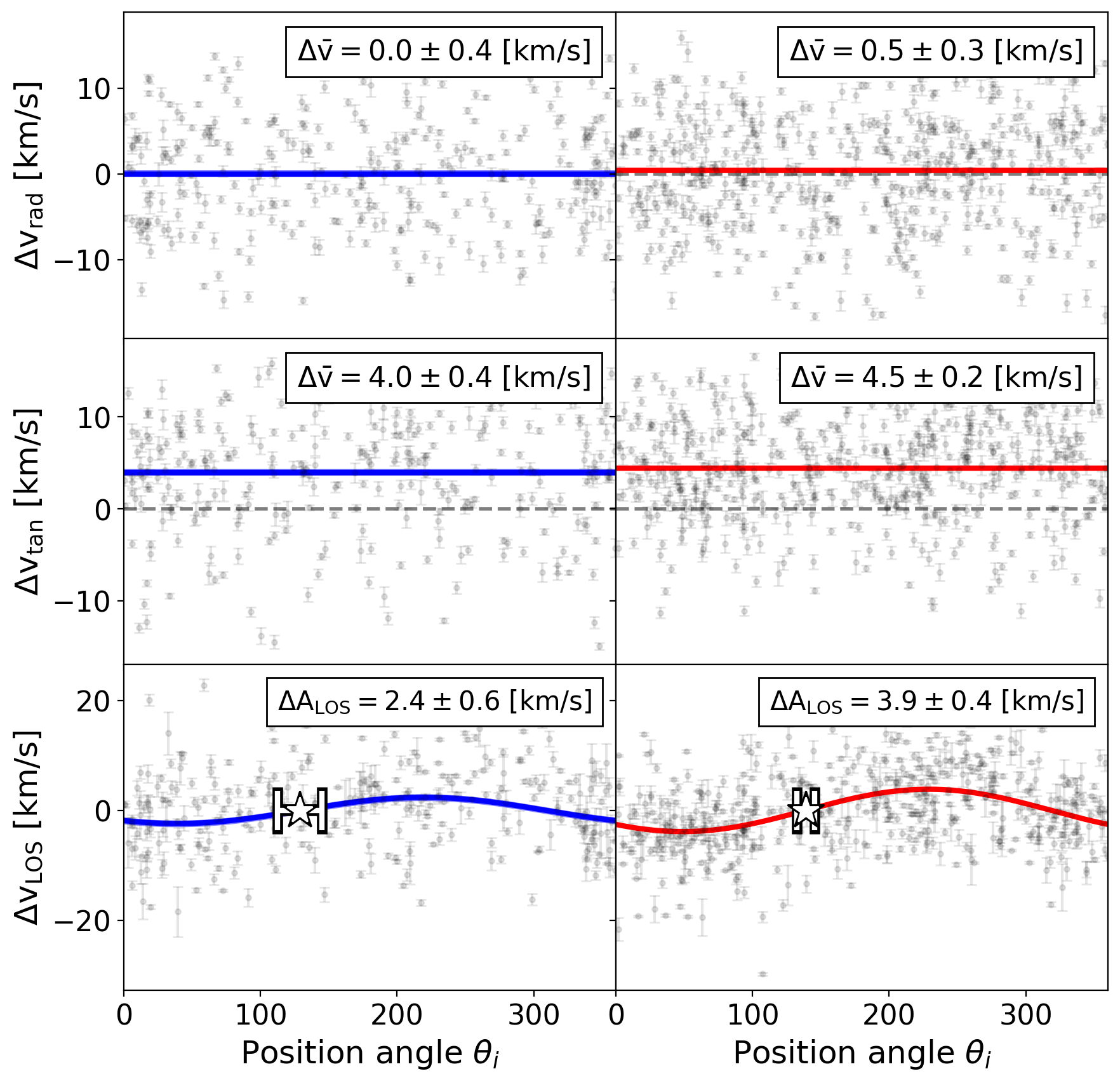}
    \vspace{-0.4cm}
    \caption{The 3-dimensional velocity components of NGC 104 as shown in Figure \ref{fig:rotation_amp_all_stars}, but for the multiple stellar populations with P1 (blue) shown in the left panels and P2 (red) shown in the right panels.}
    \label{fig:rotation_amp_pops}
\end{figure}

\begin{figure*}
    \centering
    \includegraphics[width=\textwidth]{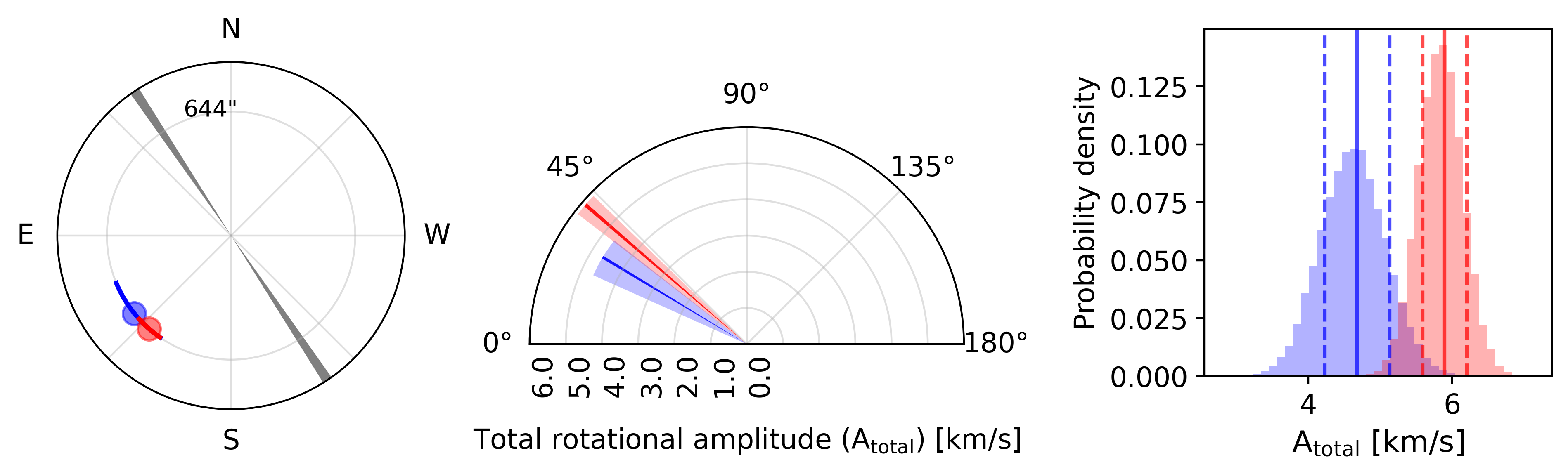}
    \vspace{-0.4cm}
    \caption{The position angles, inclination angles and total rotational amplitude probability densities of NGC 104, as shown in Figure \ref{fig:rotation_axes} for all stars, but now for the multiple stellar populations: P1 (blue) and P2 (red).}
    \label{fig:rotation_axes_populations}
\end{figure*}

The homogeneous classifications of the multiple stellar populations was completed for 30 Galactic globular clusters using a combination of ground-based \citep{2019Stetson} and space-based HST \citep{2015Piotto,2018Nardiello} photometry, as described in \cite{2023Leitinger}. The multiple stellar populations which were determined in \cite{2023Leitinger} were classified into: `P1', `P2' and in some clusters `P3'. However, for the sake of the spatial distribution analysis described in \cite{2023Leitinger}, as well as the kinematic analysis described in this paper, the `enriched stars' (P2 and P3) were combined into one population (P2) for comparisons against P1, as our main focus is on the kinematic differences between stars which show `primordial' chemical abundance patterns (P1), vs. any stars which are enriched in comparison to P1 stars.

We added NGC 104 and NGC 6656 to the original sample of 28 GCs in \cite{2023Leitinger} by using the same classification method to determine the multiple stellar populations of the RGB stars, but removed NGC 5053, NGC 6981 and NGC 6934 due to the low number of LOS velocities for each cluster. We cross-matched the photometrically classified stars of each cluster with the cleaned LOS and proper motion catalogues described in Section \ref{sec:rotation_axes} in order to assign a population to the stars in common with the velocity catalogues. We then performed the same analysis as described in Section \ref{sec:rotation_axes} on the combined catalogues and determined the position angle of the rotation axis, inclination angle and total rotational amplitude of each population in the 27 GCs.\\

The rotation signals as a function of position angle for the multiple stellar populations in NGC 104 are shown in Figure \ref{fig:rotation_amp_pops}, with P1 (P2) stars shown in the left (right) panels. The 3D velocity components exhibit a difference of $3.5\sigma$ significance in the mean velocities and rotation amplitudes between the P1 and P2 stars, with P2 stars rotating faster than P1 stars. Despite this, the left panel of Figure \ref{fig:rotation_axes_populations} shows similar position angles for P1 (blue) and P2 (red) stars, with uncertainties represented as solid lines. The middle panel shows only a slight difference in the inclination angles of the P1 and P2 stars, with the length of the vectors representing the total rotational amplitude and the width representing the uncertainties. The right panel shows the probability distribution of the calculated total rotational amplitudes $\rm{A_{total}}$ of each population, with solid lines indicating the median values and dashed lines indicating the $\pm 1\sigma$ values.

NGC 104 has a dynamical age (age/relaxation time) of $\rm{age/T_{rh}} = 2.5 \pm 0.1$, meaning it has experienced just over 2 relaxation times, so while we classify it as one of the `dynamically younger' clusters in our sample, the present day conditions may not be entirely representative of its initial conditions. With this in mind, we cannot confidently conclude the cluster was born this way - with P2 stars rotating faster than P1 stars and similar rotation axes for both populations. \textit{N}-body simulations may help to explore how much the cluster has changed from its initial conditions and investigate scenarios which can produce the same kinematic results we have observed.\\

The rotation axes plots shown in Figures \ref{fig:rotation_axes} and \ref{fig:rotation_axes_populations} were combined for the 27 clusters with adequate LOS velocities, to show the rotation axes and total rotational amplitudes of all RGB, AGB, HB and MS-TO stars (green), as well as the RGB stars of the P1 (blue) and P2 (red) populations. As the data set for all stars tends to extend to a larger projected distance than the data set with population classifications, the position angles for each data set are shown at their corresponding average distances. These figures are included in Appendix \ref{sec:Appendix_rotation} for the dynamically young clusters in our sample. Additionally, Table \ref{tab:kinematics_MPs} in Appendix \ref{sec:Appendix_pa} lists the calculated rotation parameters of the MPs, similar to Table \ref{tab:kinematics} for all stars.\\

\subsection{Velocity dispersions}
\label{sec:vdisp}

The velocity dispersions and the anisotropy for the full sample of 30 GCs were determined from the \gaia DR3 and HACKS proper motions. The number of stars with available proper motions are shown in Column 2 of Table \ref{tab:kinematics} for each cluster.\\
Since HACKS covers a square field of view, stars located close to the corners were removed by creating annuli from the centre of the cluster outwards, stopping at the radius in which there is 85\% completeness. This method is described in greater detail in Section 2.1 of \cite{2023Leitinger}, as it was also used for combining the HST and ground-based photometric catalogues. Stars in the \gaia DR3 catalogue located roughly within the HACKS field of view in the cluster centres suffered from stellar crowding, resulting in large proper motion uncertainties. For this reason, stars within the aforementioned radius limit were also removed, creating a smooth transition between the HACKS stars in the centre and \gaia stars in the outer regions. This was performed only for the velocity dispersion and anisotropy analysis, as the HACKS proper motions were not used to determine the rotation of the clusters.

\begin{figure}
    \centering
    \includegraphics[width=\hsize]{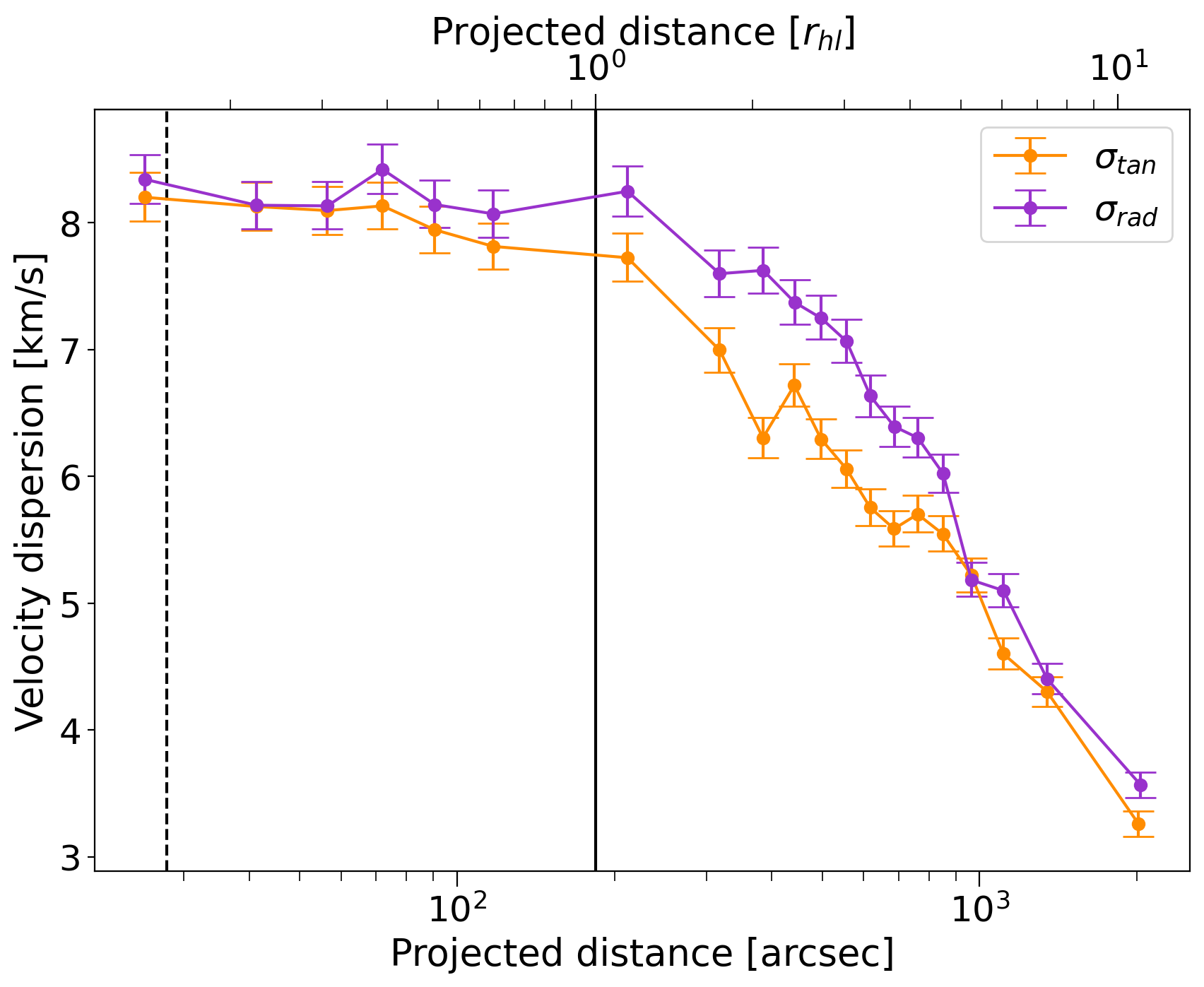}
    \vspace{-0.4cm}
    \caption{The radial (indigo) and tangential (orange) velocity dispersion as a function of projected distance for NGC 104. The dashed black line shows the core radius $r_{\rm{c}}$, while the solid black line shows the half-light radius $r_{\rm{hl}}$.}
    \label{fig:vdisp}
\end{figure}

The radial ($\rm{v_{rad}}$) and tangential ($\rm{v_{tan}}$) components of the proper motions were grouped into radial bins with equal numbers of stars, in which both the number of bins and number of stars in each bin depended on the sample available for each cluster.\\

\textit{ - A caveat on the calculation of anisotropy}: there are two ways to calculate the velocity dispersions and resulting anisotropy of stars in a GC. In the first method, the mean velocity of each component ($\mu_{\rm{rad}},\mu_{\rm{tan}}$) is subtracted from each radial bin, removing the bulk motion due to rotation and creating a reference frame in which only the intrinsic random motion between stars is observed. This reference frame is useful for observing the internal dynamical differences between stars, but does not necessarily reflect the general orbits of the stars - especially for fast rotating clusters. In the second method, the contribution of rotation is not subtracted, so that the bulk motion is also retained. This method can be useful for identifying the anisotropy of stars under the influence of rotation, which is then a combination of the bulk motion and intrinsic motion. In the case of a strongly rotating cluster, which greatly influences the orbits of the stars, this version of the anisotropy provides a more general view of these orbits. So for the purpose of identifying the internal dynamical processes, rotation-subtracted velocity dispersions are used, but for the purpose of observing the general orbits of stars in the context of energy, the rotation is included. The relative reference frame used to measure the HACKS proper motions does not include the bulk rotation of the cluster. Therefore, for the sake of consistency between the HACKS and \gaia proper motions, we used the method in which we remove the rotation to calculate the velocity dispersions and anisotropy, such that the bulk motion does not influence the anisotropy of the stars. However, we also present a version of the anisotropy results \textit{without} the rotation removed and without including the HACKS proper motions (Figure \ref{fig:gaia_anisotropy_young} in Appendix \ref{sec:Appendix_dold_anisotropy}). \\

We calculated the velocity dispersions of the radial and tangential components of the proper motions, using a modified version of the method outlined by \cite{1993Pryor}. Briefly, we initially assume a normal distribution for the individual velocity components, described by the probability density function:
\begin{equation}
    f(v_i) = \frac{1}{\sqrt{2\pi(\sigma_c^2 + \sigma_{e,i}^2)}} \rm{exp}\left( -\frac{(v_i - \bar{v})^2}{2(\sigma_c^2 + \sigma_{e,i}^2)} \right),
\end{equation}
where $\bar{v}$ is the average cluster velocity, $\sigma_c$ is the intrinsic cluster dispersion, $v_i$ are the individual stellar velocities and $\sigma_{e,i}$ are the associated errors for the individual velocities. The likelihood function of all $N$ stars was then calculated with the following set of equations:
\begin{align}
    \sum_{i=1}^N \frac{v_i}{(\sigma_c^2 + \sigma_{e,i}^2)} - \bar{v} \sum_{i=1}^N \frac{1}{(\sigma_c^2 + \sigma_{e,i}^2)} &= 0\\
    \sum_{i=1}^N \frac{(v_i-\bar{v})^2}{(\sigma_c^2 + \sigma_{e,i}^2)^2} - \sum_{i=1}^N \frac{1}{(\sigma_c^2 + \sigma_{e,i}^2)} &= 0,
\end{align}
which were numerically solved through iteration for both $\bar{v}$ and $\sigma_c$, where the initial guesses for these parameters assumed uniform uncertainties, leading to the simplified equations:
\begin{equation*}
    \bar{v} = \frac{1}{N} \sum_{i=1}^N v_i \qquad \qquad \sigma_c^2 = \frac{1}{N} \sum_{i=1}^N (v_i - \bar{v})^2 - \sigma_{e,i}^2.
\end{equation*}

During this iteration process, stars deviating beyond a 2.5$\sigma$ rejection tolerance from the median were systematically removed for each velocity dispersion bin until no further rejections were necessary, which will underestimate the true velocity dispersion by a few percent.
The upper and lower bounds of the uncertainties were estimated by iteratively determining dispersion values which corresponded to a likelihood difference of 0.5 with respect to the maximum likelihood. Figure \ref{fig:vdisp} shows the resulting radial and tangential velocity dispersion as a function of projected distance for NGC 104, with dashed(solid) lines indicating the core(half-light) radius of the cluster. 

%%%%%%%%%%%%%%%%%%%%%%%%%%%%%%%%%%%%%%%%%%%%%%%%%%%%%%%%%%%%%%%%%%%%%%%%%%%%%%%%%%%%%%%%%%
%%%%%%%%%%%%%%%%%%%%%%%%%%%%%%% ANISOTROPY %%%%%%%%%%%%%%%%%%%%%%%%%%%%%%%%%%%%%%%%%%%%%%%
%%%%%%%%%%%%%%%%%%%%%%%%%%%%%%%%%%%%%%%%%%%%%%%%%%%%%%%%%%%%%%%%%%%%%%%%%%%%%%%%%%%%%%%%%%

\subsection{Anisotropy}
\label{sec:anisotropy}

\begin{figure}
    \centering
    \includegraphics[width=\hsize]{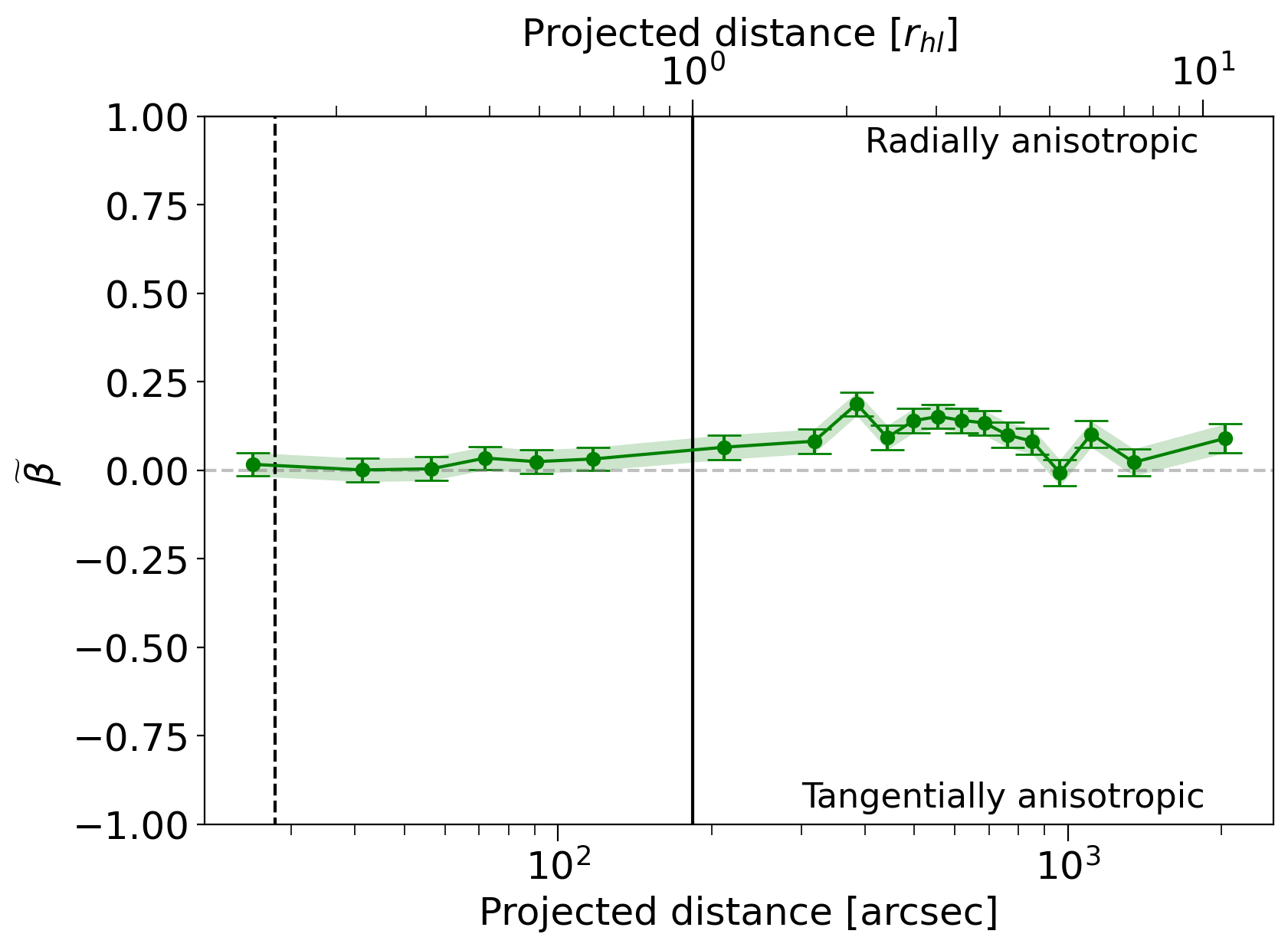}
    \vspace{-0.4cm}
    \caption{The normalised anisotropy of NGC 104 as a function of projected distance from the centre of the cluster in terms of arcseconds (bottom x-axis) and half-light radii (top x-axis). We removed the contribution due to rotation in this version of the anisotropy. The dashed black line indicates the core radius $r_{\rm{c}}$, while the solid black line indicates the half-light radius $r_{\rm{hl}}$.}
    \label{fig:anisotropy_all}
\end{figure}

Using the radial $\rm{\sigma_{rad}}$ and tangential $\rm{\sigma_{tan}}$ velocity dispersions, the anisotropy could be calculated using $\beta = 1 - \left( \frac{\rm{\sigma_{tan}}^2}{\rm{\sigma_{rad}}^2} \right)$, which provides limits of $-\infty < \beta \leq 1$ for the possible anisotropy values. In order to remove the asymmetry in the possible values, we normalised this version of the anisotropy:
\begin{equation}
    \tilde{\beta} = \frac{\beta}{(2-\beta)},
\end{equation}
in which the new limits are $-1 \leq \tilde{\beta} \leq 1$, where $\tilde{\beta} = -1$ is fully tangentially anisotropic, $\tilde{\beta} = 1$ is fully radially anisotropic and $\tilde{\beta} = 0$ is isotropic. We calculated this normalised anisotropy as a function of projected distance from the centre of the cluster for the full sample of stars in each cluster, as demonstrated in Figure \ref{fig:anisotropy_all} for NGC 104, which shows isotropic behaviour for the inner $\sim$100'' and radial anisotropy in the outer regions.\\

Since we used HACKS proper motions for the inner regions of each cluster, we compared and found consistency between our results and those of \cite{2022Libralato}, when adopting the same definition of the anisotropy as \cite{2022Libralato}: $\rm{\sigma_{tan}}$/$\rm{\sigma_{rad}}$. However, regardless of the initial conditions of cluster anisotropy, we would expect the inner regions of the cluster to become isotropic faster than the outer regions due to dynamical mixing. The outer regions of a cluster are expected to show the most significant deviations from isotropy. We especially found this to be the case for the dynamically youngest ($\rm{age/T_{rh}} <$  4.5) globular clusters in the sample, which demonstrate a significant ($>2\sigma$) central concentration of either P1 or P2 stars (see Figure 15 of \cite{2023Leitinger}). This is discussed further in Section \ref{sec:anisotropy_results}.\\

\subsection{Velocity dispersions and anisotropy of the multiple stellar populations}
\label{sec:vdisp_anisotropy_MPs}

\begin{figure}
    \centering
    \includegraphics[width=\hsize]{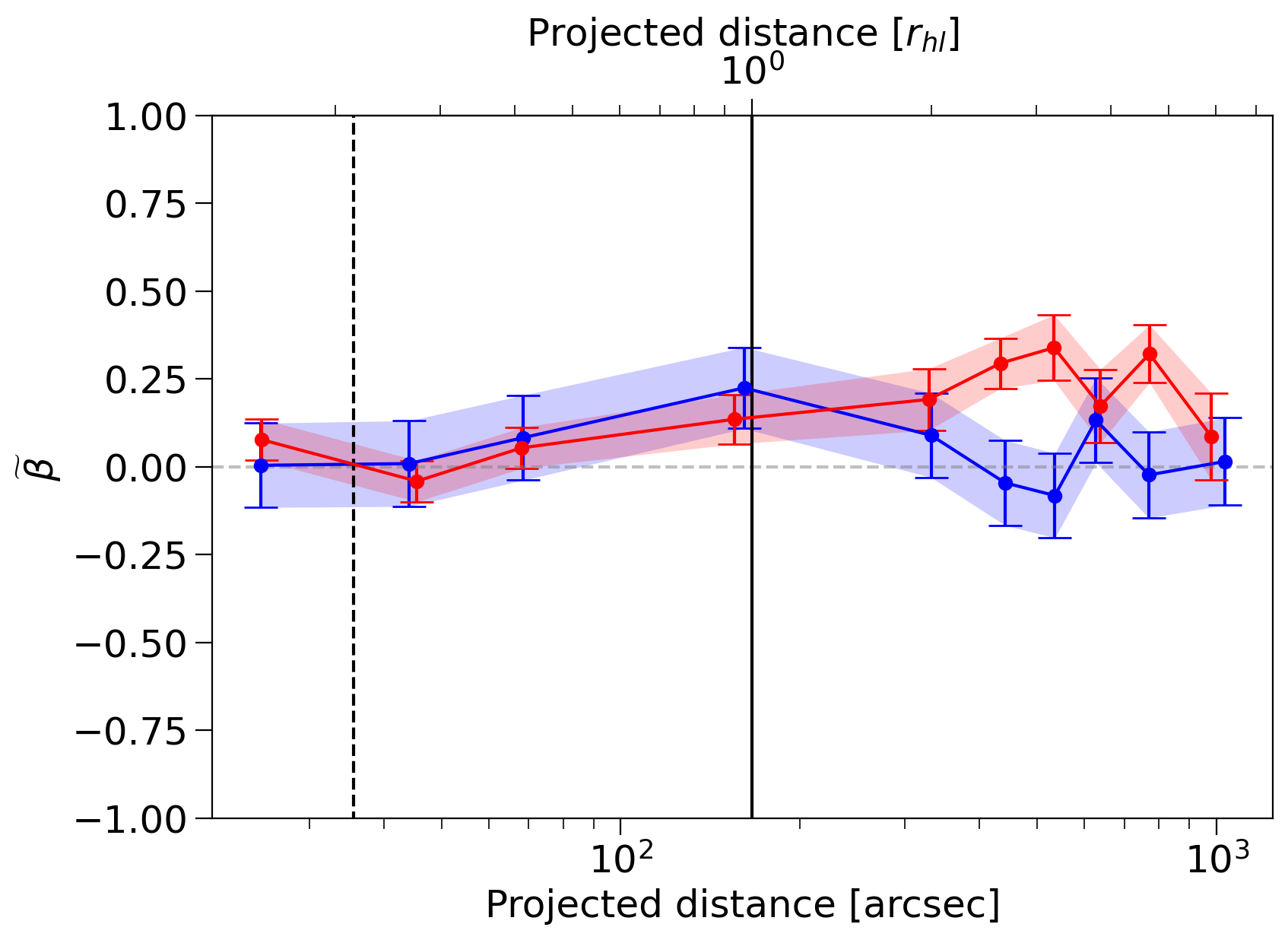}
    \vspace{-0.4cm}
    \caption{The normalised anisotropy of NGC 104 as a function of projected distance from the cluster centre in terms of arcseconds (bottom x-axis) and half-light radii (top x-axis) for the P1 (blue) and P2 (red) populations. The dashed red line indicates isotropy. The dashed black line indicates the core radius $r_{\rm{c}}$, while the solid black line indicates the half-light radius $r_{\rm{hl}}$.}
    \label{fig:ani_pop1}
\end{figure}

\begin{figure}
    \centering
    \includegraphics[width=\hsize]{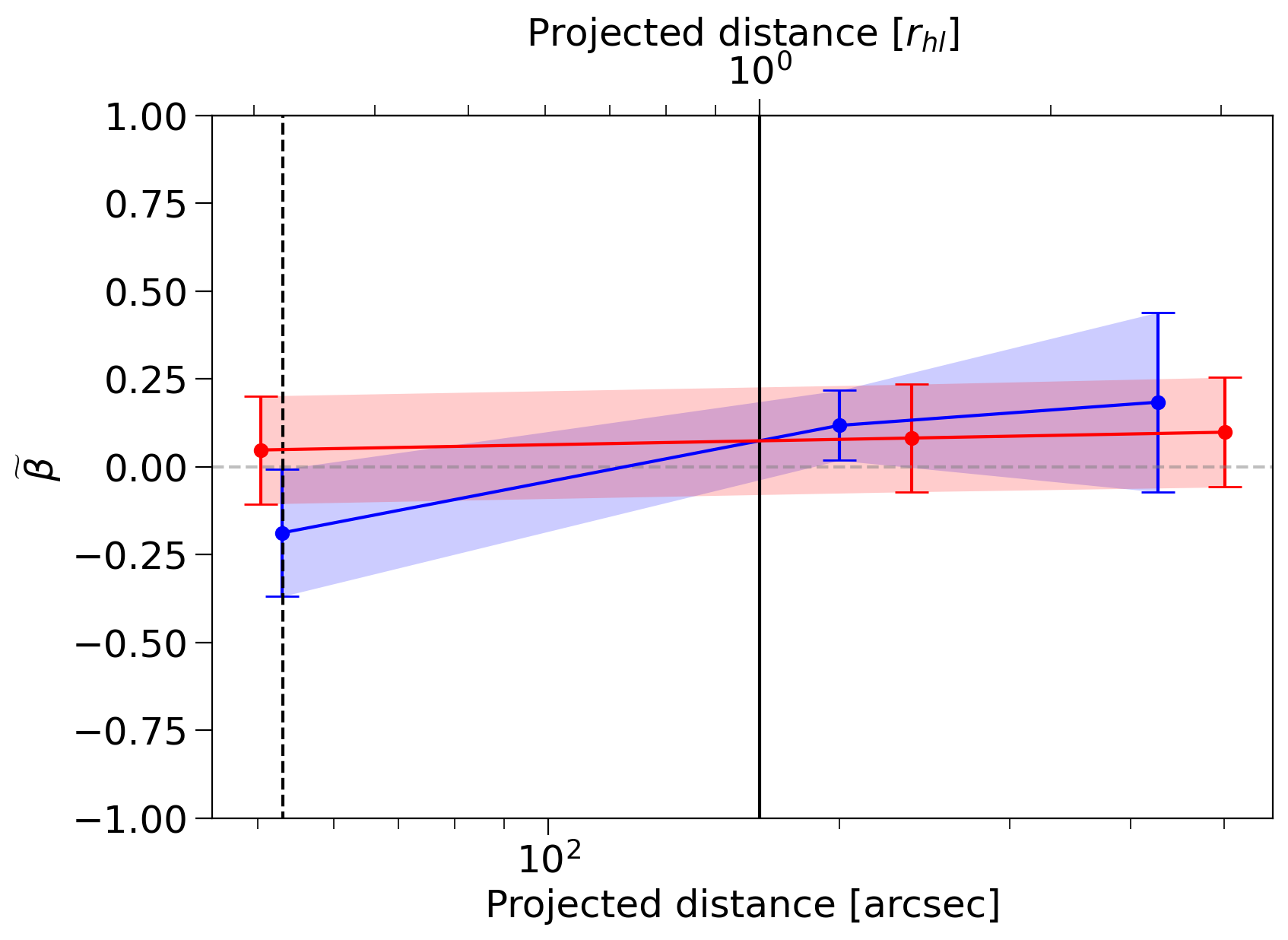}
    \vspace{-0.4cm}
    \caption{The same as Figure \ref{fig:ani_pop1} but for NGC 3201.}
    \label{fig:ani_pop2}
\end{figure}

For the multiple stellar populations, we performed the same analysis as described in Section \ref{sec:vdisp} and Section \ref{sec:anisotropy}, but for the cleaned proper motion data described in Section \ref{sec:pm_data} matched with the combined HST and ground-based photometry outlined in Section \ref{sec:photometry_data}. To ensure the two populations were comparable, we binned stars using multiple annuli from the centre of the cluster to the outermost regions. Since the anisotropy was investigated as a function of radius, it was important that the stars of each population occupy the same area in each bin. Otherwise, if we binned by equal number of stars in each population instead, we risked comparing P1 and P2 stars of different radii.\\
The median value of the projected distance for each bin varied slightly between each cluster, as shown in Figure \ref{fig:ani_pop1} for NGC 104, but the bin edges were the same for each population. As NGC 104 had the largest sample of stars, the general radial anisotropy found for stars at $\gtrsim1 r_{\rm{hl}}$ in Figure \ref{fig:anisotropy_all} was found to be driven by the P2 population more than P1. However, for many clusters, the anisotropy of the MPs was inconclusive due to the lower number of stars to match with in the photometric catalogue, leading to results such as NGC 3201 in Figure \ref{fig:ani_pop2} for example. Although we found that NGC 3201 shows radial anisotropy in the outer regions, the errorbars of Figure \ref{fig:ani_pop2} make it difficult to conclude whether one population is responsible for this, or both. Any significant anisotropy results for the MPs will be included and discussed in Section \ref{sec:Most_dynamically_young} for the relevant, dynamically young clusters.\\

%%%%%%%%%%%%%%%%%%%%%%%%%%%%%%%%%%%%%%%%%%%%%%%%%%%%%%%%%%%%%%%%%%%%%%%%%%%%%%%%%%%%%%%%%%
%%%%%%%%%%%%%%%%%%%%%%%% COMBINED RESULTS %%%%%%%%%%%%%%%%%%%%%%%%%%%%%%%%%%%%%%%%%%%%%%%%
%%%%%%%%%%%%%%%%%%%%%%%%%%%%%%%%%%%%%%%%%%%%%%%%%%%%%%%%%%%%%%%%%%%%%%%%%%%%%%%%%%%%%%%%%%

\section{Results}
\label{sec:Disc_Kinematics}

\begin{figure*}
    \centering
    \includegraphics[width=0.95\textwidth]{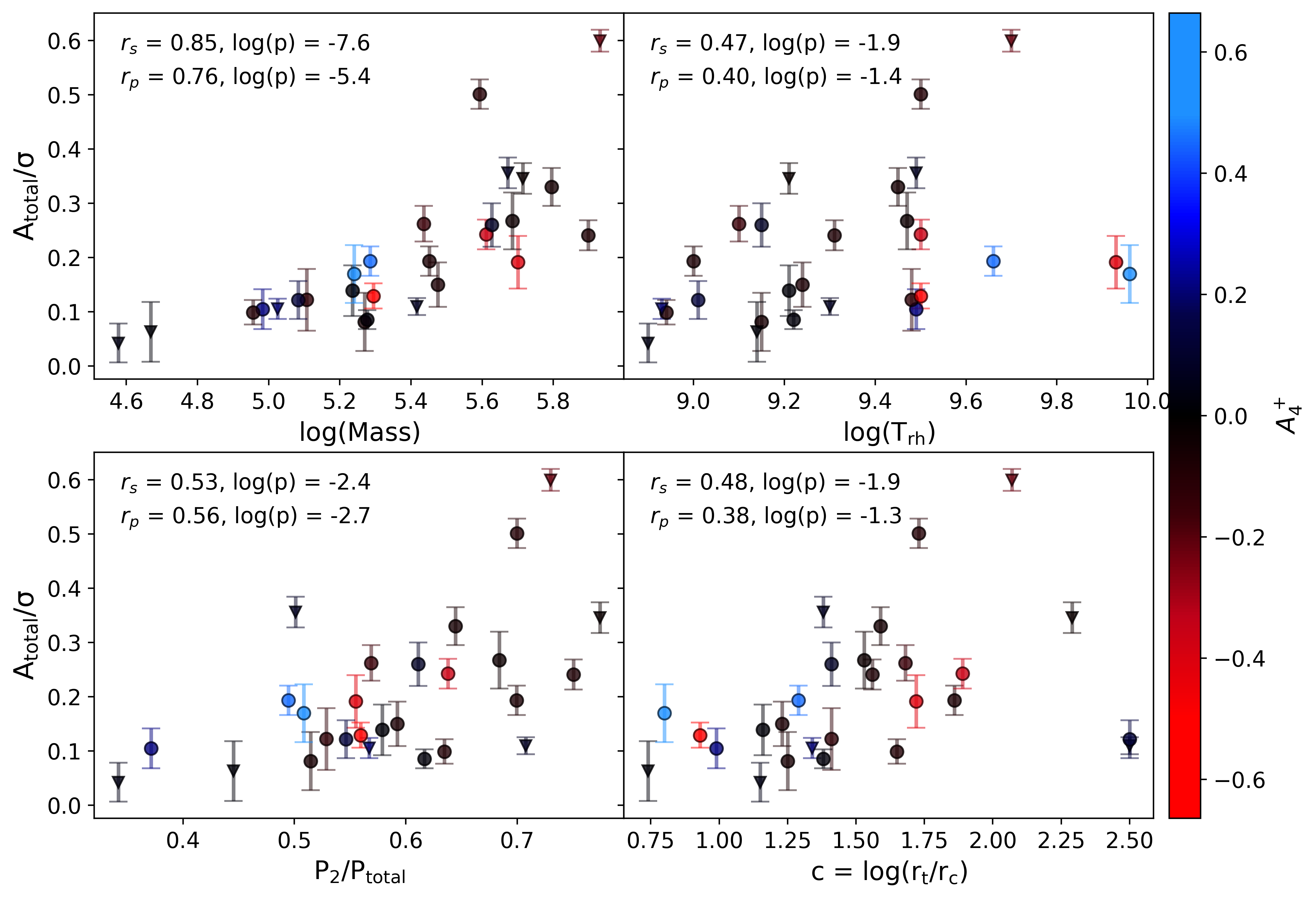}
    \vspace{-0.1cm}
    \caption{The rotation over dispersion ratio $\rm{A_{total}/\sigma}$ as a function of \textit{top left}: current mass \protect\citep{GGCD}, \textit{top right}: relaxation time \protect\citep{GGCD}, \textit{bottom left}: enriched star fraction \protect\citep{2023Leitinger} and \textit{bottom right}: concentration parameter \protect\citep{2010Harris}. Clusters are colour-coded by $A^+_4$ values \protect\citep{2023Leitinger} and each panel shows the Spearman and Pearson rank correlation coefficients for each parameter, with corresponding log p-values to show the significance of the correlations (log(p) $\leq$ -1.3 indicates a p-value $\leq$ 0.05, meaning the correlation is statistically significant and not a result of random chance).}
    \label{fig:Asigma_4panel}
\end{figure*}

This section focuses on the implications of the 3D kinematic analysis of the clusters in our sample, in terms of the rotation over dispersion ratio ($\rm{A_{total}/\sigma}$), position angle of the rotation axis ($\theta_0$) and inclination angle (\textit{i}), for the full sample of stars in Section \ref{sec:rotation_results_all} and the subset which was classified into multiple stellar populations in Section \ref{sec:rotation_results_MPs}.

\begin{table*}
\centering
\caption{The detection of rotation from literature in comparison to this work (following the style of Table 3 of \cite{2019Sollima}) in which checkmarks, crosses and tildes indicate positive, negative and $2\sigma$ detections, respectively. Clusters in our sample for which rotation could not be analysed due to a low number of LOS measurements are marked with a dash. Each column corresponds to the previous works of \cite{2010Lane} (L10), \cite{2012Bellazzini} (B12), \cite{2014Fabricius} (F14), \cite{2015Lardo} (L15), \cite{2015Kimmig} (K15), \cite{2018Kamann} (K18), \cite{2018Mikis} (F18), \cite{2018Helmi} (G18), \cite{2018Bianchini} (B18), \cite{2019Vasiliev} (V19), \cite{2019Sollima} (S19), \cite{2023Martens} (M23) and \cite{2024Petralia} (P24). Checkmarks within parenthesis in the S19 column indicate GCs with uncertain detections in the work of \cite{2019Sollima} (see details within).}
% \resizebox{\hsize}{!}{
\begin{tabular}{ccccccccccccccc}
\hline
\hline\\[-0.7em]
Cluster  & L10        & B12        & F14        & L15        & K15        & K18        & F18        & G18        & B18        & V19        & S19          & M23 & P24        & This work  \\
NGC 104  & \checkmark & \checkmark &            &            & \checkmark & \checkmark &            & \checkmark & \checkmark & \checkmark & \checkmark   & \checkmark & \checkmark & \checkmark \\
NGC 288  & X          & X          &            &            & X        &        & \checkmark &            & X          & X          & X         &   &            & X          \\
NGC 1261 &            &            &            &            &            &            & $\sim$     &            &            &            & X            &            & & X          \\
NGC 1851 &            & \checkmark &            & \checkmark &            & \checkmark &            &            & X          & X          & X            & \checkmark & \checkmark & \checkmark \\
NGC 2808 &            & \checkmark &            & \checkmark & X          & \checkmark &            &            & X          & X          & \checkmark   & \checkmark & \checkmark & \checkmark \\
NGC 3201 &            & \checkmark &            &            &            & \checkmark & \checkmark &            & $\sim$     & X          & $\sim$       & \checkmark & \checkmark& \checkmark \\
NGC 4590 & X          & \checkmark &            &            & X          &            &            &            & X          & X          & X            &            & & X          \\
NGC 4833 &            &            &            & X          &            &            &            &            & X          & X          &              &            & & X          \\
NGC 5024 & X          & X          & \checkmark &            & X          &            &            &            &            &            & $\sim$       &            & & \checkmark \\
NGC 5053 &            &            &            &            & X          &            &            &            &            &            &              &            & & -          \\
NGC 5272 &            &            & \checkmark &            & X          &            & \checkmark & \checkmark & \checkmark & $\sim$     & (\checkmark) &            & \checkmark & \checkmark \\
NGC 5286 &            &            &            &            &            &            &            &            & $\sim$     & X          & X            & \checkmark & & \checkmark \\
NGC 5904 &            & \checkmark & \checkmark &            & \checkmark & \checkmark &            & \checkmark & \checkmark & \checkmark & \checkmark   & \checkmark & \checkmark & \checkmark \\
NGC 5986 &            &            &            &            &            &            &            &            & X          & $\sim$     & $\sim$       &            & & \checkmark \\
NGC 6101 &            &            &            &            &            &            &            &            &            &            &              &            & & \checkmark \\
NGC 6121 & \checkmark & \checkmark &            &            & $\sim$     & $\sim$     &            &            & $\sim$     & X          & X            &            & \checkmark & \checkmark \\
NGC 6205 &            &            & \checkmark &            &            &            &            &            & $\sim$     & X          & \checkmark   &            & \checkmark & \checkmark \\
NGC 6218 & X          & X          & \checkmark &            & X          &            &            &            & X          & X          & (\checkmark) & \checkmark & X & \checkmark \\
NGC 6254 &            & X          & \checkmark &            &            & \checkmark & $\sim$     &            & $\sim$     & X          & X            & \checkmark & \checkmark & \checkmark \\
NGC 6341 &            &            & \checkmark &            & $\sim$     &            &            &            & X          & $\sim$     & (\checkmark) &            & \checkmark & \checkmark \\
NGC 6366 &            &            &            &            &            &            &            &            & X          & X          &              &            &  & X          \\
NGC 6656 & \checkmark & \checkmark &            &            & X          & \checkmark &            & \checkmark & \checkmark & \checkmark & \checkmark   & \checkmark & \checkmark & \checkmark \\
NGC 6752 & X          & X          &            & X          & X          & $\sim$     &            & \checkmark & \checkmark & $\sim$     & (\checkmark) & \checkmark & \checkmark & \checkmark \\
NGC 6809 & $\sim$     & $\sim$     &            &            & $\sim$     &            &            & \checkmark & \checkmark & $\sim$     & (\checkmark) &            & \checkmark & \checkmark \\
NGC 6838 &            & $\sim$     &            &            & X          &            &            &            & X          & X          & $\sim$       &            & X & X          \\
NGC 6934 &            &            & \checkmark &            & X          &            &            &            &            &            &              &            & & -          \\
NGC 6981 &            &            &            &            &            &            &            &            &            &            &              &            & & -          \\
NGC 7078 &            & \checkmark &            & \checkmark & \checkmark & \checkmark &            & \checkmark & \checkmark & \checkmark & \checkmark   & \checkmark & \checkmark & \checkmark \\
NGC 7089 &            &            &            &            & \checkmark & \checkmark &            &            & \checkmark & \checkmark & \checkmark   & \checkmark & & \checkmark \\
NGC 7099 & X          & X          &            &            & X          & $\sim$     &            &            &            & X          & (\checkmark) & \checkmark & & \checkmark\\
\hline
\label{table:rotation_comparisons}
\end{tabular}
% }
\end{table*}

\subsection{3D rotation of the clusters}
\label{sec:rotation_results_all}
We calculated the 3D rotation over dispersion ratios ($\rm{A_{total}/\sigma}$) for 27 Galactic GCs in our sample, as shown in Table \ref{tab:kinematics}, in which we found significant ($>3\sigma$) rotation for 21 of the clusters. We also investigated $\rm{A_{total}/\sigma}$ as a function of every structural and orbital parameter available on the Galactic Globular Cluster Database \citep{GGCD}, as well as parameters calculated in this work, our previous work \citep{2023Leitinger} and the catalogue of parameters for Milky Way globular clusters \citep{2010Harris}. Additionally, we divided the sample into in-situ (7 GCs) vs. accreted (23 GCs) using the classifications listed in \cite{2019Massari}. We present the Pearson and Spearman rank correlation coefficients and p-values for the strongest correlations, as shown in Figure \ref{fig:Asigma_4panel}.\\

The strongest correlations with $\rm{A_{total}/\sigma}$ were found to be with the mass of the cluster (both current and initial), the fraction of enriched stars, the relaxation time and the concentration of the cluster. We show these correlations in Figure \ref{fig:Asigma_4panel}, with in-situ clusters marked as inverse triangles and accreted clusters marked as circles. We found a very strong correlation with the current mass of the cluster ($\rm{log(Mass)}$), shown in the top left panel of Figure \ref{fig:Asigma_4panel}, which was also found when using the estimated initial mass of the cluster ($\rm{log(M_{i})}$). Each panel of Figure \ref{fig:Asigma_4panel} is colour-coded by the $A^+_4$ parameters of \cite{2023Leitinger}, indicating whether a cluster was found to be P1 centrally concentrated (blue), P2 centrally concentrated (red) or spatially mixed (black). Both the current and initial mass correlations produced Pearson and Spearman rank correlation coefficients of $r_s = 0.85, r_p = 0.76$ and $r_s = 0.66, r_p = 0.58$, respectively. Previously, \cite{2018Bianchini} analysed the rotation of 51 Galactic GCs using \gaia DR2 proper motions and compared $V/\sigma$ against the ellipticity, relaxation time, metallicity, concentration, mass-to-light ratio and total cluster mass. When isolating clusters with significant ($>3\sigma$) rotation, \cite{2018Bianchini} found moderate correlations with the mass, metallicity and relaxation time, in that order of significance. However, they note that when considering all clusters, the correlation with metallicity is erased. Our results support the correlation with mass found by \cite{2018Bianchini}, but when we also restrict our sample to only those clusters with significant ($>3\sigma$) rotation, it did not significantly increase our rank correlation coefficients for correlations with mass. We also did not find a strong correlation with metallicity in either case.

In the top right panel of Figure \ref{fig:Asigma_4panel}, we observe a strong correlation between rotation and relaxation time (log($\rm{T_{rh} [yr]}$)), with the exception of 3 dynamically young clusters in particular - NGC 3201 (blue), NGC 5024 (red) and NGC 6101 (blue) - at log($\rm{T_{rh})>9.6}$ and $\rm{A_{total}/\sigma} \approx 0.2$. The only notable difference between these 3 clusters and the remaining clusters which follow the correlation, is that they exhibit some of the largest perigalactic distances of the sample (NGC 3201: $\rm{R_{peri}} = 8.34$ kpc, NGC 5024: $\rm{R_{peri}} = 9.28$ kpc and NGC 6101: $\rm{R_{peri}} = 10.19$ kpc), alongside NGC 4590 ($\rm{R_{peri}} = 8.88$ kpc) which also exhibits low rotation with log($\rm{T_{rh})=9.48}$. This appears to simply be coincidental, as we found no correlations between rotation and eccentricity of the Galactic orbit, apo- or perigalactic distance. Regardless of these outliers, the correlation between rotation and relaxation time confirms the results of \cite{2018Kamann}, \cite{2018Bianchini} and the \textit{N}-body simulations of \cite{2017Tiongco}, which conclude that if a cluster is born with an initial rotation, the ability of that cluster to retain its rotation signal is dependent on the relaxation time. As a cluster dynamically relaxes, angular momentum is redistributed towards the outer regions, effectively slowing the total rotation of the cluster.

In the lower right panel of Figure \ref{fig:Asigma_4panel} a correlation between rotation and the concentration of a cluster can be seen, with concentration defined as the ratio between the tidal radius and core radius of a cluster (log($\rm{r_t/r_c}$)), taken from \cite{2010Harris}. This parameter is driven primarily by the density profile within a few half-light radii, indicating that core-collapsed clusters are rotating faster. In general, the P2 centrally concentrated clusters have higher concentration values ($c>1.7$), while the P1 centrally concentrated clusters have lower concentrations ($c<1.3$). The exception is NGC 6809, which has lost a significant fraction of its initial mass ($\rm{M_{current}/M_{initial}} = 0.26$ \citep{2023Leitinger}). The two outlier clusters with low rotation ($\rm{A_{total}/\sigma} < 0.3$) but high concentration ($c = 2.5$) are NGC 7099 and NGC 6752, both of which have undergone core collapse \citep{2010Harris}, alongside NGC 7078 with $c = 2.3$, which is rotating significantly. It is possible that due to the redistribution of angular momentum during core collapse, the process of core collapse has affected each cluster differently depending on its initial rotation or mass.

The strong correlation between rotation and the enriched star fraction ($\rm{P_2/P_{total}}$) of the clusters in the lower left panel of Figure \ref{fig:Asigma_4panel} was to be expected, as it traces the well established correlation between the mass of a cluster and the fraction of enriched stars it hosts \citep{2017Milone,2023Leitinger}. Similarly, more massive clusters tend to have longer relaxation times, while longer relaxation times allow clusters to retain their initial rotation more efficiently \citep{2018Bianchini}. We tested whether the strong correlation between rotation and relaxation time was also caused by a correlation with mass. To confirm this, we performed a residual analysis to remove the effect of mass from the rotation, finding that all three of the strong correlations we found ($\rm{P_2/P_{total}}$, log($\rm{T_{rh}}$) and $c$) disappeared. \\

\begin{figure}
    \centering
    \includegraphics[width=\hsize]{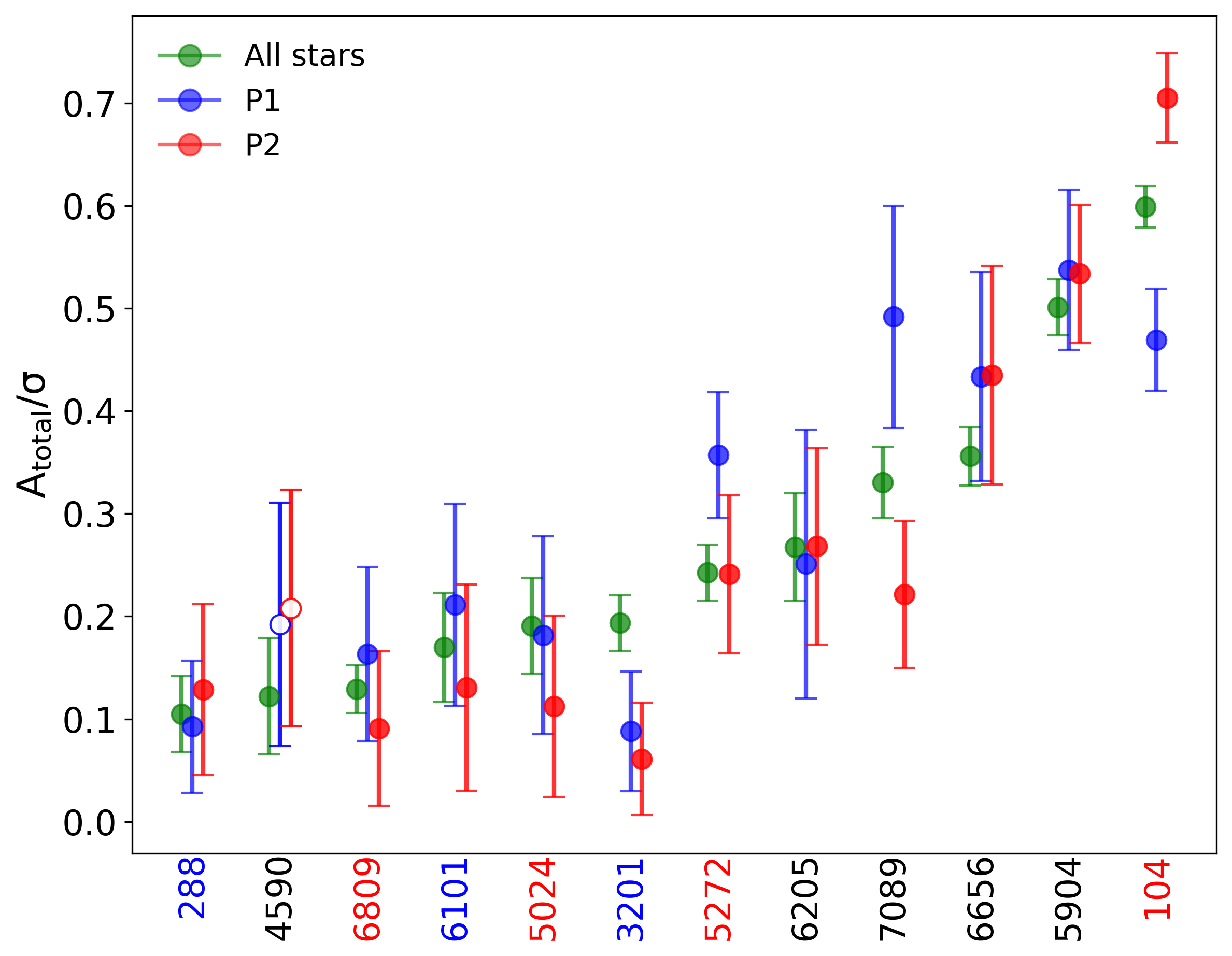}
    \vspace{-0.3cm}
    \caption{The total rotational amplitudes divided by the velocity dispersion for the P1 (blue) and P2 (red) stars for each of the dynamically young clusters in our sample, as well as for all stars (green) as described in Section \ref{sec:rotation_axes} and listed in Table \ref{tab:kinematics}. Solid circles indicate that both populations contain enough stars with LOS velocities for a statistically robust result ($N_{\rm{LOS,P1}} + N_{\rm{LOS,P2}} >$ 300), while open circles indicate the opposite. Cluster names are colour-coded based on whether the cluster has a significant P1 (blue) central concentration, P2 (red) central concentration or are spatially mixed (black). Clusters are ordered based on the $\rm{A_{total}/\sigma}$ values derived using all available stars as listed in Table \ref{tab:kinematics}, such that clusters on the left (right) have lower (higher) rotation.}
    \label{fig:rotation_young}
\end{figure}

We also present the 3D rotation over dispersion ratios ($\rm{A_{total}/\sigma}$) calculated in this work in Table \ref{table:rotation_comparisons} alongside previous literature values for comparison. The fastest rotating clusters in our sample: NGC 104, NGC 5904, NGC 6656 and NGC 7078 are also shown to exhibit fast rotation in the vast majority of previous kinematic results in Table \ref{table:rotation_comparisons}. Overall there is only one rotating cluster in our sample: NGC 5986, which somewhat disagrees with previous literature, although we can only compare to \cite{2019Sollima} and \cite{2019Vasiliev} who state the rotation is ambiguous, while \cite{2018Bianchini} found no rotation. In general, our results for the rotation of each cluster agree with at least one other previous literature result, where available (note that NGC 6101 was not included in any of the previous works listed).

Our sample of GCs has the largest overlap with the work of \cite{2023Martens}, who performed a large scale kinematic analysis of 25 Galactic GCs using LOS velocities derived from MUSE spectroscopy, 13 of which are present in our sample. Our works agree in finding unambiguous rotation for all 13 of these overlapping clusters, despite their analysis focusing only on the central regions of each cluster, without the use of proper motions. Additionally, our results agree for all except one cluster in common with the analysis of \cite{2024Petralia}, who used APOGEE DR14 spectra to determine rotational signatures from LOS velocities. For the cluster in disagreement between our works: NGC 6218, we also see contention in whether it is rotating from the previous analyses in Table \ref{table:rotation_comparisons}.

Our rotation results also agree mostly with the results of \cite{2019Sollima}, who analysed a large sample of 62 Galactic GCs using \gaia DR2 proper motions and the 2018 version of the radial velocity catalogue from \cite{2018BaumgardtHilker}. More specifically, we both found unambiguous rotation signals for 7 clusters in common: NGC 104, NGC 2808, NGC 5904, NGC 6205, NGC 6656, NGC 7078 and NGC 7089. For the majority of the significantly rotating clusters in our sample such as NGC 5272, NGC 6218, NGC 6341, NGC 6752 and NGC 6809, Table 1 of \cite{2019Sollima} lists these clusters as having a $P \geq 99.9$\% chance of rotating, even if they classify the rotation as `uncertain' (see Section 3.2 in \cite{2019Sollima} for more details). There is disagreement between our works for the clusters NGC 1851, NGC 5286, NGC 6121 and NGC 6254, however, we note that in each case it is because we found a rotation signal which was not apparent in the analysis of \cite{2019Sollima}. We also note that both kinematic data sets in this work, \gaia DR3 and the LOS velocity catalogues first produced by \cite{2018BaumgardtHilker}, are updated versions of the data sets used by \cite{2019Sollima}, and therefore allow for results with lower error bars, which may explain the discrepancies between our works.

We also found good agreement with \cite{2018Kamann}, who completed an analysis on the rotation of 22 Galactic GCs using MUSE line-of-sight velocities, in which 9 GCs show unambiguous rotation in both their work and ours, out of 12 GCs that overlap with our sample. For those other 3 overlapping clusters, we found rotation at a $>3\sigma$ significance, but \cite{2018Kamann} found rotation at only a $2\sigma$ significance, which may be due to MUSE targeting only the inner regions of each cluster, while we use velocity measurements in the outer regions as well. The position angles of the rotation axes calculated by \cite{2018Kamann} for the 12 overlapping clusters are in agreement with our results and show that the position angles are generally perpendicular to the photometric semi-major axis for most clusters. These results support the argument presented by \cite{2014Fabricius}, in which an oblate, isotropic, rotating cluster is expected to have its rotation axis perpendicular to its flattening axis. Additionally, \cite{2018Kamann} discovered that 5 of the clusters in their sample showed evidence of a decoupled core, exhibited by a change in the position angle as a function of radius. This implies the centre of the cluster is rotating differently from the outer regions and is proposed to be explained by tidal interactions influencing the outer regions, while the central regions still retain traces of the rotational signatures created at birth. Of these 5 GCs with decoupled cores, 3 are part of our sample as well: NGC 5904, NGC 6254 and NGC 7078. All 3 of these clusters are fast rotators, but only NGC 5904 (age/$\rm{T_{rh}}$ = 3.6 $\pm$ 0.2) is considered dynamically young, albeit on the older end. It is expected that fast rotating GCs which have already gone through multiple relaxation times were rotating even faster at birth \citep{2015HenaultGieles}, which may also be the case for these clusters. \\
\begin{figure*}
    \centering
    \includegraphics[width=0.9\textwidth]{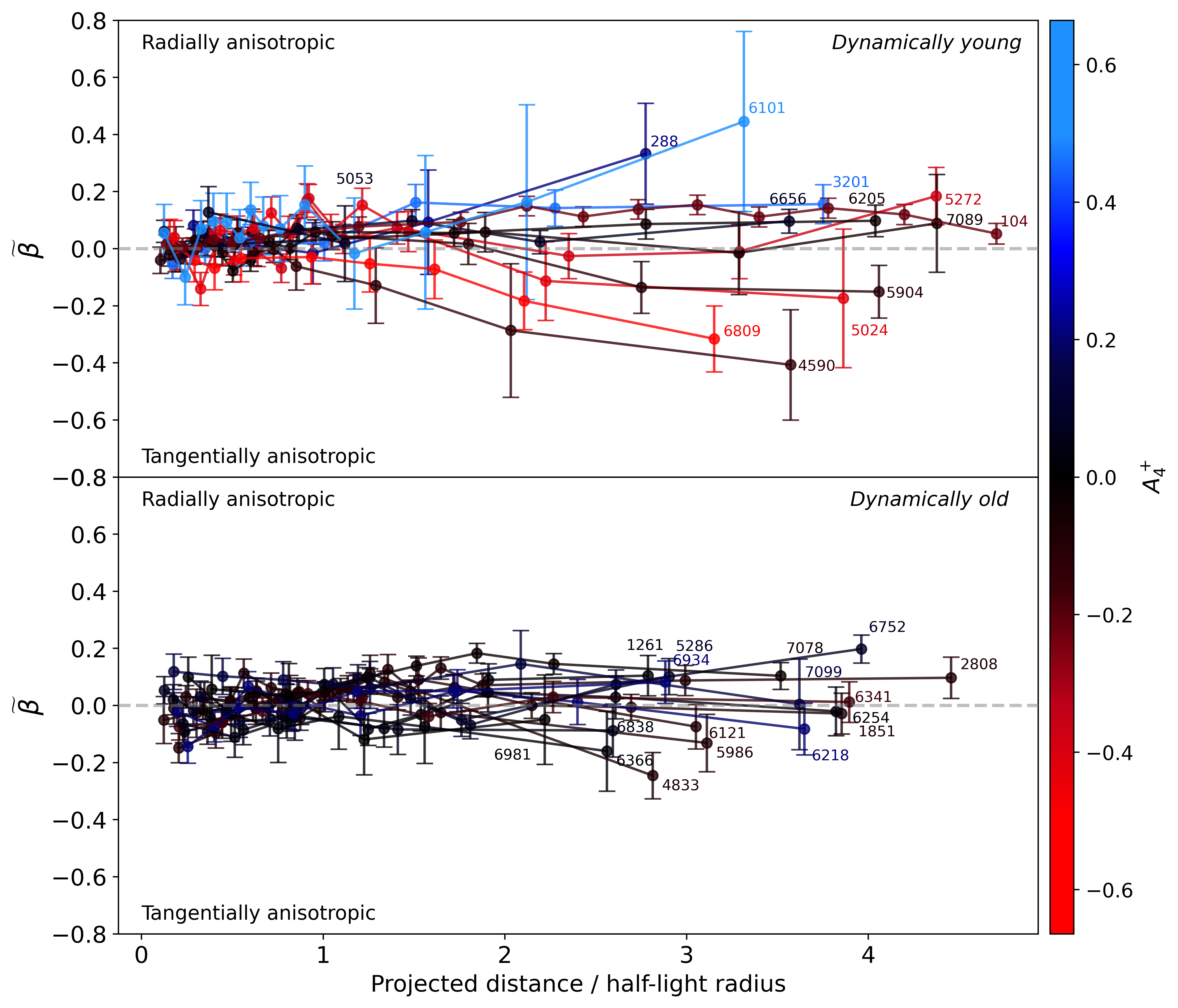}
    \vspace{-0.1cm}
    \caption{The normalised anisotropy of \textit{top}: the dynamically youngest GCs in the sample ($\rm{age/T_{rh}} <$ 4.5) and \textit{bottom}: the dynamically old clusters ($\rm{age/T_{rh}} >$ 4.5) as a function of the projected distance normalised by the half-light radius of each cluster. Each cluster is colour-coded by its $A^+$ parameter, as calculated in \protect\cite{2023Leitinger}. Clusters in blue(red) are shown to have P1(P2) more centrally concentrated, while clusters in black have spatially mixed populations.}
    \label{fig:anisotropy_young}
\end{figure*}

\subsection{3D rotation of the multiple stellar populations}
\label{sec:rotation_results_MPs}
We now focus on the dynamically youngest clusters in our sample (age/$\rm{T_{rh}}$ < 4.5), in terms of the 3D rotation over dispersion ratios ($\rm{A_{total}/\sigma}$) of their multiple stellar populations, as shown in Figure \ref{fig:rotation_young} and displayed in Table \ref{tab:kinematics_MPs} in Appendix \ref{sec:Appendix_pa}. The x-axis is ordered by the $\rm{A_{total}/\sigma}$ values using all stars, as described in Section \ref{sec:rotation_axes}, with NGC 288 showing the lowest degree of rotation ($\rm{A_{total}/\sigma} = 0.10 \pm 0.04$) and NGC 104 showing the highest ($\rm{A_{total}/\sigma} = 0.60 \pm 0.02$). The clusters in Figure \ref{fig:rotation_young} are colour-coded based on the values of $A^+_4$ calculated in \cite{2023Leitinger}, indicating whether they exhibit a significant P1 central concentration (blue), P2 central concentration (red) or are spatially mixed (black). We found no consensus between whether one population rotates faster than the other, even if one population is more centrally concentrated. For example, in the P1 centrally concentrated clusters NGC 288, NGC 3201 and NGC 6101, we do not find that the P1 stars located primarily in the inner regions rotate faster than P2 stars. NGC 3201 and NGC 6101 could be argued as having P1 stars rotating faster than P2 stars, but NGC 288 shows the opposite (ignoring error bars). Similarly, the P2 centrally concentrated clusters NGC 104, NGC 5024, NGC 5272 and NGC 6809 show no evidence of the P2 stars consistently rotating faster. One may conclude the P2 stars of NGC 104 are rotating faster than the P1 stars, but in NGC 5024, NGC 5272 and 6809 (ignoring error bars), the opposite is observed. This general lack of evidence for one population preferentially rotating faster than the other is supported by \cite{2023Martens}, but is at odds with the work of \cite{2024Dalessandro}, who investigated the 3D kinematics of MPs in 16 Galactic GCs and observed that P2 stars preferentially rotate faster than P1 stars.

The largest difference between the $\rm{A_{total}/\sigma}$ values of the P1 and P2 stars in our sample of clusters is observed in NGC 104, with P2 rotating faster than P1 at a $3.5\sigma$ significance (see Table \ref{tab:kinematics_MPs}, but no differences in the position angles of the MPs. This result is supported by the work of \cite{2024Dalessandro}, who found that P2 stars rotate faster than P1 stars in NGC 104, with no significant difference between the position angles. The only other cluster in which we found a difference of $>2\sigma$ between the populations is NGC 7089, in which P1 is rotating faster than P2, but NGC 7089 was not part of the sample in \cite{2024Dalessandro}. The results of \cite{2023Martens} for NGC 104 and NGC 7089 did not identify a difference in the rotation over dispersion, ($(v/\sigma)_{\rm{HL}}$ evaluated at the half-light radius) between the populations. However, \cite{2023Martens} did find differences between the rotation of the multiple stellar populations for NGC 2808, NGC 6093 (not in our sample) and NGC 7078, at around $\sim1-2\sigma$ significance. In NGC 2808 we found that P1 is rotating faster than P2 at a $1.8\sigma$ significance, which is supported by the result of \cite{2023Martens} with the same significance level. In NGC 7078 we found that P1 is rotating faster than P2 at a significance level of $<1\sigma$, while \cite{2023Martens} found the opposite: P2 stars rotating faster than P1 at a 2.2$\sigma$ significance. As the work of \cite{2023Martens} focuses on the centre of the cluster with MUSE and NGC 7078 has been shown to exhibit a decoupled core \citep{2018Kamann}, it is expected that the inner and outer regions of the cluster are rotating differently, which may explain the disparity between our works for NGC 7078, while for NGC 2808 we produce compatible results. The calculated rotation values for the multiple populations of all clusters in our sample is shown in Table \ref{tab:kinematics_MPs} in Appendix \ref{sec:Appendix_pa}. \\

\begin{figure}
    \centering
    \includegraphics[width=\hsize]{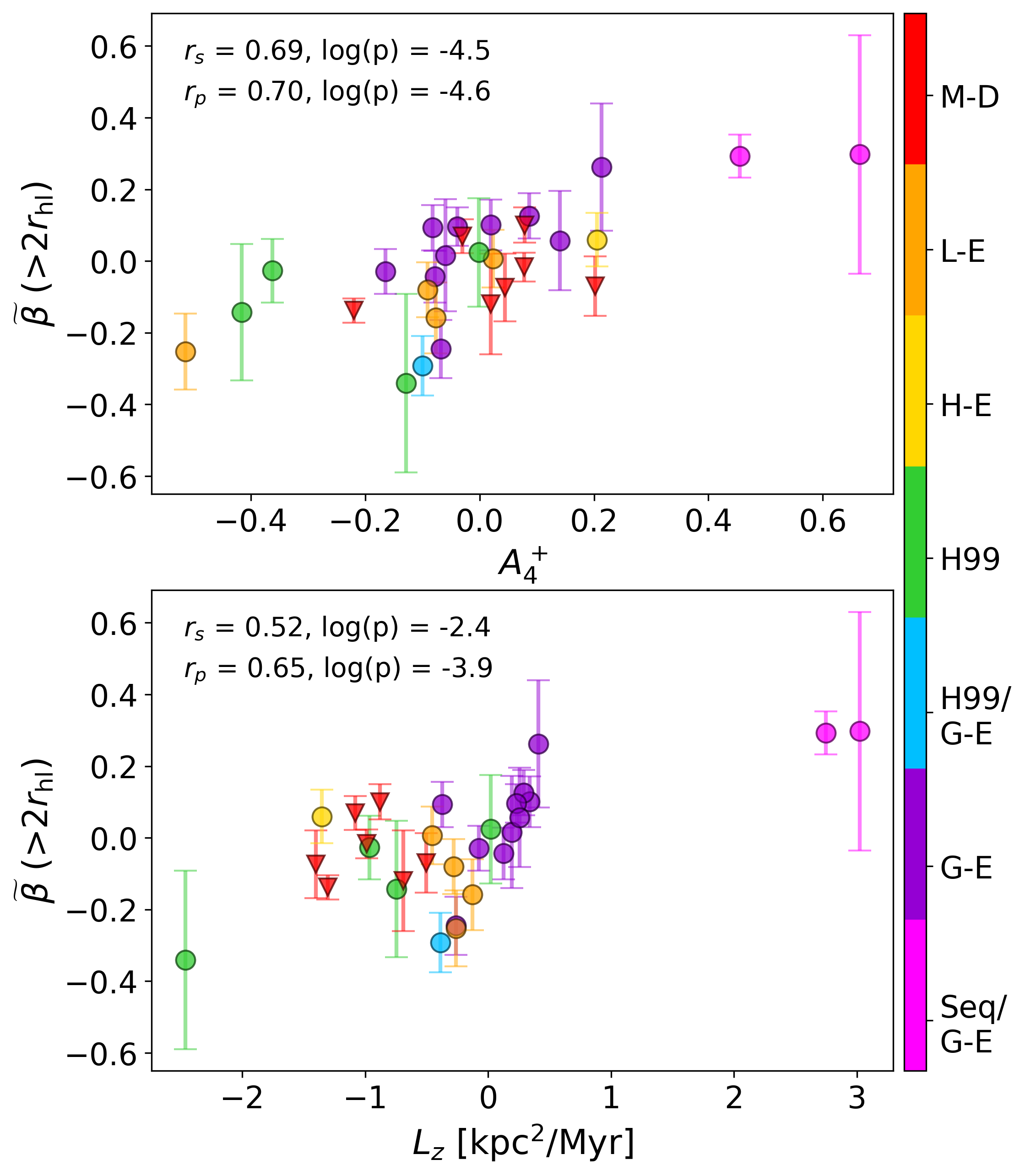}
    \vspace{-0.1cm}
    \caption{The average normalised anisotropy in the outer regions ($>2 r_{\rm{hl}}$) using only \gaia proper motions, without removing the contribution by rotation. \textit{Top panel}: the correlation with the cumulative radial distribution parameter $A^+_4$ \protect\citep{2023Leitinger}, and \textit{bottom panel}: the correlation against the z-component of the angular momentum, where positive(negative) values indicate retrograde(prograde) orbits.   Clusters are colour-coded by their progenitor structures, as classified by \protect\cite{2019Massari}. M-D: main disk of the Milky Way progenitor, L-E: unassociated low-energy group, H-E: high-energy unassociated group, H99: Helmi streams, G-E: Gaia-Enceladus, Seq: Sequoia. Each panel also shows the Spearman and Pearson rank correlation coefficients for each parameter, with corresponding log p-values to show the significance of the correlations (log(p) $\leq$ -1.3 indicates a p-value $\leq$ 0.05, meaning the correlation is statistically significant and not likely to be the result of random chance).}
    \label{fig:Nbeta_2panel}
\end{figure}

\subsection{Anisotropy}
\label{sec:anisotropy_results}

In our anisotropy analysis, we isolated the dynamically youngest clusters in our sample (age/$\rm{T_{rh}}$ < 4.5) and normalised the projected radii of each cluster to its half-light radius. We then truncated every cluster to $5 r_{\rm{hl}}$ where possible and calculated the anisotropy for all cluster stars in our sample, shown together in Figure \ref{fig:anisotropy_young}, with the contribution by rotation removed. All clusters have stars out to $5 r_{\rm{hl}}$ in Figure \ref{fig:anisotropy_young}, except for NGC 5053. Here, we show that clusters which were found to have a P1 central concentration ($A^+ > 0.2$) exhibit radial anisotropy in the outer regions, while clusters with a P2 central concentration ($A^+ < -0.2$) tend to exhibit tangential anisotropy in the outer regions, with the exception of NGC 104 and NGC 5272. The dynamically old clusters in our sample ($\rm{age/T_{rh}}$ $>$ 4.5) have mixed populations ($A^+ \sim 0$) and are largely isotropic throughout, as shown in the bottom panel of Figure \ref{fig:anisotropy_young}. But regardless of dynamical age, all clusters show approximately isotropic behaviour in their inner regions ($<1 r_{\rm{hl}}$ in Figure \ref{fig:anisotropy_young}), which is covered by the HACKS catalogue. In agreement with \cite{2015Watkins} and \cite{2022Libralato}, isotropy is expected for the denser central regions of clusters, especially those which have undergone several relaxation times in their lifetime. However, this work has uncovered a previously undiscovered result: clusters with P1 central concentrations exhibit radial anisotropy in their outskirts, while those with P2 central concentrations show tangential anisotropy or isotropy in their outer regions.

An alternative version of Figure \ref{fig:anisotropy_young} is shown in Appendix \ref{sec:Appendix_dold_anisotropy}, in which rotation is \textit{not} removed, therefore excluding stars with only HACKS proper motions due to the lack of rotation and instead only including the \gaia proper motions with rotation included. There are minimal differences in the anisotropy profiles for the majority of clusters whether the rotation is removed (Figure \ref{fig:anisotropy_young}) or not (Figure \ref{fig:gaia_anisotropy_young}), except for NGC 104 which changed from radially anisotropic in the outer regions without rotation, to tangentially anisotropic with rotation. In Figure \ref{fig:gaia_anisotropy_young} we find that all P2 centrally concentrated clusters except NGC 5272 are tangentially anisotropic in the outer regions, with NGC 5272 showing more isotropy in its outer regions than shown in Figure \ref{fig:anisotropy_young}. All P1 centrally concentrated clusters still show radial anisotropy in the outer regions when rotation is included.

NGC 5024 is the dynamically youngest cluster in our sample to contain a P2 central concentration, followed by NGC 104 and then NGC 5272 in the upper panels of Figure \ref{fig:anisotropy_young} and \ref{fig:gaia_anisotropy_young}. NGC 6809 is another cluster with a P2 central concentration, but although it is still considered dynamically young, it has undergone a high degree of mass loss (Figure \ref{fig:updated_Aplus_mlr}) and is therefore not likely to be indicative of its initial conditions despite showing significant tangential anisotropy. Even so, it is difficult to make robust conclusions regarding a correlation between P2 central concentrations and tangential anisotropy, as two of the dynamically young clusters which contain spatially mixed populations according to \cite{2023Leitinger}: NGC 4590 and NGC 5904, also exhibit clear tangential anisotropy in the outskirts, with NGC 4590 displaying the strongest tangential anisotropy in our sample. The remaining spatially mixed clusters: NGC 6205, NGC 6656, NGC 7089 are approximately isotropic throughout. Clusters with the highest P1 central concentrations: NGC 288, NGC 3201 and NGC 6101 all exhibit radial anisotropy in the outskirts, which may be a result of dynamical evolution from an initially compact cluster (e.g. \cite{2016Tiongco}), or a surviving signature from the initial conditions. We investigate these clusters in further detail in Section \ref{sec:P1conc}.\\

As the central regions of each cluster are approximately isotropic, the focus of the next section will be on the outer regions; specifically, we only use stars $>2r_{\rm{hl}}$ from the cluster centre and average the anisotropy values for these outer regions. In this way, we are focusing on the regions which have a greater chance of still being representative of the initial conditions at birth. We average the anisotropy bins in these outer regions - which corresponds to the bins with projected distance / half-light radius $>2$ in Figure \ref{fig:gaia_anisotropy_young}. For this section we only use the anisotropy profiles in which rotation is included, as the aim is to investigate trends with the general motion of the stars, as opposed to the intrinsic motion of the stars relative to one another.

We investigated the average rotation-included anisotropy in the outer regions covered only by \gaia ($\widetilde{\beta} (r_{\rm{hl}} \geq 2$)) as a function of every structural and orbital parameter available on the Galactic Globular Cluster Database \citep{GGCD},  \citep{2023Leitinger}, \citep{2010Harris} and parameters calculated in this work. The clusters were classified as in-situ (7 GCs) or accreted (23 GCs) using the classifications of \cite{2019Massari}. We present the Pearson and Spearman rank correlation coefficients and p-values for each parameter which showed the strongest correlations, shown in Figure \ref{fig:Nbeta_2panel}.\\

We found correlations with anisotropy between both the cumulative radial distribution parameter ($A^+_4$ from \cite{2023Leitinger}) and the z component of the angular momentum ($L_z$). Accreted clusters are shown as circles, while in-situ clusters are shown as inverted triangles. The clusters are colour-coded by their progenitor, as determined by \cite{2019Massari}. Clusters with the highest P2 central concentrations ($A^+_4 < -0.2$) display either tangential anisotropy or isotropy. Of these clusters, NGC 5024 and NGC 5272 originate from the Helmi streams, NGC 104 formed in-situ and NGC 6809 is part of an unclassified low-energy structure (see \cite{2019Massari} for details regarding the classifications). Clusters which formed in-situ within the Milky Way are marked as red inverted triangles and are largely isotropic with spatially mixed populations. The exception to this is the in-situ cluster NGC 104, which - when keeping the contribution of rotation in the anisotropy analysis - displays tangential anisotropy in its outer regions. The most P1 centrally concentrated clusters, NGC 3201 and NGC 6101 ($A^+_4 > 0.3$), are the most radially anisotropic clusters and are also the only 2 accreted clusters in our sample classified as originating from either Sequoia or Gaia-Enceladus, with \cite{2019Massari} stating that the progenitor is preferred to be Sequoia. It is possible that either 1) the specific star and cluster formation conditions within the progenitor system influenced both the P1 central concentration and radial anisotropy, 2) the accretion event of this progenitor influenced these parameters or 3) a combination of both.

The correlation between anisotropy in the outer regions and the z-component of the angular momentum further supports the idea that the physical conditions in the progenitor systems had an influence on the early stages of dynamical evolution in these clusters. However, there are 3 clusters in particular which seem to contribute the most to this correlation: NGC 4590 at $L_z < -2\,{\rm kpc}^2/{\rm Myr}$ and NGC 3201 and NGC 6101 both at $L_z > 2\,{\rm kpc}^2/{\rm Myr}$. In order to test the validity of this correlation, we experimented with removing certain clusters from the sample. The results of this experiment are shown in Table \ref{table:correlation_test}, which show the Pearson and Spearman rank correlation coefficients and associated log p-values for each test. We began by removing all in-situ clusters (red inverted triangles in Figure \ref{fig:Nbeta_2panel}) to focus only on accreted clusters, which strengthened the correlation. Removing the extremes in $L_z$: NGC 4590 \textit{or} both NGC 3201 and NGC 6101 still produced strong correlations. Removing the in-situ clusters, NGC 4590, NGC 3201 and NGC 6101 still produced a strong correlation for the Spearman coefficient, but the p-value for the Pearson coefficient suggested the result could be a product of random chance. Finally, keeping the in-situ clusters, but removing the extreme $L_z$ clusters: NGC 4590, NGC 3201 and NGC 6101, produced p-values for both coefficients that suggest the result is not statistically significant. Regardless, these extreme clusters are a part of our sample and the full sample of clusters does produce a strong correlation with $L_z$ which, when coupled with the $A^+_4$ parameter correlation, suggests that the physical conditions of star cluster formation could be different  in the former host galaxies, and thus influence the early stages of the dynamical evolution of their star clusters, leaving present-day clues in their outer regions. Future investigations would benefit from the inclusion of more clusters with extreme $L_z$ values. For example, adding the anisotropy of the clusters NGC 7006, NGC 4499, NGC 1758 and Pal 13 - all of which exhibit $L_z>1\,{\rm kpc}^2/{\rm Myr}$, along with Pal 10, Pal 12, NGC 5824, NGC 2419, Pal 4, Pal 3 and Pal 1 with $L_z<-2\,{\rm kpc}^2/{\rm Myr}$, would serve to fill in the gaps in Figure \ref{fig:Nbeta_2panel} and confirm whether the correlation is still significant.

\begin{table}
\label{table:correlation_test}
\caption{Tests to check the validity of the correlation found in the bottom panel of Figure \ref{fig:Nbeta_2panel}. The first column is the condition tested, the second and third are the Spearman rank correlation coefficients and associated log p-values, while the fourth and fifth columns are the Pearson rank correlation coefficients and associated log p-values. Green log p-values indicate $p<0.05$, in which the result is statistically significant, while red values indicate the opposite.}
\resizebox{\hsize}{!}{
\begin{tabular}{ccccc}
\hline
\hline\\[-0.7em]
Condition                    & $r_s$ & log(p)                      & $r_p$ & log(p)                       \\[0.4em]
\hline
All clusters                 & 0.52  & {\color[HTML]{32CB00} -2.4} & 0.65  & {\color[HTML]{32CB00} -3.9}  \\
In-situ removed              & 0.70  & {\color[HTML]{32CB00} -3.5} & 0.71  & {\color[HTML]{32CB00} -3.7}  \\
Only NGC 4590 removed                  & 0.47  & {\color[HTML]{32CB00} -1.9} & 0.59  & {\color[HTML]{32CB00} -3.1}  \\
NGC 3201 and NGC 6101 removed                  & 0.41  & {\color[HTML]{32CB00} -1.5} & 0.45  & {\color[HTML]{32CB00} -1.7}  \\
In-situ, NGC 4590, NGC 3201 \\and NGC 6101 removed & 0.53  & {\color[HTML]{32CB00} -1.7} & 0.36  & {\color[HTML]{FE0000} -0.8}  \\
NGC 4590, NGC 3201 \\and NGC 6101 removed          & 0.34  & {\color[HTML]{FE0000} -1.0} & 0.27  & {\color[HTML]{FE0000} -0.75}\\
\hline
\end{tabular}
}
\end{table}

%%%%%%%%%%%%%%%%%%%%%%%%%%%%%%%%%%%%%%%%%%%%%%%%%%%%%%%%%%%%%%%%%%%%%%%%%%%%%%%%%%%%%%%%%%
%%%%%%%%%%%%%%%%%%%%%%%% DYNAMICALLY YOUNG %%%%%%%%%%%%%%%%%%%%%%%%%%%%%%%%%%%%%%%%%%%%%%%
%%%%%%%%%%%%%%%%%%%%%%%%%%%%%%%%%%%%%%%%%%%%%%%%%%%%%%%%%%%%%%%%%%%%%%%%%%%%%%%%%%%%%%%%%%

\section{The dynamically youngest clusters}
\label{sec:Most_dynamically_young}

The focus in this section will be on the dynamically youngest clusters (age/$\rm{T_{rh}}$ < 4.5) in our sample, combining all available information from our spatial analysis in \cite{2023Leitinger} and kinematic analyses in this work, comparing with previous literature results where possible. The dynamically young clusters are separated into three sections: those which exhibit a P2 central concentration are discussed in Section \ref{sec:P2conc}, P1 centrally concentrated clusters are discussed in Section \ref{sec:P1conc} and those with an approximately homogeneous mix of the populations are discussed in Section \ref{sec:Apluszero}. 

\subsection{P2 centrally concentrated clusters}
\label{sec:P2conc}
In this section we discuss the individual results of the clusters NGC 104, NGC 5024, NGC 5272 and NGC 6809, which contain a significant central concentration of the enriched stars. The majority of these clusters also exhibit tangential anisotropy in the outer regions, with the exception of the isotropic cluster NGC 5272 (See Figure \ref{fig:anisotropy_young}) and NGC 104 when removing the effect of rotation (see the caveat in Section \ref{sec:vdisp} and discussion in Appendix \ref{sec:Appendix_dold_anisotropy}). Additionally, each cluster was found to have higher concentration parameters as calculated by \cite{2010Harris}, with the exception of NGC 6809, which has undergone significant mass loss in comparison to the other clusters.

\subsubsection{NGC 104 (47 Tucanae)}
NGC 104 is one of the brightest, closest and most massive GCs in the Milky Way and has thus been the focus of many photometric and kinematic studies. In this work, we identified a P2 central concentration in NGC 104 which has also been well established in previous literature \citep{2012Milone,2014Cordero,2022Jang}, with a high enriched star fraction of $\rm{P_2 / P_{total}} = 0.73\pm0.02$. It is also the fastest rotating cluster in our sample with $\rm{A_{total}/\sigma} = 0.60 \pm 0.02$, which is supported by many previous kinematic analyses \citep{2010Lane,2012Bellazzini,2015Kimmig,2018Kamann,2018Helmi,2018Bianchini,2019Vasiliev,2019Sollima,2023Martens}. We found clear and high rotation of NGC 104, but with no differences between the P1 and P2 stars in terms of position angle of the rotation axis or inclination angle. However, we did find a difference between the rotation ($\rm{A_{total}/\sigma}$ shown in Table \ref{tab:kinematics_MPs} in Appendix \ref{sec:Appendix_pa}) of the MPs, in which P2 rotates faster than P1 at a $3.5\sigma$ significance, which supports the same result found by \cite{2024Dalessandro}. However, this result is not supported by \cite{2023Martens} using MUSE LOS velocities, suggesting the contribution to rotation given by the outer regions and/or the tangential component of the proper motions contributes to this significant difference in rotation between the populations. A lack of significant rotational differences between the populations of NGC 104 was also reported by \cite{2018Milone} using \gaia DR2 proper motions, in which they observed P1 stars with higher tangential velocities than P2 stars, to a $\sim2\sigma$ level of significance - the opposite of our results. Similarly, \cite{2020Cordoni} combined \gaia DR2 proper motions with VLT LOS velocities and found no significant differences between the rotation of the MPs. An analysis by \cite{2023Scalco} using \gaia DR3 proper motions and magnitudes converted into stellar masses through isochrone fitting, found a rotation-mass relation in which rotational velocity increases with stellar mass. They claim this is not the result of mass segregation for NGC 104, but is instead due to long term dynamical evolution causing more massive stars to rotate faster around the core of the cluster than low-mass stars. \cite{2023Scalco} did not assign population classifications to the stars in their sample, but since rotational velocity is a function of radius, the P2 central concentration in NGC 104 suggests that the P2 stars are on average located closer to the peak of the rotation curve than P1 stars.\\

In terms of anisotropy in NGC 104, we found an approximately isotropic central region, but two different results for the outer regions, depending on whether we remove the contribution by rotation or not (see the discussion in Appendix \ref{sec:Appendix_dold_anisotropy}). There is significant tangential rotation shown in the middle panel of Figure \ref{fig:rotation_amp_all_stars}, which greatly affects the motion of the stars in the cluster, as the bulk motion of the stars is tangential motion. Removing this rotation when calculating the velocity dispersion reveals that the intrinsic random motion of the stars shows radial anisotropy in the outer regions. This result is consistent with the anisotropy analysis of  \cite{2018Milone}, \cite{2018Bianchini}, \cite{2020Cordoni} and \cite{2022Libralato}, who also remove the contribution by rotation. Additionally, we found that the P2 stars are contributing to this radial anisotropy (Figure \ref{fig:ani_pop1}), with P1 stars exhibiting isotropy throughout the cluster. This approach allows us to observe the intrinsic motion of the stars in a non-rotating reference frame, such that the motion is only showing the intrinsic differences between stars, allowing assumptions to be made about the internal dynamical processes between stars. Because of this reference frame, \cite{2018Milone} concluded that the combination of rotation with the radial anisotropy of P2 stars is indicative of a scenario in which the P2 stars were born more centrally concentrated and are diffusing outwards due to the radial anisotropy, while P1 stars are isotropic throughout. However, we found that when considering the general orbits of stars in the context of the bulk motion due to rotation \textit{and} the intrinsic motions (Appendix \ref{sec:Appendix_dold_anisotropy}), NGC 104 is tangentially anisotropic in the outer regions. Furthermore, we found this tangential anisotropy was driven by the P1 stars, whereas the faster rotating P2 stars are isotropic throughout. To understand the complex interplay between rotation and anisotropy observed in NGC 104, detailed dynamical modelling of the cluster will be needed, which is beyond the scope of this paper. Overall, NGC 104 shows high rotation with tangential anisotropy and no significant differences between the rotation axes of its MPs, but with P2 stars rotating faster than P1 stars.\\

\subsubsection{NGC 5024 (M 53)}
NGC 5024 is one of the dynamically youngest clusters in our sample, exhibiting a relatively low rotation over dispersion ratio with $\rm{A_{total}/\sigma} = 0.19 \pm 0.05$. \cite{2017Boberg} studied the rotation of NGC 5024 using spectra from the Hydra sample on the WIYN 3.5 m telescope for 245 stars, finding it has a peak rotation amplitude of 1.4 $\pm$ 0.1 km/s. These stars are a subset of the LOS catalogue we have analysed in this work, with our full sample of 344 stars exhibiting a lower rotation of $\Delta A = $0.7 $\pm$ 0.3 km/s for the LOS amplitude alone, but a total rotational amplitude of $\rm{A_{total}} = 1.4 \pm 0.3$. Additionally, \cite{2017Boberg} state that their discovery of a radial gradient of the position angle of the rotation axis suggests that the inner region of NGC 5024 is still reflective of the cluster formation and early dynamics, while the outer region is more affected by the Milky Way tidal field and interactions with the nearby cluster NGC 5053.

\begin{figure}
    \centering
    \includegraphics[width=\hsize]{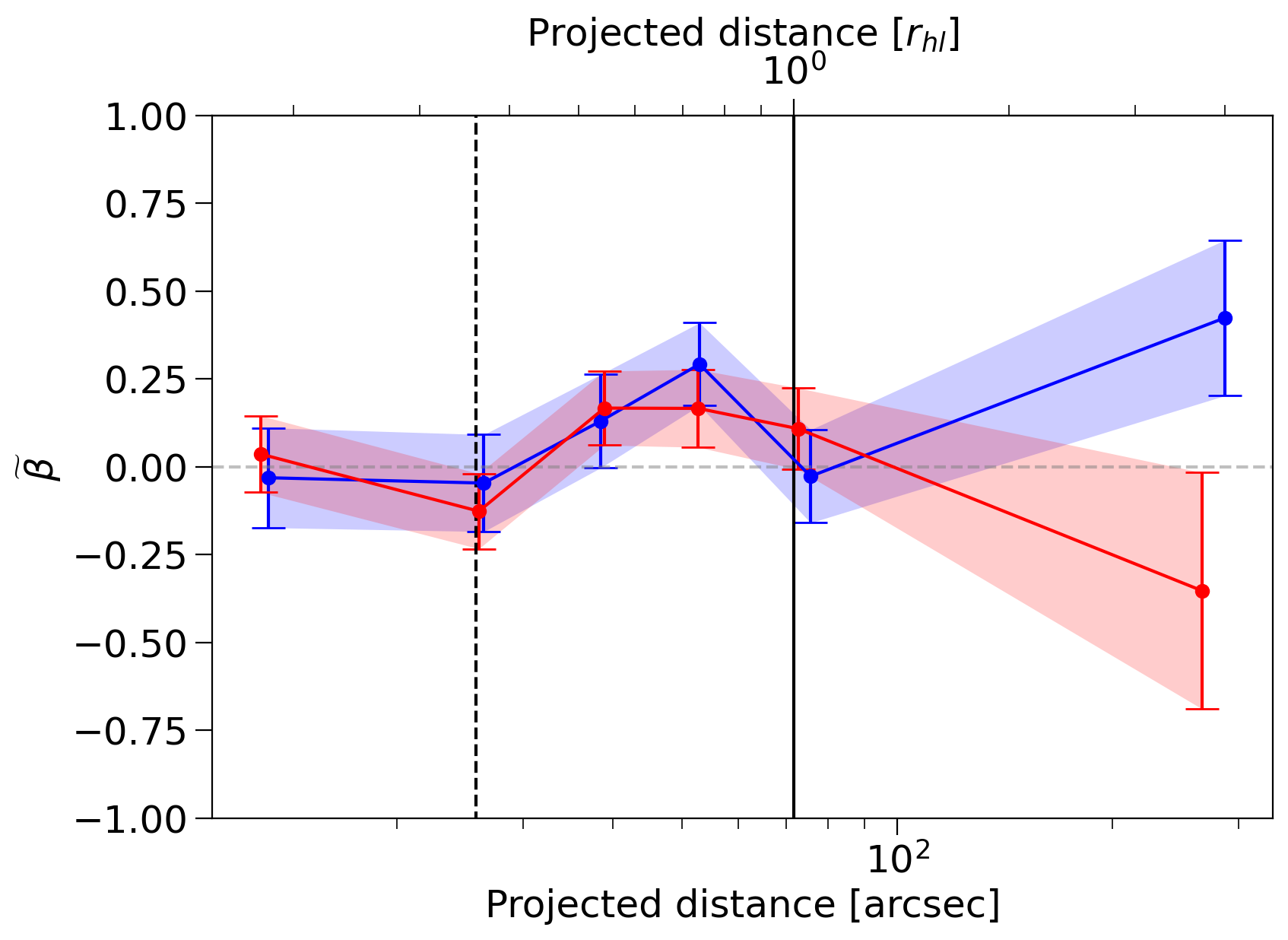}
    \vspace{-0.4cm}
    \caption{The normalised anisotropy of the P1 stars (blue) and P2 stars (red) in NGC 5024. The dashed vertical line indicates the core radius, while the solid vertical line indicates the half-light radius.}
    \label{fig:NGC5024_anisotropy}
\end{figure}

We found that the inner regions ($<30$ arcseconds) of NGC 5024 are slightly tangential, before shifting to radial anisotropy, with a tangential anisotropy again developing beyond $\sim 2 r_{\rm{hl}}$. Separating the sample into MPs shows that the two populations follow the same degree of anisotropy until the outer region $> 1 r_{\rm{hl}}$, where the P1 stars become radially anisotropic, while P2 stars show tangential anisotropy, as shown in Figure \ref{fig:NGC5024_anisotropy}. It is unclear whether the radial anisotropy of P1 stars in the outer regions supports a scenario in which P1 stars were preferentially lost by the cluster. Given that NGC 5024 is considered to be associated with the cluster NGC 5053 due to an angular separation of only 0.96 degrees and a physical separation of $\sim$ 1 kpc \citep{2010Jordi}, the claim that interactions are altering the outer regions of NGC 5024 could imply that the outer regions are no longer representative of their initial conditions. Using SDSS photometry, \cite{2010Chun} claimed to identify a tidal bridge between the two clusters, but the analysis of \cite{2010Jordi}, also using SDSS photometry, was unable to support this. If the outer regions have indeed been affected by interactions, there should also be evidence of stripped stars in the region surrounding the cluster; however, only 7 stars were identified as potential members of NGC 5024 amongst the extratidal field star population by \cite{2020Hanke}. A caveat included in the work of \cite{2020Hanke} is that the field between NGC 5024 and NGC 5053 was only covered by $\sim$ 10 fibres, making it unlikely to unveil evidence of a potential tidal bridge.

Nevertheless, the overall tangential anisotropy in the outer regions of NGC 5024 found in this work could be the result of: 1) the cluster being born with tangential anisotropy which has been preserved until the present day, or 2) the tidal field and interactions with NGC 5053 preferentially causing stars on radial orbits in the outer regions to be stripped, leaving only those stars which are on tangential orbits (possibly preferentially stripping P1 stars). Considering that NGC 5024 is a relatively compact cluster, ($c = 1.72$ \citep{2010Harris}), the stellar density of the cluster drops rapidly from the centre outwards, so it may be less likely for the outer regions to lose a significant enough fraction of stars to produce an observable tidal bridge within $\sim 1.49$ relaxation times. Future work could focus on identifying a tidal bridge between NGC 5024 and NGC 5053 or extratidal structures around NGC 5024 and, if identified, determining which population the stripped stars belong to.

\subsubsection{NGC 5272 (M 3)}
We found significant rotation for NGC 5272, which has been established by many previous works to be rotating at a significance level of $\geq2\sigma$ \cite{2014Fabricius,2018Mikis,2018Helmi,2018Bianchini,2019Sollima}. We found that P1 stars appear to be rotating slightly faster ($\sim1\sigma$ significance) than P2 stars, which is an unexpected result for a P2 centrally concentrated cluster, considering most formation theories predict the population in the centre to have a higher rotational amplitude. However, our results are at odds with \cite{2024Dalessandro}, who found the opposite: P2 stars rotate slightly faster than P1 stars. There are many peculiarities surrounding this cluster, not only within our analysis, but also from previous literature. For example, NGC 5272 is the only dynamically young, P2 centrally concentrated cluster we have found to be approximately isotropic throughout, with slight radial anisotropy developing beyond $\sim3r_{\rm{hl}}$. It has been proposed that NGC 5272 was either accreted from the Helmi stream \citep{2019Massari} or the result of a merger between two GCs in a dwarf galaxy environment and then accreted \citep{2021lee}. For a cluster which is proposed to have had a tumultuous introduction to the Milky Way, there is no evidence of a large scale tidal structure surrounding the cluster \citep{2010Jordi,2020Hanke}, which also agrees with the low mass loss fraction we reported in \cite{2023Leitinger} ($\rm{M_c/M_i} = 0.48$). Another peculiarity is that the P1 stars are extended in the HST pseudo-colour index used for chromosome maps: $\Delta C_{\rm{F275W,F814W}}$ \citep{2019Marino,2022Jang,2023Leitinger}, which may be driven by small variations in Fe/H \citep{2022Lardo}.
The combination of these elements creates great difficulty in tracing back the formation history of this cluster, despite its young dynamical age.

\subsubsection{NGC 6809 (M 55)}
In \cite{2023Leitinger} we discussed the structural properties of NGC 6809, in which the main focus was that for a dynamically young cluster ($\rm{age/T_{rh}} = 4.1 \pm 0.2$), it currently only has $\sim 1/4$ of its initial mass. Within this remaining mass, we identified a P2 central concentration in \cite{2023Leitinger}, with $\rm{P_2/P_{total} = 0.56 \pm 0.04}$.
In this paper we have found NGC 6809 to be tangentially anisotropic in the outer regions with a low overall rotation $\rm{A_{total}/\sigma} = 0.13\pm 0.02$. The combination of these observables may suggest that the outer regions of NGC 6809 exhibit such strong tangential anisotropy because all stars on radial orbits were stripped away, leaving only those on tangential orbits.

\subsection{P1 centrally concentrated clusters}
\label{sec:P1conc}
In this section we focus on the P1 centrally concentrated clusters: NGC 288, NGC 3201 and NGC 6101. We note that in the case of NGC 3201, the $A^+_4$ parameter in \cite{2023Leitinger} suggests the cluster is P1 centrally concentrated, but the enriched star fraction $\rm{P_2/P_{total}}$ in both \cite{2023Leitinger} (Figure A3) and \cite{2024Cadelano} (Figure 5) suggests there are primarily P2 stars within $\sim100''$ of the cluster centre, primarily P1 stars between $\sim 100''$ and $\sim 300''$ and finally, primarily P2 stars again beyond $\sim 300''$. This result could suggest that NGC 3201 formed P2 centrally concentrated, with many of the inner P2 stars then distributed to the outer regions on radial orbits. But regardless of whether P2 stars are preferentially in the inner $\sim100''$, we still observe a greater fraction of P2 stars in the outer regions than P1 stars, which is why we still label the current state of this cluster as P1 centrally concentrated. In both NGC 288 and NGC 6101, the $\rm{P_2/P_{total}}$ fraction does not exhibit this trend \citep{2023Leitinger}, so we still consider them as fully P1 centrally concentrated clusters. The focus of this section is therefore to investigate any common features between these 3 clusters, which all primarily contain P2 stars in their outer regions.\\

NGC 288, NGC 3201 and NGC 6101 show some of the highest radial anisotropy in the outer regions when considering the intrinsic random motion of stars relative to each other (Figure \ref{fig:anisotropy_young}), as well as the anisotropy including rotation (see Figure \ref{fig:gaia_anisotropy_young}). All 3 clusters have low rotation ($\rm{A_{total}/\sigma} < 0.25$), lower concentration values than the P2 centrally concentrated clusters according to \cite{2010Harris} (with the exception of NGC 6809), which could be explained by \cite{2016Tiongco}, who used \textit{N}-body simulations to conclude that initially isotropic and tidally underfilling clusters (in terms of the half-mass radius to Jacobi radius ratio) can produce a build-up of radially anisotropic stars in the outer regions. In these cases, complete isotropy is not expected to be observable until they have undergone significant mass loss and dynamical evolution, which has not yet occurred for these dynamically young clusters. Interestingly, all 3 clusters were also found by \cite{2021Ibata} to have tidal tails, whereas no tidal tails were identified for any of the P2 centrally concentrated clusters.

\subsubsection{NGC 288}
NGC 288 has retained only a fraction $\rm{M_c/M_i} \sim 0.24$ of its initial mass, with an enriched fraction of $\rm{P_2} / \rm{P_{total}} = 0.37 \pm 0.04$ \citep{2023Leitinger}. From our kinematic analysis, we found that the P2 stars rotate slightly faster ($<1\sigma$) than the P1 stars, which is consistent with the result of \cite{2024Dalessandro}. The outer regions of the cluster exhibit strong radial anisotropy, which can be expected for a cluster with such high mass loss. However, NGC 6809 has also experienced a similar degree of mass loss and exhibits tangential anisotropy in its outer regions. Significant tidal tails have been identified surrounding NGC 288 \citep{2021Ibata} and the cluster has likely recently experienced a tidal shock from the disk and bulge \citep{2000Leon}, supporting a history in which NGC 288 has lost a large amount of its initial mass. As there were too few stars in the cross-match between our kinematic catalogues and our photometric catalogue, both P1 and P2 stars appeared approximately isotropic throughout, due to the size of the errorbars, so we cannot identify whether a specific population is responsible for the radial anisotropy. 

\subsubsection{NGC 3201}
We found no significant differences between the rotation parameters of the MPs in NGC 3201, in terms of position angle, inclination angle, or total rotational amplitude, which is consistent with the results of both \cite{2023Martens} and \cite{2024Dalessandro}. In our analysis of the anisotropy, we discovered NGC 3201 is radially anisotropic in the outskirts, both when the anisotropy is defined in terms of the intrinsic motion and, to a greater degree, when the anisotropy also includes the rotation (see Appendix \ref{sec:Appendix_dold_anisotropy} for details).

\cite{2019Bianchini} and \cite{2021Wan} investigated the peculiar kinematics in the outskirts of NGC 3201, which contains tidal tails and exhibits flattened velocity dispersions. The fact that P2 stars are primarily located in the outer regions of NGC 3201 could indicate that the peculiar kinematics found by \cite{2019Bianchini} and \cite{2021Wan} is driven by the P2 stars. By that same logic, the radial anisotropy we found in the outer regions of NGC 3201 has a higher chance of being driven by the P2 stars as well. Although our sample has too few stars to determine which population drives the radial anisotropy, the recent kinematics analysis of \cite{2024Cadelano} supports this idea of enriched stars driving the radial anisotropy. They classified the multiple stellar populations using the same method and photometric data as we did in \cite{2023Leitinger}, obtaining the same results in terms of the cumulative radial distribution and enriched star fraction. They concluded that the bimodal distribution of the enriched star fraction is a result of an irregular 2D distribution of the populations, rather than a symmetrical distribution. \cite{2024Cadelano} then used \gaia DR3 proper motions to investigate the velocity dispersions and anisotropy of 297 P1 and 325 P2 stars. They discovered the P1 stars in their sample are isotropic throughout the cluster, while P2 stars are isotropic in the centre and become radially anisotropic beyond the half-mass radius. Such a result, coupled with the U-shaped $\rm{P_2/P_{total}}$ distribution (Figure A3 of \cite{2023Leitinger} and Figure 5 of \cite{2024Cadelano}), seems to provide evidence for a formation scenario in which the P2 stars were initially more concentrated before the expansion of the cluster ejected stars on radial orbits. If this is the case, one would expect to find a significant amount of P2 stars in the observed tidal tails of NGC 3201.

NGC 3201 has also been investigated in terms of hosting a black hole population, which could contribute to the peculiar structural and dynamical features observed. \cite{2019Giesers} concluded in their analysis of NGC 3201 that the cluster shows evidence of an extensive subpopulation of black holes, with \cite{2018Askar}, \cite{2019Kremer} and \cite{2020Weatherford} estimating the number of black holes in NGC 3201 is between $100<N_{BH}<200$, although \cite{2020Weatherford} also concludes that this number is significantly reduced to $N_{BH} = 41^{+40}_{-34}$ if the effects of mass segregation are accounted for, making it less likely that the black holes in NGC 3201 have significantly contributed to the peculiar current day observations.

\subsubsection{NGC 6101}
\begin{figure}
    \centering
    \includegraphics[width=\hsize]{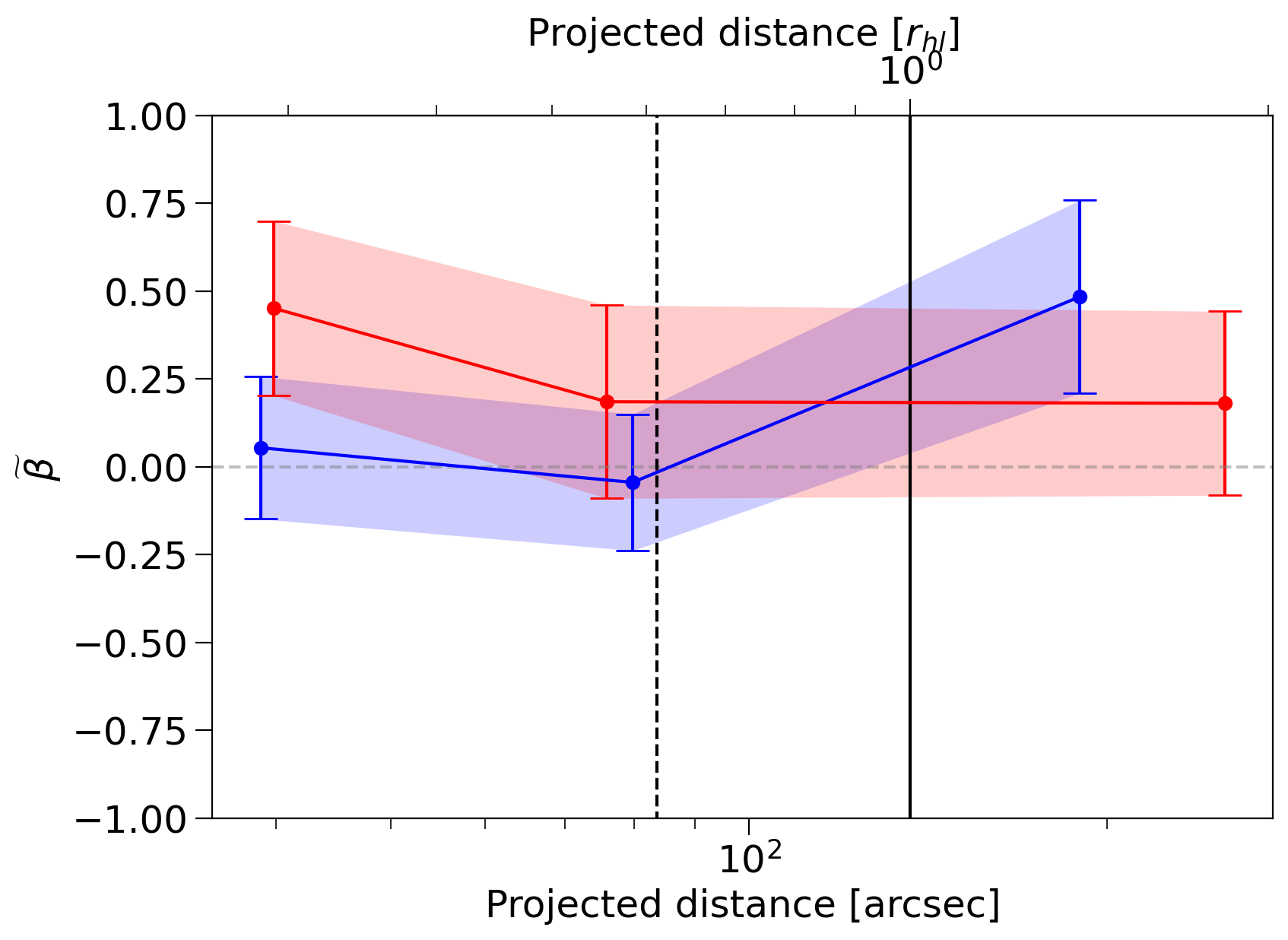}
    \vspace{-0.4cm}
    \caption{The normalised anisotropy of the P1 stars (blue) and P2 stars (red) in NGC 6101. The dashed vertical line indicates the core radius, while the solid vertical line indicates the half-light radius.}
    \label{fig:NGC6101_anisotropy}
\end{figure}
The rotation of NGC 6101 is mainly representative of the rotation of the inner regions covered by MUSE, as the majority of LOS velocities were taken from the ESO programme(s) 099.D-0824(A), while only 26 stars were taken from the LOS catalogue compiled by \cite{2018BaumgardtHilker}. We found low rotation for NGC 6101, with no significant differences between the rotation amplitudes or position angles between the multiple populations. Our anisotropy analysis was performed using 2109 proper motions for all stars, showing the strongest radial anisotropy in the outskirts, of any cluster in our sample. Combining these proper motion measurements with the photometric sample of 481 stars in \cite{2023Leitinger} meant that the anisotropy of the MPs included only 3 bins, as shown in Figure \ref{fig:NGC6101_anisotropy}. Changing the number of bins to 2 and 4 produced the same result - the P2 stars within the core radius were preferentially on radial orbits, while both populations were contributing to the radial anisotropy in the outer regions.

NGC 6101 has lost a very low amount of mass within its lifetime, but nevertheless, \cite{2021Ibata} discovered the presence of tidal tails for this cluster. \cite{2016peuten} and \cite{2015dalessandro} found no evidence of mass segregation within the cluster and, using \textit{N}-body simulations, \cite{2016peuten} concluded that a population of stellar mass black holes may be responsible for this. 
The analysis of \cite{2018Askar} performing simulations of GCs using the \texttt{MOCCA} code concluded that NGC 6101 is likely to be host to a large subsystem of 100 to 250 black holes, with \cite{2020Weatherford} supporting this idea with their confident estimation of 75 - 236 black holes. It may be possible that these stellar mass black holes have influenced the kinematics of the cluster, but additional observations and more data are required to further investigate this.

\subsection{Spatially mixed populations}
\label{sec:Apluszero}
This section focuses on the dynamically young clusters that have retained most of their initial conditions, but are nevertheless spatially mixed in terms of their populations. These clusters include NGC 4590, NGC 5053, NGC 5904, NGC 6205 and NGC 7089. These spatially mixed clusters show varying degrees of mass loss and rotation over dispersion ratios. We also observe variations between isotropy or tangential anisotropy in their outer regions. 

\subsubsection{NGC 4590 (M 68)}
We found NGC 4590 to have very low rotation, which is supported by the results of \cite{2010Lane}, \cite{2015Kimmig}, \cite{2018Bianchini}, \cite{2019Vasiliev} and \cite{2019Sollima}, who also did not detect significant rotation. Interestingly, it exhibits the highest degree of tangential anisotropy in the outer regions, of any cluster in our sample. The 361 stars classified into MPs in \cite{2023Leitinger} were matched with the proper motion catalogues included in this work, and the resulting anisotropy of the MPs showed that the tangential anisotropy in the outer regions is driven mainly by the P2 stars. Changing the number of bins for this anisotropy analysis did not change the result of P2 stars showing tangential anisotropy, but the P1 stars varied between showing isotropy or slight radial anisotropy in the outer regions. This was also the case for the anisotropy of P1 and P2 stars in NGC 5024. Overall, the tangential anisotropy in the outer regions suggests that NGC 4590 is a dynamically stable cluster despite its young dynamical age.

From previous literature on NGC 4590, \cite{2019baumgardthilker} found that NGC 4590 has large perigalactic ($8.95 \pm 0.06$ kpc) and apogalactic ($29.51 \pm 0.42$ kpc) distances, \cite{2019Massari} predicts that the Helmi streams is the progenitor of NGC 4590, and it has been reported to have very long tidal tails \citep{2021Ibata}. The large peri- and apogalactic distances suggest tidal stripping was unlikely to have removed a significant fraction of stars, but it is currently unclear how the accretion of the progenitor system of NGC 4590 into the Milky Way affected the kinematics of this cluster. NGC 5024, NGC 5272 and NGC 6981 were accreted from the same progenitor, but none of these clusters have tidal tails like NGC 4590. However, NGC 5024 and NGC 6981 exhibit tangential anisotropy in the outer regions similar to NGC 4590, while NGC 5272 exhibits slight radial anisotropy.

\subsubsection{NGC 5053}
We found NGC 5053 to have consistent radial anisotropy from the centre of the cluster to $\sim 1 r_{\rm{hl}}$, but there were too few stars in the photometry and LOS velocity catalogues for NGC 5053 in order for the anisotropy of MPs or the rotational analysis to be performed. Previous work has found NGC 5053 to be dynamically complicated as it contains significant tidal tails \citep{2006lauchner,2010Jordi} and it is postulated that NGC 5053 and NGC 5024 were accreted into the Milky Way from the same progenitor, possibly undergoing interactions with one another. Despite this common history, the two clusters show no similarities to one another. NGC 5053 shows slight radial anisotropy and has a concentration of $c = 0.7$, making it far less compact compared to NGC 5024 ($c = 1.7$), which shows tangential anisotropy in the outer regions. Additionally, there is a P2 central concentration within NGC 5024, while NGC 5053 is spatially mixed, although both clusters have approximately similar enriched star fractions \citep{2023Leitinger}.

\subsubsection{NGC 5904 (M 5)}
NGC 5904 is a very interesting cluster in terms of kinematics and dynamics. We found NGC 5904 to be the second-fastest rotator in our sample ($\rm{A_{total}/\sigma} = 0.50 \pm 0.03$), consistent with previous kinematic analyses \citep{2012Bellazzini,2014Fabricius,2015Kimmig,2018Kamann,2018Helmi,2018Bianchini,2019Vasiliev,2019Sollima,2020Cordoni,2023Martens}. However, we found no differences in the rotation amplitudes nor rotation axes of the spatially mixed populations within NGC 5904, which is supported by the results of \cite{2023Martens}, but partially at odds with the work of \cite{2024Dalessandro} who found P2 rotates faster than P1 ($\sim2\sigma$ significance), with no differences between the rotation axes of the MPs. \cite{2020Cordoni} also analysed the MPs of NGC 5904, finding the two populations exhibit the same rotational amplitude, but different rotation axes in terms of only the position angles. NGC 5904 has been found by \cite{2018Kamann} to exhibit a decoupled core, in which the position angle changes from the central region outwards, so this may account for the differences between works. The rotation-mass relation found for NGC 104 by \cite{2023Scalco} was also discovered for NGC 5904, meaning higher mass stars rotate faster around the cluster than low mass stars (unrelated to mass segregation), but the spatially mixed populations within this cluster suggest that this may not necessarily be attributed to a specific population. There is also evidence of tidal tails surrounding NGC 5904 \citep{2010Jordi,2019Grillmair,2021Ibata}, which is interesting as NGC 4590 is the only other cluster in our sample which exhibits significant tangential anisotropy in its outer regions and also has tidal tails. The progenitor system of NGC 5904 may be the Helmi streams, similar to NGC 4590, but the classification by \cite{2019Massari} is unclear and suggests NGC 5904 could belong to either the Helmi streams or Gaia-Enceladus.

\subsubsection{NGC 6205 (M 13) and NGC 7089 (M 2)}
We have combined the results of NGC 6205 and NGC 7089 into the same section due to the similarities between their masses, mass loss ratios, dynamical ages, progenitor systems, concentration values, rotation and anisotropy profiles. We found that NGC 6205 and NGC 7089 have the same anisotropy profiles, with both clusters exhibiting approximately isotropic behaviour in the inner regions, with slight radial anisotropy in the outer regions. We also found that the MPs of both clusters are approximately isotropic throughout. Both clusters share very similar concentration values ($c=1.5$ for NGC 6205 and $1.6$ for NGC 7089) and are classified as originating from Gaia-Enceladus \citep{2019Massari}. However, only NGC 7089 was observed to have tidal tails \citep{2010Jordi,2020Hanke,2021Ibata}, with one extratidal giant found to be a P2 star, which was argued to suggest there is a large fraction of P2 stars that have escaped the cluster (see Section 4.4 of \cite{2020Hanke} for caveats).

In terms of rotation, both NGC 6205 and NGC 7089 exhibit a high degree of rotation in our analysis. The same result for NGC 6205 has been found by \cite{2014Fabricius} and \cite{2019Sollima}, as well as \cite{2018Bianchini} who discovered rotation in NGC 6205 to a $2\sigma$ level, while \cite{2019Vasiliev} found no evidence of rotation in this cluster. For NGC 7089, our results agree with every literature result in common in Table \ref{table:rotation_comparisons}. Previously, \cite{2017Cordero} separated NGC 6205 into 3 populations, discovering the most enriched population rotates faster than the intermediate population, which they see as a signature of the initial formation. On the other hand, our results show the populations rotating to the same degree. We compared our sample with that of \cite{2017Cordero} and found agreement in the classification of the multiple stellar populations. We then found that we obtain the same result as \cite{2017Cordero} if we perform our rotation analysis on the same stars common between our data sets (113 stars), but that those differences disappear when using our full sample (855 stars), suggesting their analysis included too few stars. Using a larger sample of stars with 313 LOS velocities and 1201 proper motions, \cite{2024Dalessandro} found P2 stars rotating faster than P1 stars at $\sim 1\sigma$ significance, which is consistent with our results for the MPs. In NGC 7089 we have found $\sim 2\sigma$ differences in the rotation of the MPs, with P1 stars rotating faster than P2 stars despite the cluster being spatially mixed, but this cluster was not included in the analysis of \cite{2024Dalessandro} for comparisons.

\subsubsection{NGC 6656 (M 22)}
We discovered significant rotation in NGC 6656, which is supported by previous results \citep{2010Lane,2012Bellazzini,2018Kamann,2018Helmi,2018Bianchini,2019Vasiliev,2019Sollima,2023Martens}. However, we did not detect any significant differences in the rotation of the MPs, in terms of rotational amplitude or rotation axes, which cannot be compared with the results of \cite{2023Martens} as they could not find differences due to a low number of stars in each population with associated MUSE velocities. NGC 6656 exhibits the most consistent isotropy amongst the dynamically young clusters in our sample, which is also supported by \cite{2020CordoniNGC6656} using \gaia DR2 and HST data; however, they classify the populations as Fe-rich and Fe-poor, finding the populations share similar spatial distributions, with both populations isotropic. This classification of the populations is not directly comparable with our classifications, but considering we found isotropy for all stars, P1 and P2, while they found isotropy for Fe-rich and Fe-poor, it is reasonable to assume the populations are indeed isotropic in both analyses. \\

\section{Conclusions}

We have combined Hubble Space Telescope and \gaia DR3 proper motions together with a comprehensive set of line-of-sight velocities to determine the 3D rotational amplitudes, rotation axes and anisotropy profiles of 30 Galactic globular clusters and their MPs. The focus of this paper was on the dynamically young clusters in our sample: NGC 104, NGC 288, NGC 3201, NGC 4590, NGC 5053, NGC 5024, NGC 5272, NGC 5904, NGC 6101, NGC 6205, NGC 6656, NGC 6809 and NGC 7089, as these clusters have the highest chance of retaining the initial conditions of formation.\\

We discovered significant ($>$3$\sigma$) rotation in 21 GCs, consistent with previous rotational analyses (e.g. \cite{2018Kamann,2018Bianchini,2019Sollima,2023Martens}). We found no significant differences between the total rotational amplitudes of the MPs for the clusters in our sample, with the exception of NGC 104 in which P2 rotates faster than P1 at a $\sim 3.5\sigma$ significance. We found no significant differences in the rotation axes of the MPs, in terms of position angle or inclination angle. For the clusters in our sample which demonstrated differences in the 3D rotational amplitudes of the MPs and exhibited a P1 or P2 central concentration, it was not necessarily the centrally concentrated population which displayed higher rotation. There has been increasing evidence in previous literature that there are very few clusters which exhibit \textit{significant} differences in the rotation of the multiple populations \citep{2012Milone,2019Libralato,2020Cordoni,2021Szigeti,2023Martens}. While for clusters in which differences are found, there has not been consistency between which population is rotating faster (i.e. P2 always rotating faster than P1) \citep{2018Bellini,2020kamann,2021Dalessandro,2021Szigeti,2023Martens}. The exception to this is the recent work of \cite{2024Dalessandro}, who concluded that P2 stars preferentially rotate faster than P1 stars for the 16 clusters in their sample. Although our results are mostly consistent with \cite{2024Dalessandro} for the 9 clusters in common between our works, we cannot draw the same conclusion based on the clusters in our sample and the significance of the differences we find between the rotation of the MPs. A general lack of differences between the rotation of the MPs in our sample contradicts the expectations of some formation theories which place P2 stars as more centrally concentrated and with higher angular momentum compared to P1, therefore expecting that P2 stars will rotate faster \citep[e.g.,][]{2015HenaultGieles}.

We found that 3D rotation strongly correlates with the current and initial mass, relaxation time, enriched star fraction and concentration of the clusters in our sample, but that these correlations are mainly attributed to correlations with mass. This implies that either the degree of initial rotation or the ability of a cluster to retain its initial rotation depends strongly on the mass.\\

We determined the anisotropy of each cluster, as well as the anisotropy of the MPs where possible, investigating correlations with the structural parameters, orbital parameters and accretion history of the clusters from their progenitor systems. We discovered that the dynamically young clusters with the highest central concentration of primordial stars showed significant radial anisotropy in the outer regions ($>2$ half-light radii) of the cluster. Of these clusters, all were previously confirmed to exhibit tidal tails, while two were accreted from the same progenitor system. The dynamically young clusters with a central concentration of enriched stars displayed isotropy or tangential anisotropy in their outer regions, when also considering the rotation in the anisotropy analysis. These clusters generally displayed a higher mass and concentration (tidal radius/core radius) than the primordially concentrated clusters, with no confirmed tidal tails. We also found a correlation between the anisotropy in the outer regions of the clusters, with the z-component of the angular momentum of their orbits, suggesting the ancient host galaxy of a cluster not only 'donors' the orbit of the cluster around the Milky Way, but may also determine the kinematics of stars within the GC. This could also imply that the present day observed kinematics of a GC were determined by the physical conditions of star and star cluster formation within their progenitor systems, but further investigation into this correlation is necessary.\\

\begin{acknowledgements}

This study was supported by the Klaus Tschira Foundation.\\
Based on observations collected at the European Organisation for Astronomical Research in the Southern Hemisphere under ESO programme(s) 099.D-0824(A). Based on observations made with the Gran Telescopio Canarias (GTC), installed in the Spanish Observatorio del Roque de los Muchachos of the Instituto de Astrofísica de Canarias, in the island of La Palma. This work is partly based on data obtained with MEGARA instrument, funded by European Regional Development Funds (ERDF), through Programa Operativo Canarias FEDER 2014-2020.\\
MG acknowledges support from the Ministry of Science and Innovation (PID2021-125485NB-C22, CEX2019-000918-M funded by MCIN/AEI/10.13039/501100011033) and from AGAUR (SGR-2021-01069).\\
EL acknowledges support from the ERC Consolidator Grant funding scheme (project ASTEROCHRONOMETRY, \url{https://www.asterochronometry.eu}, G.A. n. 772293).\\
\end{acknowledgements}

  \bibliographystyle{aa} % style aa.bst
  \bibliography{references} % your references Yourfile.bib

\begin{appendix}
\onecolumn
{\parindent0pt

\section{Updated cumulative radial distributions of the 30 Galactic GCs as function of dynamical age and mass loss ratio}
\label{Appendix_updated_Aplus}
Since \cite{2023Leitinger} was published, some structural parameters of the globular clusters in our sample have been updated on the Galactic Globular Cluster Database \citep{GGCD} from which they were taken. In particular, the relaxation times and current cluster masses have been updated, due to the inclusion of additional HST mass functions.\\
Following this update, the cumulative radial distribution plots presented in \cite{2023Leitinger} (Figure 15 as a function of the dynamical age and Figure 16 as a function of the mass loss ratio) also required an update in order to reflect the clusters which are still considered `dynamically young', matching the criteria of $\rm{age/T_{rh}} < 4.5$.\\
The most notable changes in the dynamical ages occur for NGC 2808 (previously: $\rm{age/T_{rh}} = 3.6 \pm 0.1$ and currently: $\rm{age/T_{rh}} = 5.3 \pm 0.2$), NGC 3201 (previously: $\rm{age/T_{rh}} = 3.5 \pm 0.2$ and currently: $\rm{age/T_{rh}} = 2.5 \pm 0.2$) and NGC 7078 (previously: $\rm{age/T_{rh}} = 4.1 \pm 0.1$ and currently: $\rm{age/T_{rh}} = 8.0 \pm 0.2$). There are no substantial changes to the mass loss ratios ($\rm{M_{current}/M_{initial}}$) of each cluster. \\

\begin{figure}[H]
\centering
\includegraphics[width=0.7\textwidth]{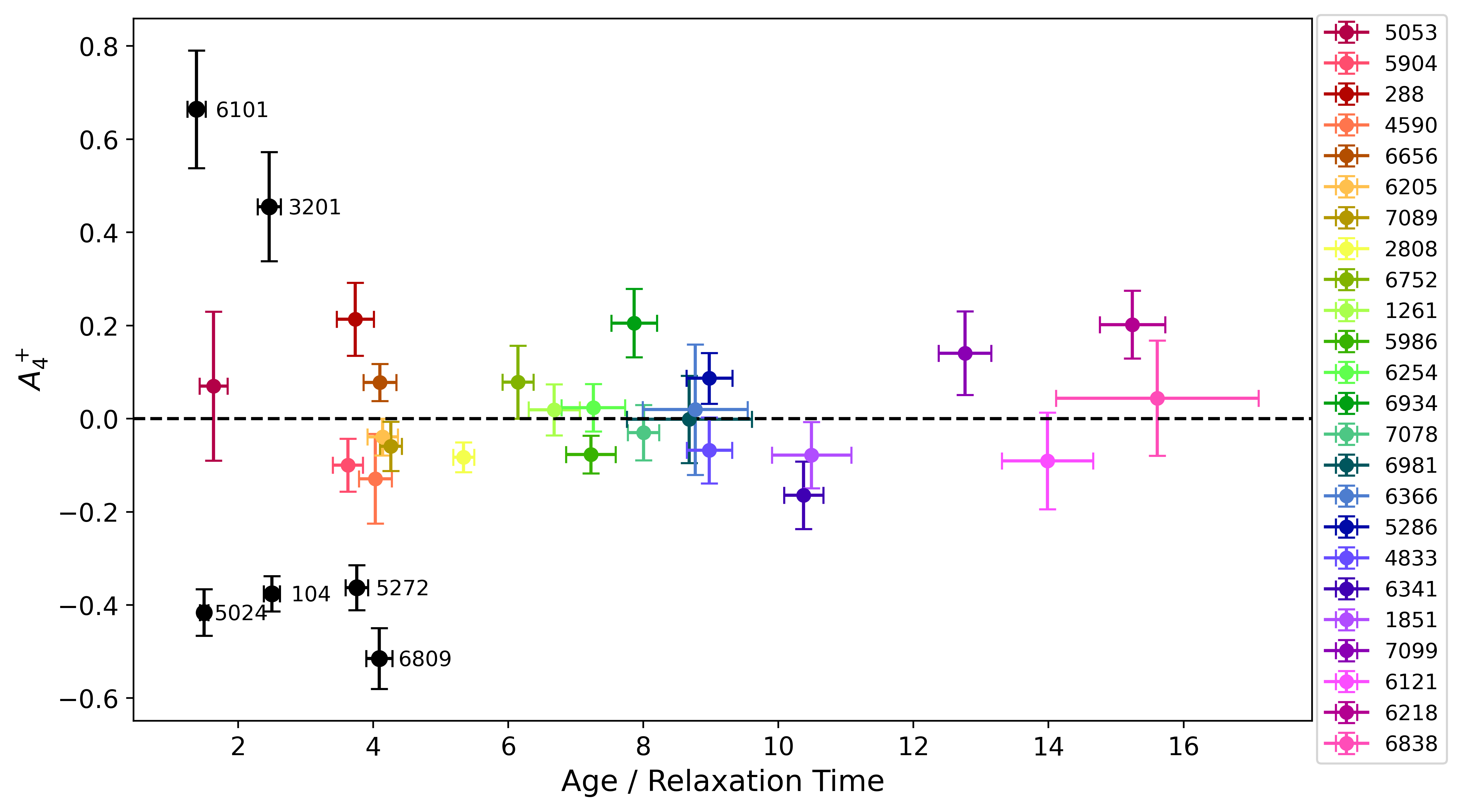}
\caption{The updated version of the normalised, cumulative radial distributions as a function of dynamical age (Figure 15 from \protect\cite{2023Leitinger}), including the 2 additional GCs: NGC 104 and NGC 6656, as well as updated dynamical ages from \protect\cite{GGCD}, which change the sample of the `dynamically young' clusters of our sample by removing NGC 2808 and NGC 7078, as the updated dynamical ages of these clusters now exceed age/$\rm{T_{rh}}$ > 4.5. \textit{Please note that the colour-coding of the clusters in this figure do not correspond to the same colour-coding in Figure 15 of \protect\cite{2023Leitinger}.}}
\label{fig:updated_Aplus_ttrh}
\end{figure}

\begin{figure}[H]
\centering
\includegraphics[width=0.7\textwidth]{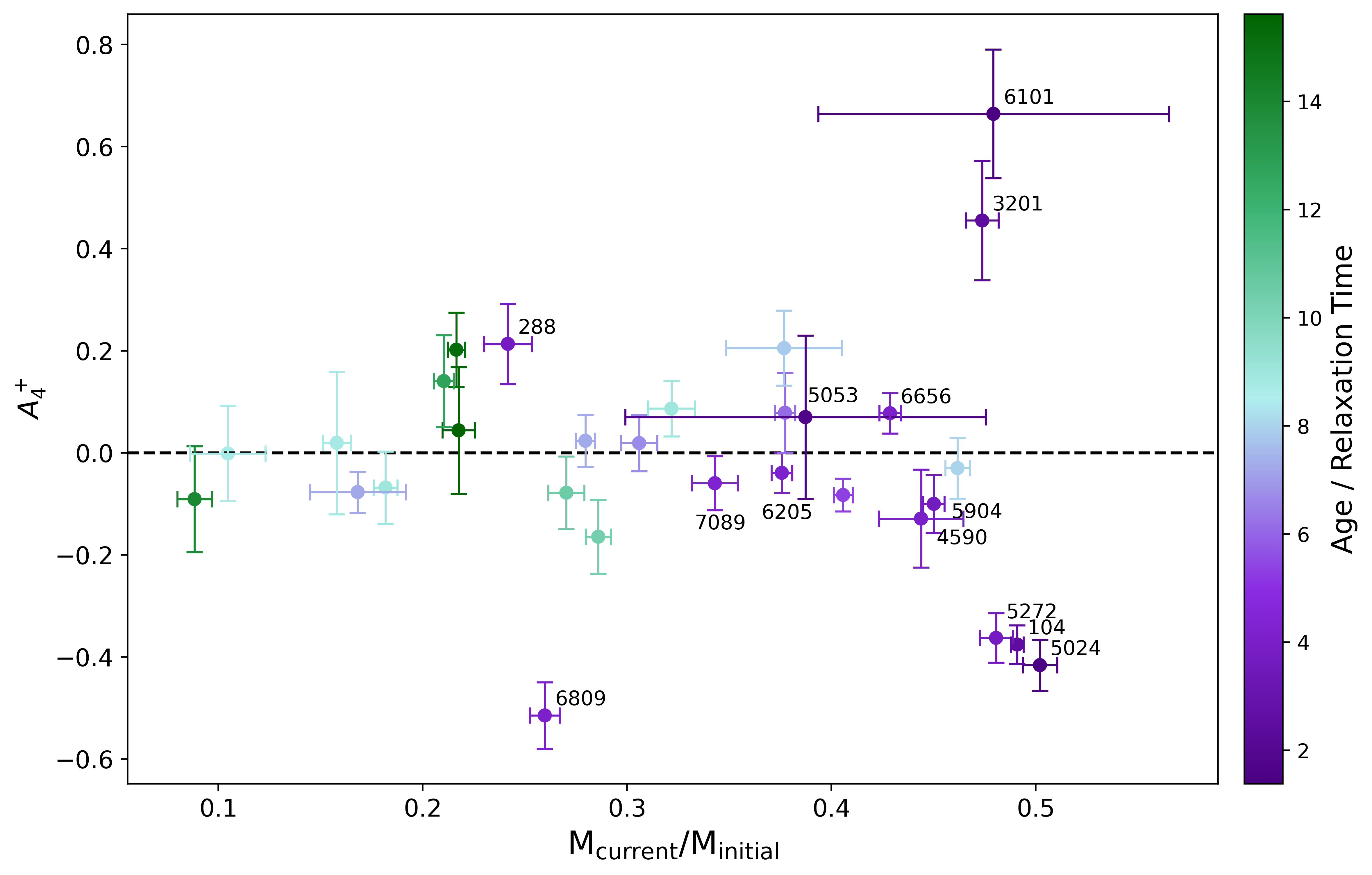}
\caption{The updated version of the normalised, cumulative radial distributions as a function of mass loss ratio (Figure 16 from \protect\cite{2023Leitinger}), including the 2 additional GCs: NGC 104 and NGC 6656, as well as updated mass loss ratios from \protect\cite{GGCD}.}
\label{fig:updated_Aplus_mlr}
\end{figure}

\FloatBarrier
\twocolumn

\onecolumn
\section{Rotation axes of dynamically youngest Galactic GCs}
\label{sec:Appendix_rotation}

Here we show the combined rotation analysis for the dynamically youngest ($\rm{age/T_{rh}} <$  4.5) clusters in our sample, as well as notable clusters such as NGC 2808 and NGC 7078. The left panel shows the position angle in the plane of the sky for all stars (green), P1 stars (blue) and P2 stars (red), with the values shown as circles and the errorbars shown as lines. In grey is the photometric semi-major axis as calculated in Section \ref{sec:rotation_axes}, with the errors represented as the width of the line.\\
The middle panel shows the calculated inclination angle of all stars (green), P1 stars (blue) and P2 stars (red), with the width of the lines corresponding to the error in the inclination angle and the extent of the line corresponding to the total rotational amplitude of the cluster, as shown on the x-axis.\\
Finally, the right panel shows the probability distributions of the total rotational amplitudes for all stars (green), P1 stars (blue) and P2 stars (red), with the corresponding median values and uncertainties displayed in the legend. In this way, we are able to visually identify the rotational differences between the different samples.\\
A caveat for these plots is that if we encounter a non-rotating cluster with too few stars in the sample, there can appear to be a difference between the probability distributions of all stars, in comparison to P1 and P2 (i.e. in Figure \ref{fig:rotationappendix3201} for NGC 3201). However, this difference merely reflects the difficulty in fitting a rotation curve to a non-rotating cluster with too few stars in the P1 and P2 samples, which is reflected in the uncertainties.

\begin{figure}[H]
    \centering
    \includegraphics[width=\hsize]{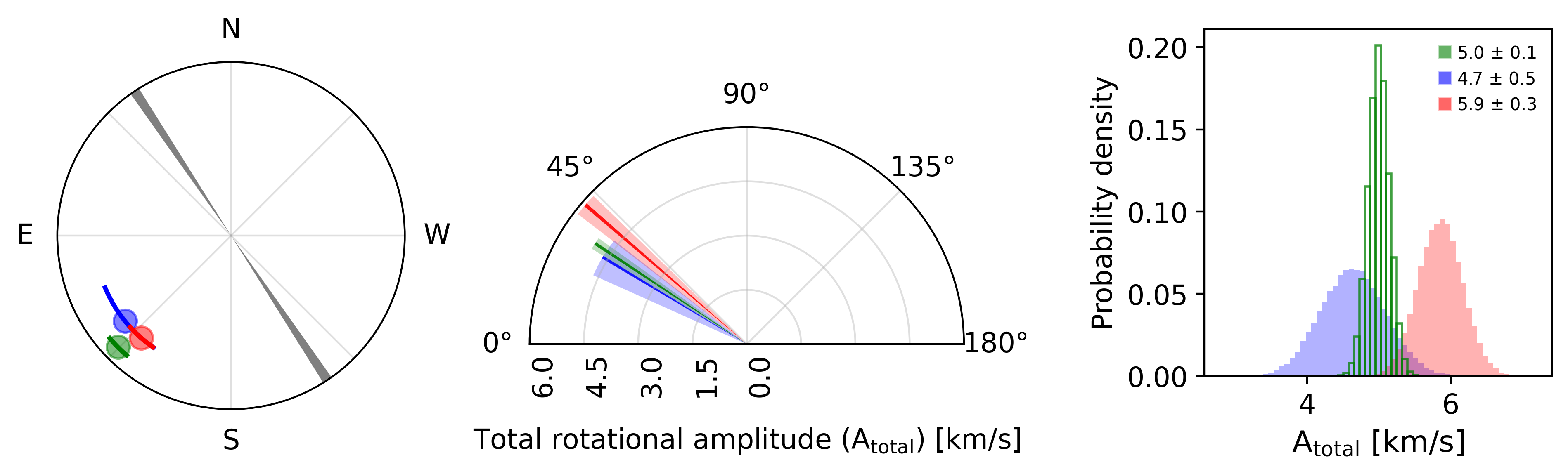}
    \caption{The rotation axis, inclination angle and total rotational amplitude of NGC 104 as shown in Figures \ref{fig:rotation_axes} and \ref{fig:rotation_axes_populations}, but now combined to show the differences for all stars (green), P1 stars (blue) and P2 stars (red).}
    \label{fig:rotationappendix104}
\end{figure}
\begin{figure}[H]
    \centering
    \includegraphics[width=\hsize]{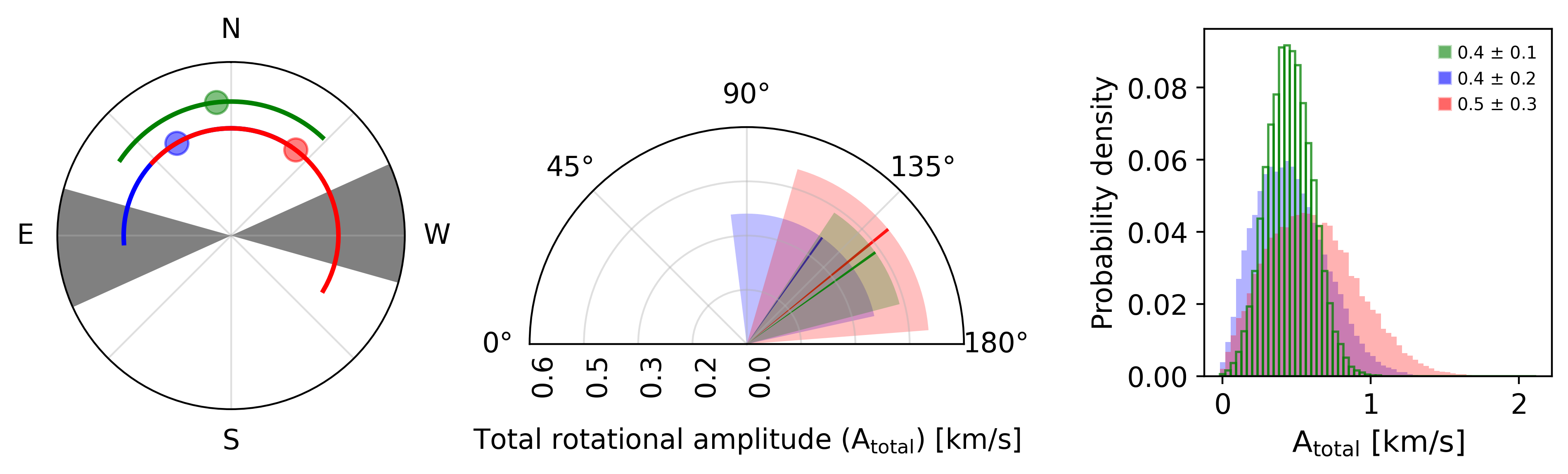}
    \caption{As in Figure \ref{fig:rotationappendix104}, but for NGC 288.}
    \label{fig:rotationappendix288}
\end{figure}
\begin{figure}[H]
    \centering
    \includegraphics[width=\hsize]{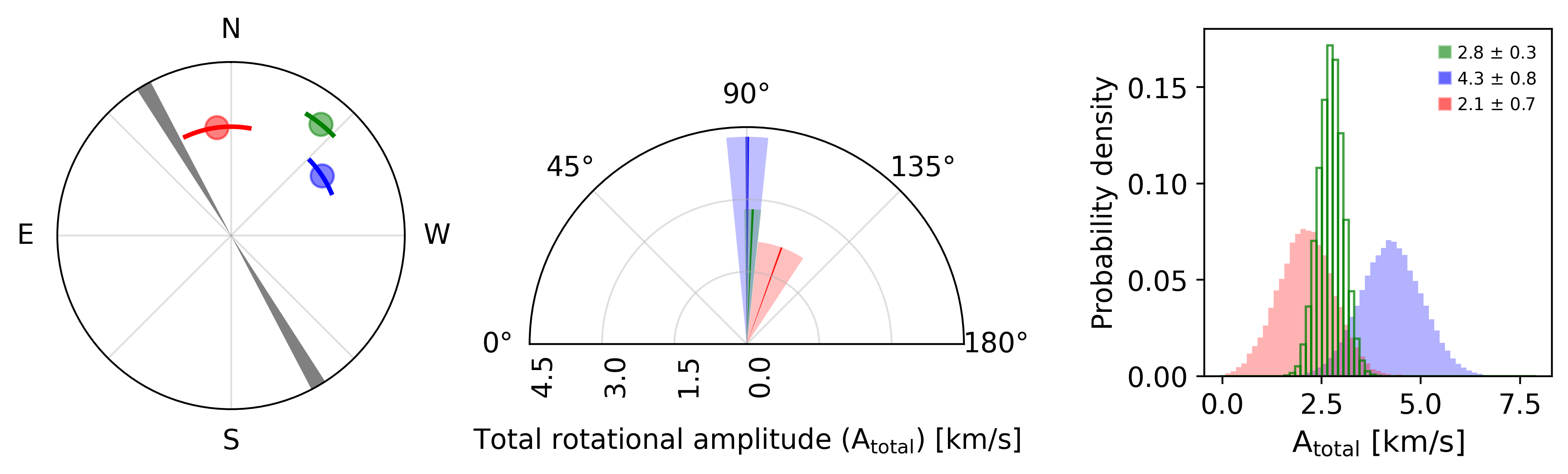}
    \caption{As in Figure \ref{fig:rotationappendix104}, but for NGC 2808.}
    \label{fig:rotationappendix2808}
\end{figure}
\begin{figure}[H]
    \centering
    \includegraphics[width=\hsize]{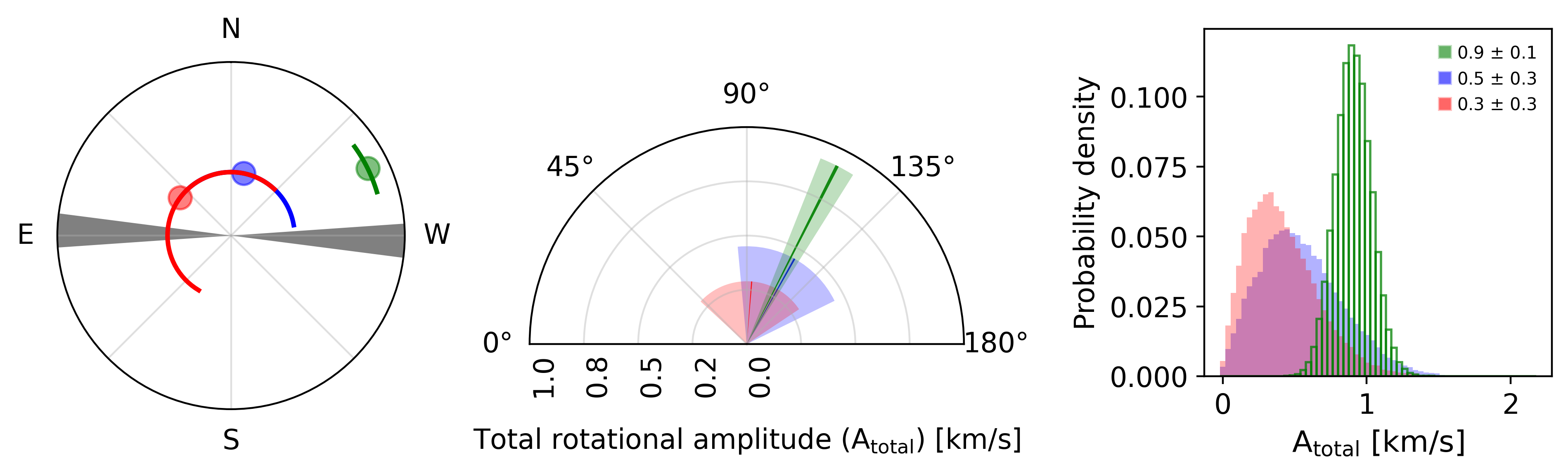}
    \caption{As in Figure \ref{fig:rotationappendix104}, but for NGC 3201.}
    \label{fig:rotationappendix3201}
\end{figure}
\begin{figure}[H]
    \centering
    \includegraphics[width=\hsize]{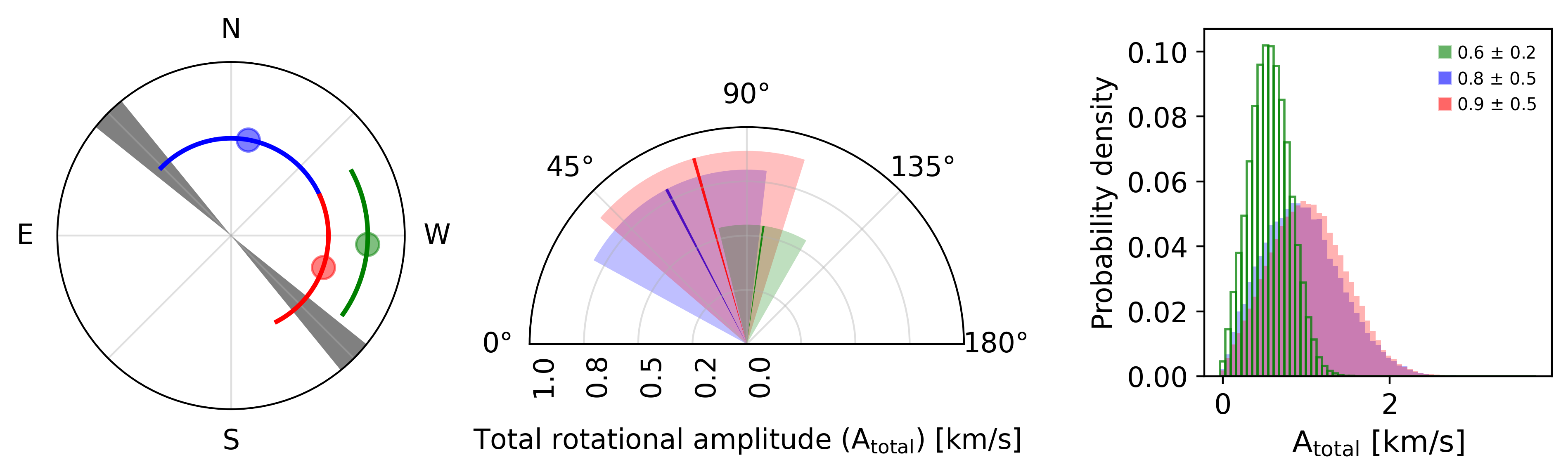}
    \caption{As in Figure \ref{fig:rotationappendix104}, but for NGC 4590.}
    \label{fig:rotationappendix4590}
\end{figure}
\begin{figure}[H]
    \centering
    \includegraphics[width=\hsize]{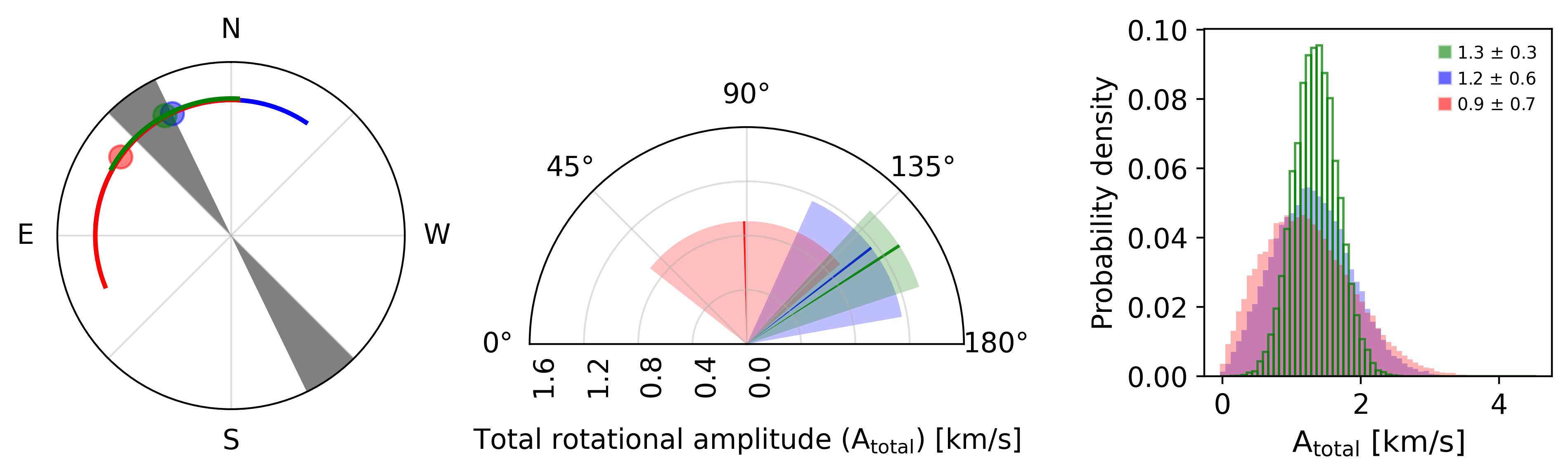}
    \caption{As in Figure \ref{fig:rotationappendix104}, but for NGC 5024.}
    \label{fig:rotationappendix5024}
\end{figure}
\begin{figure}[H]
    \centering
    \includegraphics[width=\hsize]{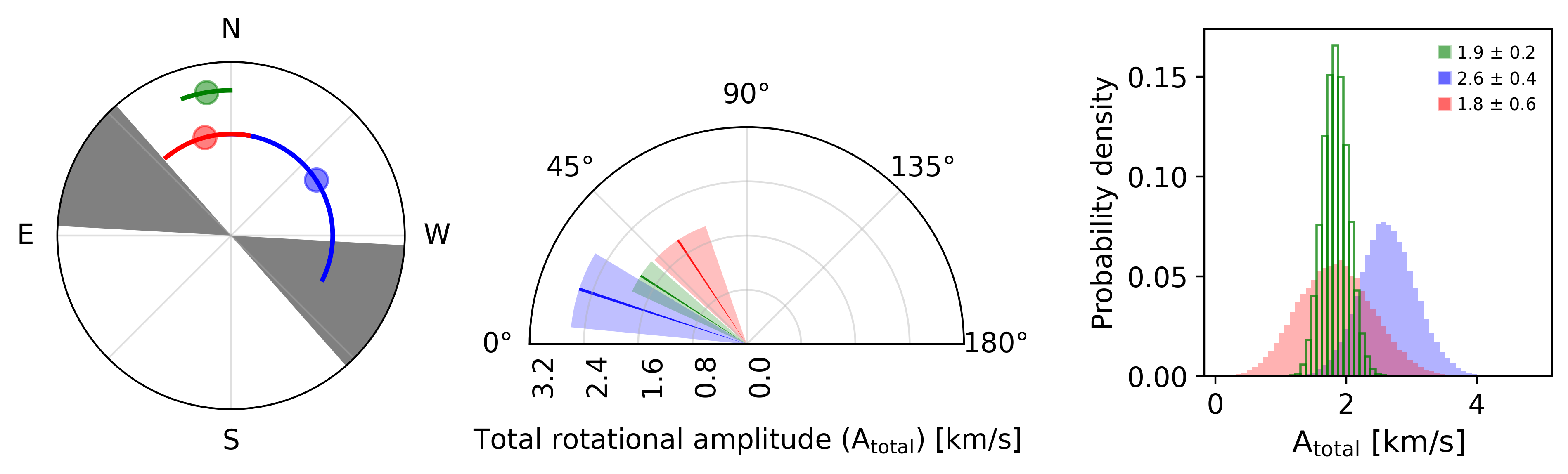}
    \caption{As in Figure \ref{fig:rotationappendix104}, but for NGC 5272.}
    \label{fig:rotationappendix5272}
\end{figure}
\begin{figure}[H]
    \centering
    \includegraphics[width=\hsize]{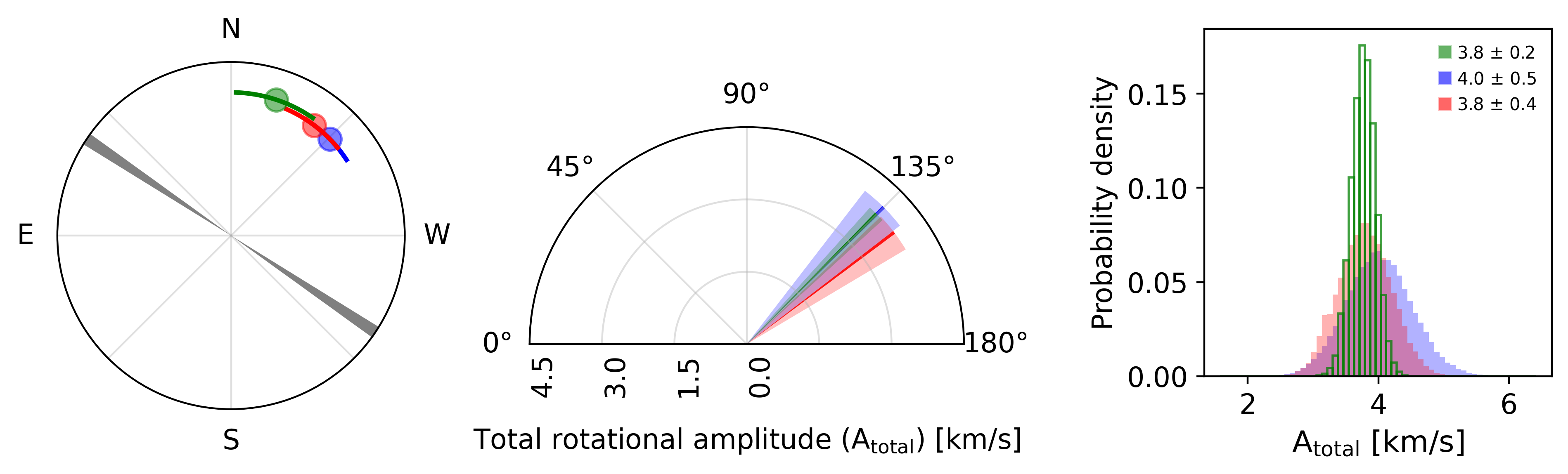}
    \caption{As in Figure \ref{fig:rotationappendix104}, but for NGC 5904.}
    \label{fig:rotationappendix5904}
\end{figure}
\begin{figure}[H]
    \centering
    \includegraphics[width=\hsize]{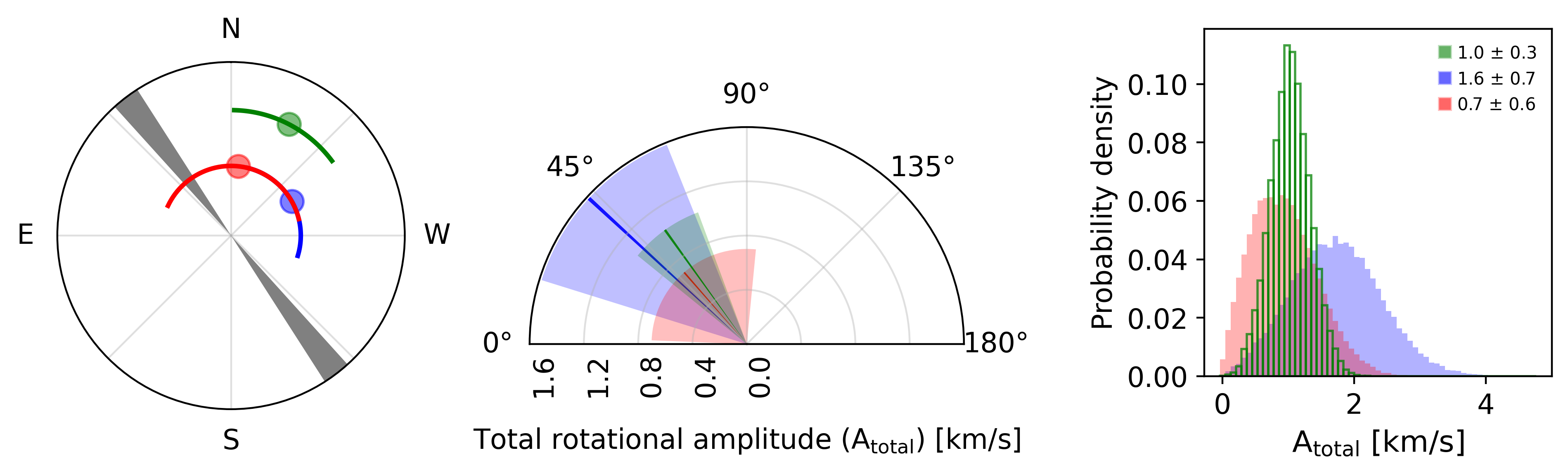}
    \caption{As in Figure \ref{fig:rotationappendix104}, but for NGC 6101.}
    \label{fig:rotationappendix6101}
\end{figure}
\begin{figure}[H]
    \centering
    \includegraphics[width=\hsize]{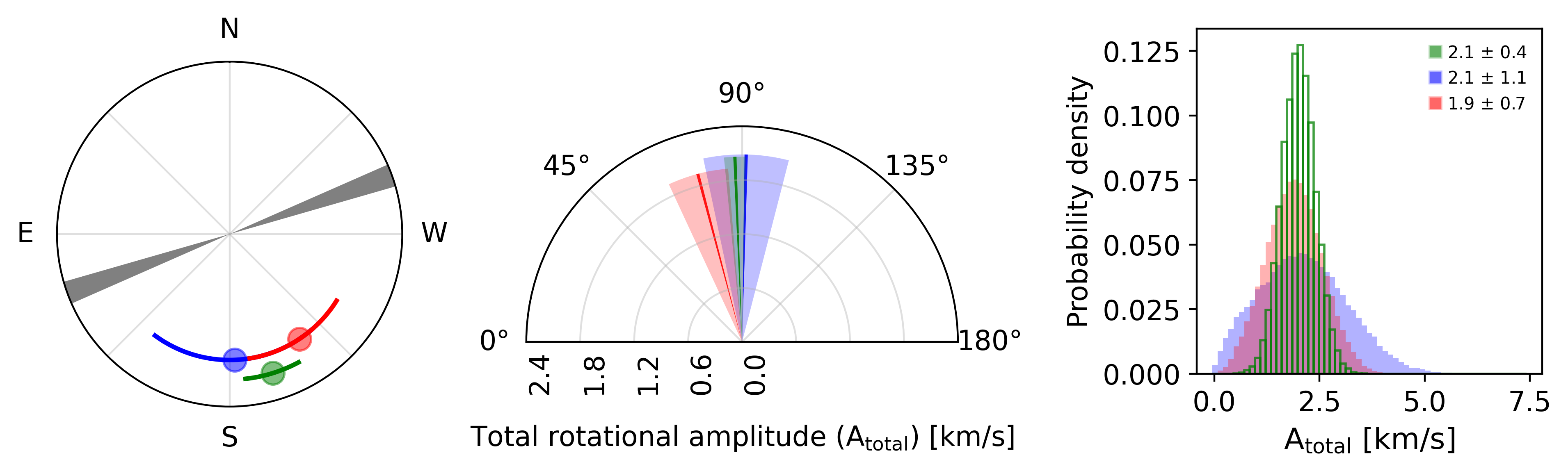}
    \caption{As in Figure \ref{fig:rotationappendix104}, but for NGC 6205.}
    \label{fig:rotationappendix6205}
\end{figure}
\begin{figure}[H]
    \centering
    \includegraphics[width=\hsize]{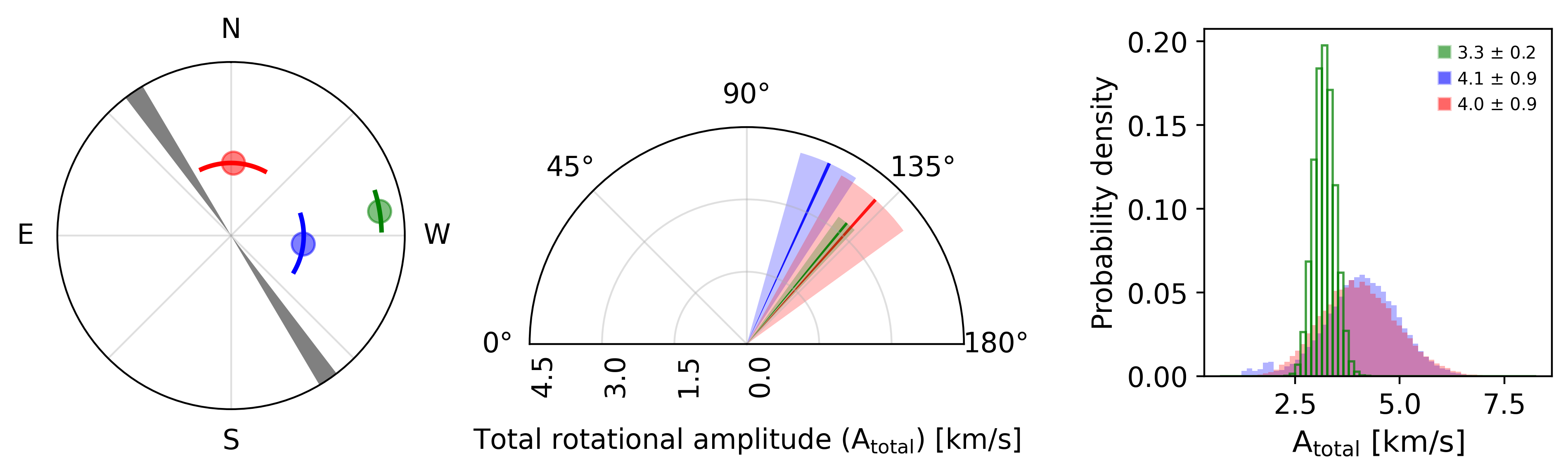}
    \caption{As in Figure \ref{fig:rotationappendix104}, but for NGC 6656.}
    \label{fig:rotationappendix6656}
\end{figure}
\begin{figure}[H]
    \centering
    \includegraphics[width=\hsize]{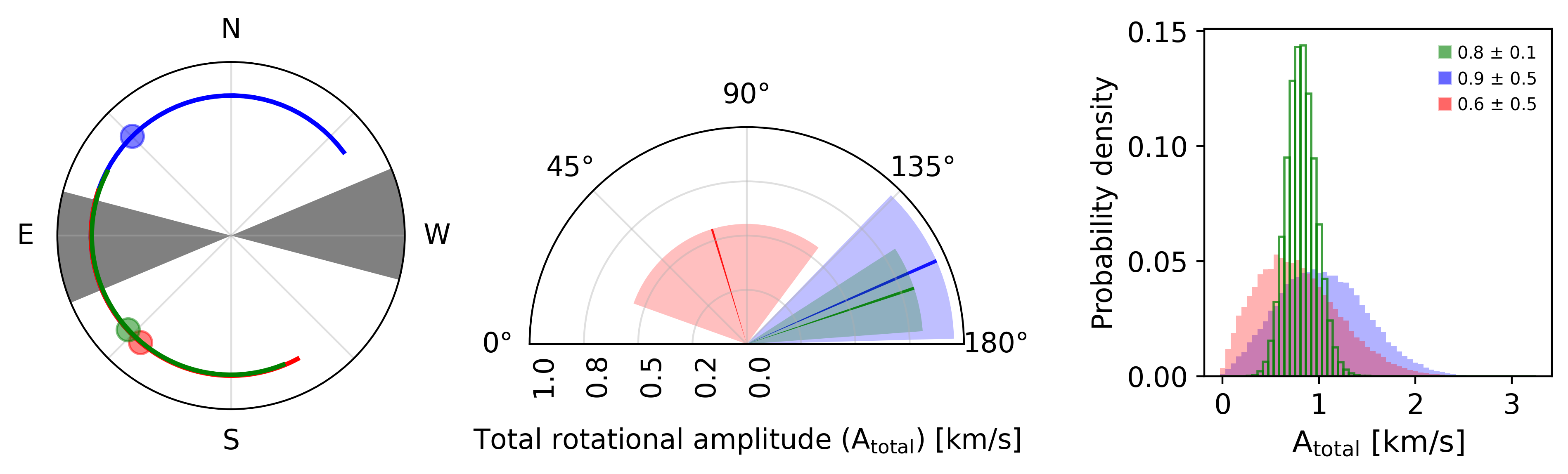}
    \caption{As in Figure \ref{fig:rotationappendix104}, but for NGC 6809.}
    \label{fig:rotationappendix6809}
\end{figure}
\begin{figure}[H]
    \centering
    \includegraphics[width=\hsize]{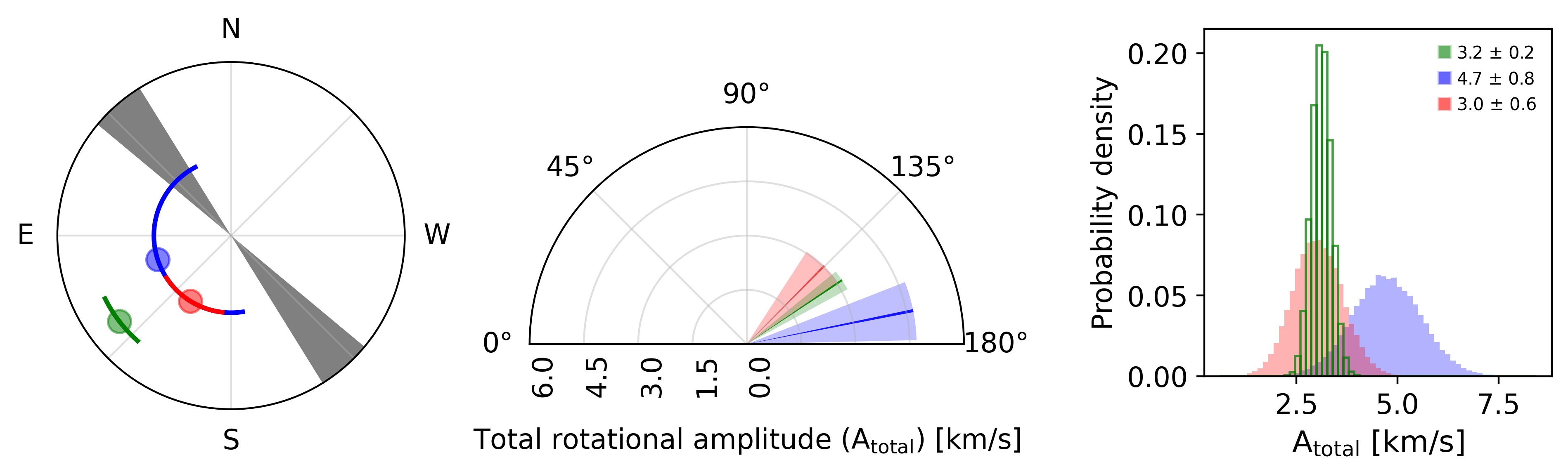}
    \caption{As in Figure \ref{fig:rotationappendix104}, but for NGC 7078.}
    \label{fig:rotationappendix7078}
\end{figure}
\begin{figure}[H]
    \centering
    \includegraphics[width=\hsize]{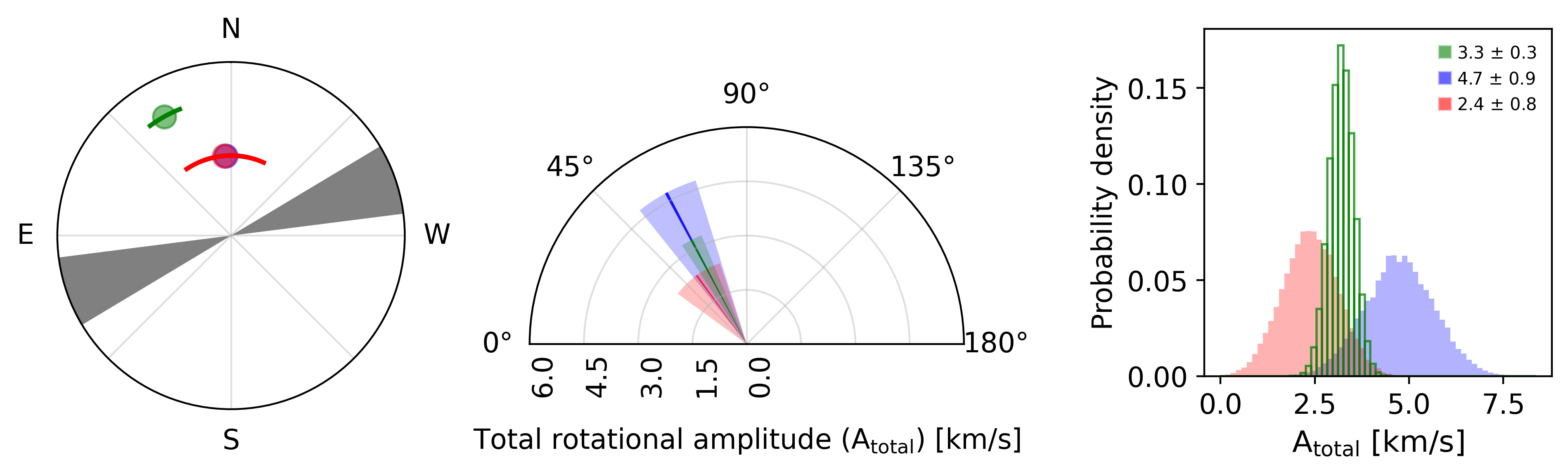}
    \caption{As in Figure \ref{fig:rotationappendix104}, but for NGC 7089.}
    \label{fig:rotationappendix7089}
\end{figure}
\newpage
\section{The combined anisotropy}
\label{sec:Appendix_dold_anisotropy}

In Section \ref{sec:vdisp}, we calculated the velocity dispersion using the combination of HACKS and \gaia DR3 proper motions. As the HACKS proper motions were measured in a relative reference system which does not allow rotation to be observed, removing the rotational contribution from the \gaia proper motions allowed the two data sets to be compatible, as in Figure \ref{fig:anisotropy_young}.\\
If the contribution by rotation is not first removed from the \gaia radial bins, there can be a discontinuity between the velocity dispersions for the HACKS radial range and the \gaia range. Due to this incompatibility, the HACKS proper motions cannot be used for the anisotropy calculation, which includes the contribution of rotation. We therefore included a second version of Figure \ref{fig:anisotropy_young} which shows the anisotropy using only the \gaia proper motions, in which rotation is not removed. Removing this bulk tangential motion as in Figure \ref{fig:anisotropy_young} reveals the intrinsic random motion of the stars, but this result complicates the physical meaning of the orbits of the stars we wished to use to investigate the correlations in Section \ref{sec:anisotropy_results}.\\ 
In the \gaia only version of the anisotropy vs radius plot (Figure \ref{fig:gaia_anisotropy_young}) the most notable differences are in the clusters NGC 104 and NGC 3201. NGC 104 exhibits the fastest rotation of all clusters in our sample, with the tangential component of the rotation especially contributing to the orbits of the stars. Under the influence of rotation, the stars move preferentially on tangential orbits as expected. There are no other clusters in our sample which switch between radial and tangential anisotropy after the rotation is removed. The next most significant change is NGC 3201, which becomes less radially anisotropic after the rotation is removed. However, in general the anisotropy of the clusters do not significantly change when including or removing the rotation, nor do the other fast rotating clusters - NGC 5904 and NGC 6656 - show a significant difference between Figure \ref{fig:anisotropy_young} and Figure \ref{fig:gaia_anisotropy_young}.\\

\begin{figure}[H]
    \centering
    \includegraphics[width=\textwidth]{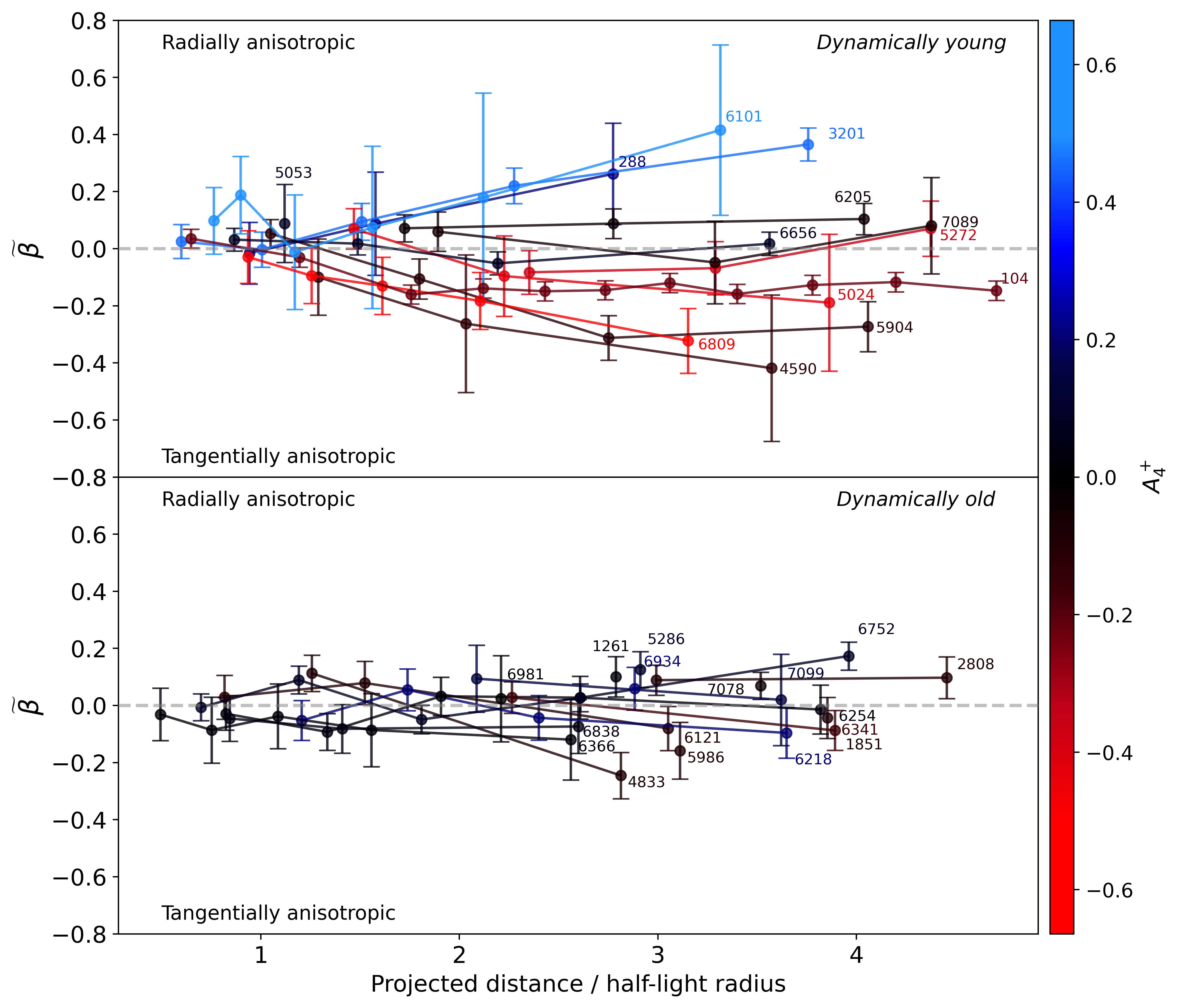}
    \vspace{-0.4cm}
    \caption{The same as Figure \ref{fig:anisotropy_young}, but using only \gaia proper motions, with no HACKS proper motions. We used the mean square velocity to calculate the anisotropy, such that the bulk motion and intrinsic random motion are both considered. For more details on this method, see the discussion above in Appendix \ref{sec:Appendix_dold_anisotropy}.}
    \label{fig:gaia_anisotropy_young}
\end{figure}

\FloatBarrier
\twocolumn
\onecolumn
\section{Tables of the photometric semi-major axis orientation and rotation parameters of the multiple stellar populations}
\label{sec:Appendix_pa}

\begin{table}[h!]
    \centering
    \caption{The estimated position angle (PA) of the photometric semi-major axis (and uncertainty) and eccentricities of the clusters in our sample, as described in Section \ref{sec:rotation_axes}.}
    \begin{tabular}{lcccc}
        \hline
        \hline\\[-0.7em]
        Cluster & PA & $e$\\
          & [deg] & \\[0.4em]
        \hline
        NGC 104  &  34 $\pm$  3 & 0.46 $\pm$ 0.03\\
        NGC 288  &  94 $\pm$ 40 & 0.31 $\pm$ 0.07\\
        NGC 1261 & -20 $\pm$ 17 & 0.57 $\pm$ 0.09\\
        NGC 1851 &  80 $\pm$ 43 & 0.27 $\pm$ 0.08\\
        NGC 2808 &  30 $\pm$  5 & 0.45 $\pm$ 0.04\\
        NGC 3201 &  88 $\pm$ 11 & 0.44 $\pm$ 0.07\\
        NGC 4590 &  45 $\pm$ 12 & 0.51 $\pm$ 0.08\\
        NGC 4833 &  60 $\pm$ 37 & 0.39 $\pm$ 0.09\\
        NGC 5024 &  35 $\pm$ 19 & 0.40 $\pm$ 0.07\\
        NGC 5053 &  64 $\pm$ 29 & 0.59 $\pm$ 0.09\\
        NGC 5272 &  64 $\pm$ 45 & 0.26 $\pm$ 0.07\\
        NGC 5286 &  48 $\pm$ 15 & 0.51 $\pm$ 0.08\\
        NGC 5904 &  56 $\pm$  4 & 0.53 $\pm$ 0.03\\
        NGC 5986 & 117 $\pm$ 40 & 0.33 $\pm$ 0.10\\
        NGC 6101 &  37 $\pm$  9 & 0.52 $\pm$ 0.07\\
        NGC 6121 &  83 $\pm$ 20 & 0.43 $\pm$ 0.10\\
        NGC 6205 & 110 $\pm$  8 & 0.46 $\pm$ 0.05\\
        NGC 6218 &  61 $\pm$ 18 & 0.40 $\pm$ 0.07\\
        NGC 6254 & -25 $\pm$ 47 & 0.30 $\pm$ 0.07\\
        NGC 6341 &  30 $\pm$  9 & 0.46 $\pm$ 0.06\\
        NGC 6366 & -10 $\pm$ 16 & 0.39 $\pm$ 0.08\\
        NGC 6656 &  34 $\pm$  7 & 0.45 $\pm$ 0.05\\
        NGC 6752 & 112 $\pm$ 40 & 0.26 $\pm$ 0.07\\
        NGC 6809 &  94 $\pm$ 38 & 0.30 $\pm$ 0.08\\
        NGC 6838 &  69 $\pm$ 15 & 0.48 $\pm$ 0.08\\
        NGC 6934 &  58 $\pm$ 19 & 0.51 $\pm$ 0.10\\
        NGC 6981 &  19 $\pm$ 33 & 0.52 $\pm$ 0.13\\
        NGC 7078 &  41 $\pm$ 18 & 0.32 $\pm$ 0.08\\
        NGC 7089 & 110 $\pm$ 24 & 0.36 $\pm$ 0.09\\
        NGC 7099 & 113 $\pm$ 14 & 0.46 $\pm$ 0.07\\
        \hline
    \end{tabular}
    \label{tab:pa}
\end{table}

\begin{landscape}
\begin{table}
    \centering
    \caption{The kinematic properties of the multiple stellar populations in the 30 Galactic globular clusters in our sample. For general column descriptions, see Table \protect\ref{tab:kinematics}. Subscripts in the column headers describe whether the value is associated with the P1 stars (blue) or the P2 stars (red).}
    \begin{tabular}{lcccccccccccc}
    \hline
    \hline\\[-0.7em]
    Cluster  & {\color[HTML]{3531FF} $\rm{\Delta A_{P1}}$} & {\color[HTML]{3531FF} $\rm{\Delta v_{tan,P1}}$} & {\color[HTML]{3531FF} $\rm{\Delta v_{rad,P1}}$} & {\color[HTML]{3531FF} $\rm{A_{total}/\sigma}_{P1}$} & {\color[HTML]{3531FF} $\theta_{0,\rm{P1}}$} & {\color[HTML]{3531FF} $i_{\rm{P1}}$} & {\color[HTML]{FE0000} $\rm{\Delta A_{P2}}$} & {\color[HTML]{FE0000} $\rm{\Delta v_{tan,P2}}$} & {\color[HTML]{FE0000} $\rm{\Delta v_{rad,P2}}$} & {\color[HTML]{FE0000} $\rm{A_{total}/\sigma}_{P2}$} & {\color[HTML]{FE0000} $\theta_{0,\rm{P2}}$} & {\color[HTML]{FE0000} $i_{\rm{P2}}$} \\
     & {\color[HTML]{3531FF} [km/s]} & {\color[HTML]{3531FF} [km/s]} & {\color[HTML]{3531FF} [km/s]} & & {\color[HTML]{3531FF} [deg]} & {\color[HTML]{3531FF} [deg]} & {\color[HTML]{FE0000} [km/s]} & {\color[HTML]{FE0000} [km/s]} & {\color[HTML]{FE0000} [km/s]} & & {\color[HTML]{FE0000} [deg]} & {\color[HTML]{FE0000} [deg]} \\[0.4em]
    \hline
NGC 104  & 2.4 $\pm$ 0.6        & 4.0 $\pm$ 0.4            & 0.0 $\pm$ 0.4            & 0.47 $\pm$ 0.05              & 128 $\pm$ 16    & 30 $\pm$ 6   & 3.9 $\pm$ 0.4        & 4.5 $\pm$ 0.2            & 0.5 $\pm$ 0.3            & 0.70 $\pm$ 0.04              & 139 $\pm$ 6     & 40 $\pm$ 3   \\
NGC 288  & 0.3 $\pm$ 0.2        & -0.2 $\pm$ 0.3           & 0.1 $\pm$ 0.3            & 0.09 $\pm$ 0.06              & 30 $\pm$ 63     & 125 $\pm$ 42 & 0.3 $\pm$ 0.3        & -0.4 $\pm$ 0.4           & 1.3 $\pm$ 0.3            & 0.13 $\pm$ 0.08              & 323 $\pm$ 83    & 140 $\pm$ 34 \\
NGC 1261 & 0.8 $\pm$ 0.6        & -0.4 $\pm$ 1.4           & 0.9 $\pm$ 1.0            & 0.10 $\pm$ 0.09              & 91 $\pm$ 66     & 115 $\pm$ 83 & 0.9 $\pm$ 0.5        & 0.6 $\pm$ 0.8            & 0.6 $\pm$ 0.8            & 0.18 $\pm$ 0.10              & 131 $\pm$ 36    & 58 $\pm$ 37  \\
NGC 1851 & 1.1 $\pm$ 0.6        & -0.4 $\pm$ 0.6           & 0.1 $\pm$ 0.7            & 0.15 $\pm$ 0.08              & 9 $\pm$ 41      & 112 $\pm$ 29 & 1.1 $\pm$ 0.5        & 0.7 $\pm$ 0.7            & -0.2 $\pm$ 0.7           & 0.16 $\pm$ 0.07              & 3 $\pm$ 34      & 58 $\pm$ 28  \\
NGC 2808 & 4.3 $\pm$ 0.8        & -0.0 $\pm$ 0.4           & 0.6 $\pm$ 0.4            & 0.37 $\pm$ 0.07              & 303 $\pm$ 10    & 90 $\pm$ 5   & 2.0 $\pm$ 0.8        & -0.7 $\pm$ 0.5           & 0.2 $\pm$ 0.5            & 0.19 $\pm$ 0.07              & 7 $\pm$ 17      & 109 $\pm$ 13 \\
NGC 3201 & 0.4 $\pm$ 0.3        & -0.2 $\pm$ 0.3           & -0.1 $\pm$ 0.3           & 0.09 $\pm$ 0.06              & 348 $\pm$ 69    & 119 $\pm$ 34 & 0.3 $\pm$ 0.3        & -0.0 $\pm$ 0.3           & -0.7 $\pm$ 0.3           & 0.06 $\pm$ 0.05              & 53 $\pm$ 96     & 94 $\pm$ 51  \\
NGC 4590 & 0.7 $\pm$ 0.5        & 0.4 $\pm$ 0.5            & 0.2 $\pm$ 0.6            & 0.19 $\pm$ 0.12              & 349 $\pm$ 56    & 62 $\pm$ 33  & 0.9 $\pm$ 0.5        & 0.2 $\pm$ 0.5            & -0.4 $\pm$ 0.4           & 0.21 $\pm$ 0.12              & 251 $\pm$ 43    & 74 $\pm$ 33  \\
NGC 4833 & 1.9 $\pm$ 1.1        & -0.0 $\pm$ 0.4           & -0.3 $\pm$ 0.3           & 0.34 $\pm$ 0.19              & 298 $\pm$ 39    & 90 $\pm$ 10  & 1.7 $\pm$ 0.8        & -0.1 $\pm$ 0.4           & 0.4 $\pm$ 0.3            & 0.31 $\pm$ 0.14              & 86 $\pm$ 34     & 93 $\pm$ 15  \\
NGC 5024 & 0.8 $\pm$ 0.5        & -0.9 $\pm$ 0.6           & 0.8 $\pm$ 0.7            & 0.19 $\pm$ 0.10              & 27 $\pm$ 55     & 139 $\pm$ 27 & 1.0 $\pm$ 0.7        & -0.0 $\pm$ 0.8           & -1.1 $\pm$ 0.8           & 0.13 $\pm$ 0.10              & 55 $\pm$ 56     & 90 $\pm$ 49  \\
NGC 5053 & -                    & -                        & -                        & -                            & -               & -            & -                    & -                        & -                        & -                            & -               & -            \\
NGC 5272 & 0.8 $\pm$ 0.6        & 2.5 $\pm$ 0.4            & -0.2 $\pm$ 0.5           & 0.36 $\pm$ 0.06              & 303 $\pm$ 58    & 18 $\pm$ 12  & 1.5 $\pm$ 0.7        & 1.0 $\pm$ 0.3            & 0.1 $\pm$ 0.4            & 0.24 $\pm$ 0.08              & 14 $\pm$ 24     & 56 $\pm$ 14  \\
NGC 5286 & 2.8 $\pm$ 0.9        & 1.2 $\pm$ 0.7            & -0.6 $\pm$ 0.8           & 0.30 $\pm$ 0.09              & 275 $\pm$ 20    & 67 $\pm$ 13  & 3.0 $\pm$ 0.7        & -0.2 $\pm$ 0.6           & -1.2 $\pm$ 0.7           & 0.30 $\pm$ 0.07              & 273 $\pm$ 14    & 93 $\pm$ 11  \\
NGC 5904 & 2.8 $\pm$ 0.4        & -2.8 $\pm$ 0.6           & 0.1 $\pm$ 0.5            & 0.54 $\pm$ 0.08              & 314 $\pm$ 11    & 134 $\pm$ 7  & 2.3 $\pm$ 0.4        & -3.1 $\pm$ 0.4           & -0.5 $\pm$ 0.5           & 0.53 $\pm$ 0.07              & 322 $\pm$ 13    & 142 $\pm$ 6  \\
NGC 5986 & 1.3 $\pm$ 1.0        & -1.6 $\pm$ 0.8           & 0.9 $\pm$ 0.8            & 0.26 $\pm$ 0.12              & 303 $\pm$ 62    & 140 $\pm$ 25 & 1.1 $\pm$ 0.9        & -1.6 $\pm$ 0.7           & 0.0 $\pm$ 0.6            & 0.21 $\pm$ 0.09              & 144 $\pm$ 69    & 145 $\pm$ 24 \\
NGC 6101 & 1.1 $\pm$ 0.6        & 1.2 $\pm$ 0.8            & -0.1 $\pm$ 0.7           & 0.21 $\pm$ 0.10              & 299 $\pm$ 45    & 42 $\pm$ 25  & 0.5 $\pm$ 0.4        & 0.5 $\pm$ 0.6            & -1.2 $\pm$ 0.6           & 0.13 $\pm$ 0.10              & 351 $\pm$ 70    & 47 $\pm$ 44  \\
NGC 6121 & 0.7 $\pm$ 0.6        & 1.5 $\pm$ 0.7            & 0.6 $\pm$ 1.3            & 0.35 $\pm$ 0.15              & 206 $\pm$ 72    & 25 $\pm$ 20  & 0.8 $\pm$ 0.5        & -0.9 $\pm$ 1.1           & 1.0 $\pm$ 1.0            & 0.22 $\pm$ 0.17              & 233 $\pm$ 55    & 139 $\pm$ 39 \\
NGC 6205 & 2.1 $\pm$ 1.1        & -0.0 $\pm$ 0.5           & 0.9 $\pm$ 0.4            & 0.25 $\pm$ 0.13              & 182 $\pm$ 38    & 91 $\pm$ 13  & 1.9 $\pm$ 0.7        & 0.5 $\pm$ 0.3            & 0.7 $\pm$ 0.3            & 0.27 $\pm$ 0.10              & 213 $\pm$ 24    & 75 $\pm$ 10  \\
NGC 6218 & 0.4 $\pm$ 0.4        & -0.8 $\pm$ 0.4           & 0.1 $\pm$ 0.3            & 0.16 $\pm$ 0.07              & 138 $\pm$ 71    & 152 $\pm$ 22 & 0.6 $\pm$ 0.3        & -0.6 $\pm$ 0.3           & -0.5 $\pm$ 0.3           & 0.16 $\pm$ 0.06              & 224 $\pm$ 43    & 135 $\pm$ 22 \\
NGC 6254 & 0.8 $\pm$ 0.5        & -0.9 $\pm$ 0.4           & 0.1 $\pm$ 0.3            & 0.16 $\pm$ 0.06              & 178 $\pm$ 50    & 136 $\pm$ 21 & 0.4 $\pm$ 0.3        & -0.6 $\pm$ 0.3           & 0.1 $\pm$ 0.3            & 0.12 $\pm$ 0.05              & 357 $\pm$ 71    & 147 $\pm$ 23 \\
NGC 6341 & 1.5 $\pm$ 0.8        & 1.0 $\pm$ 0.5            & 1.0 $\pm$ 0.6            & 0.27 $\pm$ 0.10              & 335 $\pm$ 38    & 56 $\pm$ 18  & 0.7 $\pm$ 0.6        & 1.4 $\pm$ 0.7            & -1.0 $\pm$ 0.6           & 0.20 $\pm$ 0.09              & 31 $\pm$ 80     & 25 $\pm$ 22  \\
NGC 6366 & 0.3 $\pm$ 0.3        & 0.1 $\pm$ 0.2            & 0.2 $\pm$ 0.2            & 0.11 $\pm$ 0.09              & 190 $\pm$ 92    & 80 $\pm$ 29  & 0.3 $\pm$ 0.3        & -0.1 $\pm$ 0.2           & -0.1 $\pm$ 0.2           & 0.12 $\pm$ 0.10              & 334 $\pm$ 97    & 109 $\pm$ 33 \\
NGC 6656 & 3.7 $\pm$ 1.0        & -1.7 $\pm$ 0.5           & -0.2 $\pm$ 0.6           & 0.43 $\pm$ 0.10              & 263 $\pm$ 23    & 114 $\pm$ 8  & 3.0 $\pm$ 1.1        & -2.7 $\pm$ 0.6           & -0.0 $\pm$ 0.6           & 0.43 $\pm$ 0.11              & 358 $\pm$ 25    & 131 $\pm$ 12 \\
NGC 6752 & 0.5 $\pm$ 0.5        & 1.2 $\pm$ 0.5            & 0.5 $\pm$ 0.5            & 0.21 $\pm$ 0.08              & 230 $\pm$ 86    & 24 $\pm$ 20  & 0.3 $\pm$ 0.3        & 0.4 $\pm$ 0.3            & -0.3 $\pm$ 0.3           & 0.08 $\pm$ 0.05              & 323 $\pm$ 104   & 35 $\pm$ 30  \\
NGC 6809 & 0.4 $\pm$ 0.3        & -0.9 $\pm$ 0.5           & 0.1 $\pm$ 0.4            & 0.16 $\pm$ 0.08              & 47 $\pm$ 97     & 157 $\pm$ 21 & 0.5 $\pm$ 0.4        & 0.2 $\pm$ 0.5            & -0.8 $\pm$ 0.6           & 0.09 $\pm$ 0.08              & 137 $\pm$ 68    & 71 $\pm$ 53  \\
NGC 6838 & 0.4 $\pm$ 0.3        & -0.1 $\pm$ 0.2           & -0.1 $\pm$ 0.2           & 0.14 $\pm$ 0.10              & 328 $\pm$ 70    & 107 $\pm$ 26 & 0.4 $\pm$ 0.4        & 0.2 $\pm$ 0.3            & 0.1 $\pm$ 0.2            & 0.13 $\pm$ 0.11              & 219 $\pm$ 97    & 68 $\pm$ 37  \\
NGC 6934 & -                    & -                        & -                        & -                            & -               & -            & -                    & -                        & -                        & -                            & -               & -            \\
NGC 6981 & -                    & -                        & -                        & -                            & -               & -            & -                    & -                        & -                        & -                            & -               & -            \\
NGC 7078 & 0.9 $\pm$ 0.8        & -4.6 $\pm$ 0.8           & 0.1 $\pm$ 1.1            & 0.48 $\pm$ 0.10              & 108 $\pm$ 80    & 168 $\pm$ 10 & 2.2 $\pm$ 0.8        & -2.1 $\pm$ 0.4           & 0.1 $\pm$ 0.6            & 0.38 $\pm$ 0.08              & 148 $\pm$ 25    & 134 $\pm$ 11 \\
NGC 7089 & 4.2 $\pm$ 1.0        & 2.2 $\pm$ 0.9            & 0.5 $\pm$ 0.8            & 0.49 $\pm$ 0.11              & 3 $\pm$ 14      & 62 $\pm$ 10  & 1.9 $\pm$ 0.8        & 1.4 $\pm$ 0.7            & -0.3 $\pm$ 0.8           & 0.22 $\pm$ 0.07              & 4 $\pm$ 28      & 53 $\pm$ 17  \\
NGC 7099 & 0.7 $\pm$ 0.5        & -1.0 $\pm$ 0.8           & -0.4 $\pm$ 0.9           & 0.19 $\pm$ 0.12              & 305 $\pm$ 59    & 145 $\pm$ 29 & 1.0 $\pm$ 0.6        & 0.5 $\pm$ 0.6            & 0.0 $\pm$ 0.6            & 0.19 $\pm$ 0.10              & 300 $\pm$ 42    & 66 $\pm$ 29 \\
    \hline
    \end{tabular}
\label{tab:kinematics_MPs}
\end{table}
\end{landscape}

}
\end{appendix}

\end{document}